\documentclass[acmtog, authorversion]{acmart}

\usepackage{booktabs} 
\usepackage{subcaption}

\usepackage{dblfloatfix}
\usepackage{color}
\usepackage{algorithmicx}

\algnewcommand{\LineComment}[1]{\State{\emph{// #1}}}

\newcommand{\algspace}[0]{\vspace{.05in}}

\algrenewcommand\alglinenumber[1]{
	{\sf\footnotesize\addfontfeatures{Colour=888888,Numbers=Monospaced}#1}}

\newcommand{\eqdef}{:=}
\newcommand{\source}[0]{{M_1}}
\newcommand{\target}[0]{{M_2}}
\newcommand{\drm}{\mathrm{d}}

\newcommand{\R}[0]{\mathbb{R}}
\newcommand{\xR}{\mathbb{R}}
\newcommand{\mV}{\mathcal{V}}
\newcommand{\mE}{\mathcal{E}}
\newcommand{\mF}{\mathcal{F}}
\newcommand{\mP}{\mathcal{P}}

\newcommand{\massV}{A}

\newcommand{\nos}[1]{\lvert\mathcal{#1}\rvert}

\newcommand{\Tr}{\mathrm{tr}}

\setcitestyle{square}

\usepackage[ruled]{algorithm2e} 

\SetAlFnt{\small}
\SetAlCapFnt{\small}
\SetAlCapNameFnt{\small}
\SetAlCapHSkip{0pt}
\IncMargin{-\parindent}

\acmJournal{TOG}
\acmVolume{0}
\acmNumber{0}
\acmArticle{0}
\acmYear{20XX}
\acmMonth{0}

\acmDOI{0000001.0000001_2}

\begin{document}
\title{Reversible Harmonic Maps between Discrete Surfaces} 

\author{Danielle Ezuz}
\affiliation{%
  \institution{Technion - Israel Institute of Technology}}
\email{dandan@cs.technion.ac.il}
\author{Justin Solomon}
\affiliation{%
  \institution{MIT}}
\author{Mirela Ben-Chen} 
\affiliation{%
  \institution{Technion - Israel Institute of Technology}}

\begin{abstract}
Information transfer between triangle meshes is of great importance in computer graphics and geometry processing. To facilitate this process, a \emph{smooth and accurate map} is typically required between the two meshes. 
While such maps can sometimes be computed between nearly-isometric meshes, the more general case of meshes with diverse geometries remains challenging. We propose a novel approach for \emph{direct} map computation between triangle meshes without mapping to an intermediate domain, which optimizes for the \emph{harmonicity} and \emph{reversibility} of the forward and backward maps. Our method is general both in the information it can receive as input, e.g.\ point landmarks, a dense map or a functional map, and in the diversity of the geometries to which it can be applied. 
We demonstrate that our maps exhibit lower conformal distortion than the state-of-the-art, while succeeding in correctly mapping key features of the input shapes. 
\end{abstract}

%
%
%

%
%

\maketitle


\newtheorem{prop}{Proposition}

\section{Introduction}

Mapping 3D shapes to one another is a basic task in computer graphics and geometry processing. Correspondence is needed, for example, to transfer artist-generated assets such as texture and pose from one mesh to another~\cite{sumner2004deformation}, to compute in-between shapes using shape interpolation~\cite{heeren2012time,von2015real}, and to carry out statistical shape analysis~\cite{munsell2008evaluating}. In these applications, desirable correspondences satisfy some basic key properties: They should be \emph{smooth} to avoid introducing geometric noise during transfer; they should \emph{preserve semantic features} to ensure that key features are put in correspondence; and they should be \emph{reversible}, namely invariant to which of the two shapes is chosen as the source. 

Many exiting approaches to shape mapping focus on generating maps with low \emph{global} distortion (e.g. preserving pairwise distances~\cite{sahillioǧlu2011coarse}) at the expense of large local distortion, which reduces the quality of the correspondence and hinders downstream applications. 
On the other hand, approaches that minimize local distortion measures mostly require an intermediate domain and construct the final map as a composition through this domain (e.g.~\cite{aigerman2016hyperbolic}). 
While such methods minimize distortion of the maps into the intermediate domain, the distortion of the composed map can be large. This problem is exacerbated when the input shapes have significantly different geometric features, such as four-legged animals with different dimensions, e.g.\ a cat and a giraffe. In this case, the \emph{isometric distortion} of the optimal map is expected to be large, and thus minimizing the distortion of the two maps into an intermediate domain is quite different from minimizing the distortion of the composition.

We propose a novel approach for computing a smooth and reversible map between surfaces that are not isometric to each other, without requiring an intermediate domain. We incorporate semantic information by starting from some user guidance given in the form of sparse landmark constraints or a functional correspondence. Our main contribution is the formulation of an optimization problem whose objective is to minimize the \emph{geodesic Dirichlet energy} of the forward and backward maps, while maximizing their reversibility. We compute an approximate solution to this problem using a high-dimensional Euclidean embedding and an optimization technique known as \emph{half-quadratic splitting}~\cite{geman1995nonlinear}. 
We demonstrate that our maps have considerably lower local distortion than those from state-of-the-art methods for the difficult case of non-isometric deformations. We further show that our maps are semantically accurate by measuring their adherence to self-symmetries of the input shapes, their agreement with ground-truth when the deformation is known, and their compatibility with human-generated segmentations. 
\subsection{Related Work}
The shape correspondence literature is vast, and we refer the reader to recent surveys for a thorough review of the state-of-the-art in geometry-driven~\cite{van2011survey,tam2013registration} and data-driven~\cite{xu2016data} shape correspondence. We will focus our related work overview on methods for computing maps between triangular meshes that can handle shapes that are far from being related by an isometric deformation. In this realm, we characterize methods by the type of input they require and the type of output they generate. We therefore distinguish between \emph{vertex-to-vertex} maps that yield a correspondence between the source and target vertices of the triangular meshes, \emph{precise} maps that map every vertex on the source to a \emph{point} on the triangulated surface of the target, and \emph{generalized} maps that put in correspondence functions or probability distributions. 

\paragraph*{Fully-automatic methods.}
\citet{kim2011blended} suggested one of the first fully automatic methods for non-isometric shape matching, that consistently generated high-quality outputs on a benchmark of shapes. This approach, denoted by BIM, generates a \emph{precise} map as a blending of conformal maps, with blending weights optimized to minimize isometric distortion. While providing excellent results in many cases, BIM is limited to genus-zero surfaces and can introduce large distortions for some shapes. Recently, ~\citet{zheng2017surface} suggested to map between high-genus surfaces with the same genus, by decomposing the surfaces using a pants decomposition, and then computing harmonic maps between a set of intermediate cylindrical domains. This leads to a piecewise harmonic map between the input surfaces, which is further relaxed using geodesic heat flow. While this approach can be used without user intervention, if a globally semantic map is needed then accurate input landmarks are required, and it is limited to shapes of the same genus. Finally,~\citet{LahnerVBLSRRBBK17} suggested a method that computes \emph{vertex-to-vertex} maps based on a set of matching descriptors and pairwise distances. While the method is robust to topological changes, it may leave areas unmapped, and is therefore less appropriate for applications such as texture transfer.

The functional map approach, introduced by Ovsjanikov et al.\ \shortcite{ovsjanikov2012functional}, was originally designed for nearly-isometric shapes but has since been extended to non-isometric matching~\cite{kovnatsky2013coupled,kovnatsky2015functional,burghard2017embedding}. This method generates a generalized map which puts in correspondence the function spaces on the mapped shapes.~\citet{nogneng2017informative} recently suggested a method that computes functional maps using fewer input descriptors by formulating commutativity constraints, and~\citet{huang2017adjoint} suggested to use \emph{adjoint} functional maps to improve and analyze correspondences.
Using a hybrid approach, \citet{maron2016point} optimize jointly for a functional map and sparse correspondences, which are then extended to a dense vertex-to-vertex map. Finally, \citet{solomon2016entropic} compute a generalized map, which puts in correspondence probability distributions on the input shapes, by minimizing the Gromov--Wasserstein distance between the shapes. 

While generalized maps are beneficial for challenging mapping problems, such as mapping between shapes of different genus, extracting a precise map is a necessary post-processing step if the output map is to be used for transferring high-frequency data such as textures, normals, or deformations. Furthermore, when the shapes are geometrically different and the optimal map is not expected to be isometric, the shape correspondence problem is ill-posed without additional \emph{semantic} information. Such information can be given in the form of landmark constraints or an initial generalized map, from which a refined dense map can be computed. 

\paragraph*{Input: landmarks.}
Parameterization-based approaches compute bijective smooth maps to a common intermediate domain, and define precise maps between arbitrary shapes as the composition of the maps to the common domain. A variety of intermediate domains have been used in the literature, e.g. the plane~\cite{aigerman2015seamless}, the sphere~\cite{gu2004genus}, the hyperbolic disk~\cite{shi2017hyperbolic} and orbifolds~\cite{tsui2013globally,aigerman2015orbifold,aigerman2016hyperbolic,aigerman2017spherical}, to mention a few. These methods optimize the distortion of the map from the shapes to the target domain, but the composed map is not guaranteed in general to have low distortion. Furthermore, mapping through an intermediate domain places a topological restriction on the type of mapped shapes, as they should be topologically equivalent.
Alternatively, \citet{panozzo2013weighted} compute a direct map between two triangle meshes without requiring an intermediate domain. Their method computes on-surface barycentric coordinates with respect to the source landmarks, and then uses them with respect to the target landmarks to compute the corresponding point on the target shape. Similar to our approach, they use a high-dimensional Euclidean embedding to speed up the computation. Despite excellent results, in some cases a large number of landmarks is required to compute a correct correspondence.
More recently, \citet{mandad2017variance} use landmarks or extrinsic alignment for initialization and then optimize simultaneously for a generalized map and a precise map. We compare with their approach and demonstrate that our maps achieve better conformal distortion and adherence to the shape semantics.

\paragraph*{Input: generalized maps.}
Several methods for recovery of vertex-to-vertex maps from generalized maps have been suggested.~\citet{shtern2014iterative} suggest a refinement technique that is based on heat kernel alignment, \citet{rodola2015point} use a probabilistic model, and~\citet{vestner2017pmf} solve a linear assignment problem. In general, vertex-to-vertex map have higher conformal distortion than precise maps. Alternatively, \citet{ezuz2017deblurring} suggest a pointwise recovery method that generates \emph{precise} maps using a smoothness prior, based on a spectral approach. Our method does not rely on spectral representations, and thus exhibits fewer artifacts in complex cases.

\subsection{Contributions}
We present an algorithm for shape correspondence between non-isometric triangular meshes, that has the following advantages:
\begin{itemize}
	\item The algorithm is widely applicable, and the resulting maps are semantic and exhibit low conformal distortion.
	\item The formulation is simple and efficient to optimize and thus can be combined with additional energy terms and various initializations.
	\item The maps are accurate enough for downstream applications, such as shape interpolation and quad mesh transfer.
\end{itemize}

\section{Background: Harmonic Maps and Local Distortion}

Suppose $\source,\target\subseteq\R^3$ are smooth, compact surfaces with or without boundary.  Given a map $\phi_{12}\!:\!\source\!\rightarrow\!\target$ from one into another, a natural task is to measure the distortion of $\source$ as it is mapped via $\phi_{12}$ onto $\target$; this distortion measure eventually will serve as an objective function for optimization problems whose unknown is the correspondence $\phi_{12}$.  The basic role of these distortion measures is to evaluate whether nearby points are mapped to nearby points under $\phi_{12}$, at least differentially, a common proxy for the quality of the map.

In the theory of differential geometry, a key distortion measure is the \emph{Dirichlet energy} $E[\cdot]$ (defined below) of $\phi_{12}$; minimizers of $E[\cdot]$ are called \emph{harmonic maps}.  Intuitively, if we think of $\source$ as a rubber sheet, a harmonic map represents an equilibrium position of the sheet after stretching it over $\target$ and letting it compress.  The Dirichlet energy and its minimizers find many roles in the geometry processing literature, most prominently in surface parameterization~\cite{levy2002least}, due to its intuitive measurement of distortion and connections to notions of conformality.  At the same time, theory and practice of harmonic mapping become considerably more challenging when $\target$ has areas of positive curvature; intuitively these can cause the rubber sheet to slip or bunch, yielding singularities in gradient flow procedures designed to uncover harmonic maps.

\begin{figure}[b]
	\centering
	\begin{subfigure}[b]{.19\linewidth}
		\centering
		\includegraphics[width=\textwidth]{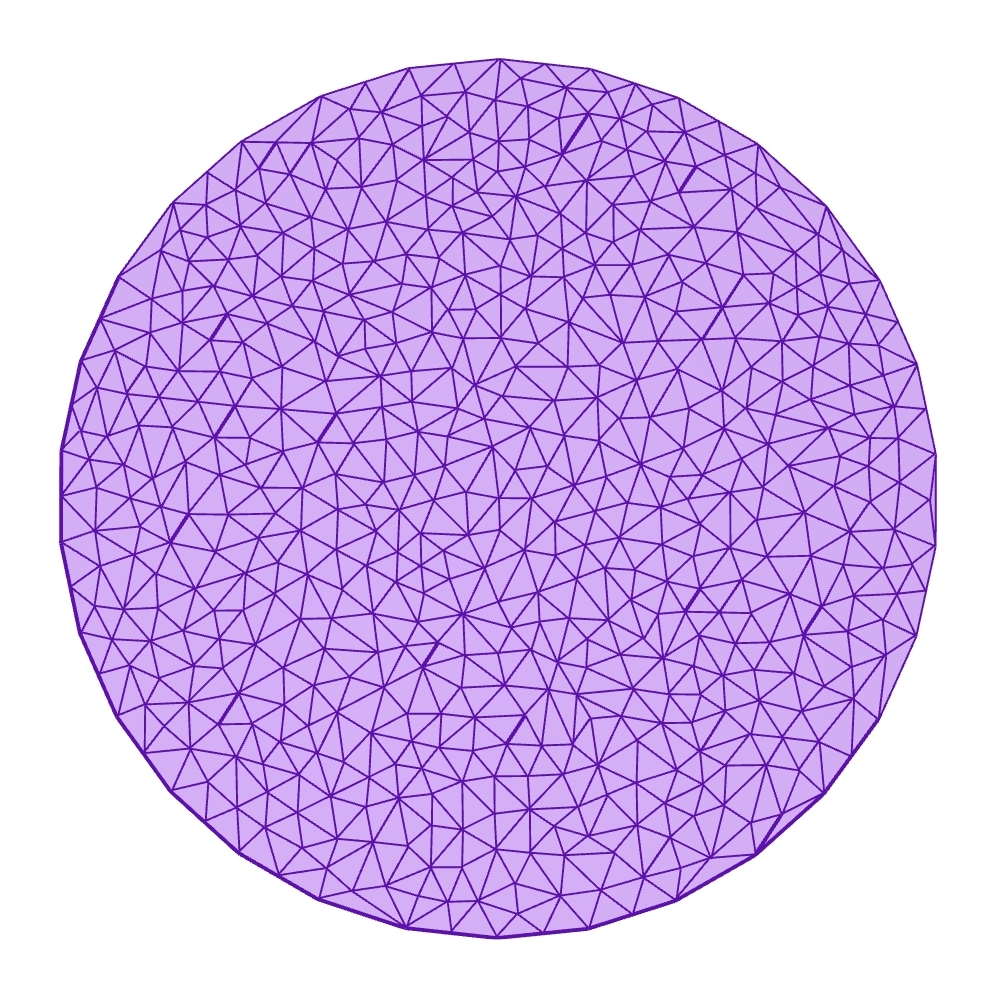}
		\caption{}
	\end{subfigure}
	\begin{subfigure}[b]{.19\linewidth}
		\centering
		\includegraphics[width=\textwidth]{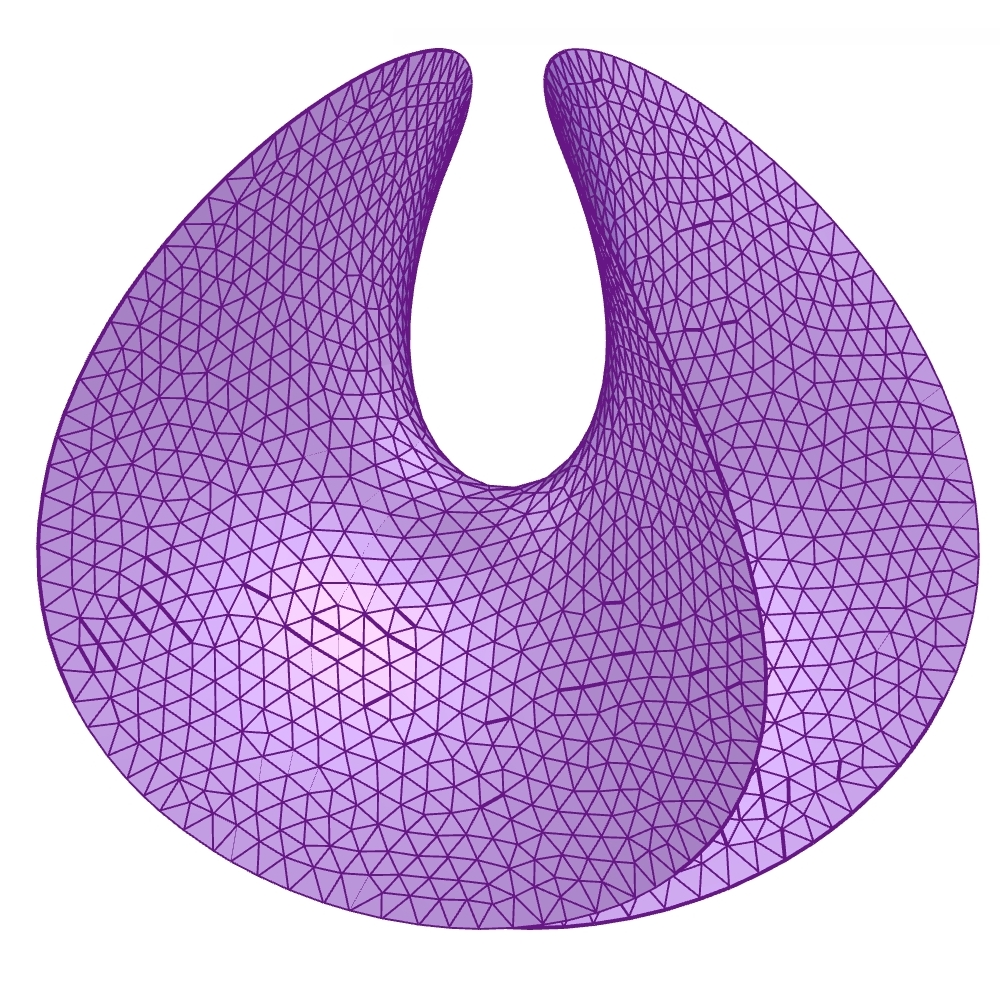}
		\caption{}
	\end{subfigure}
	\begin{subfigure}[b]{.19\linewidth}
		\centering
		\includegraphics[width=\textwidth]{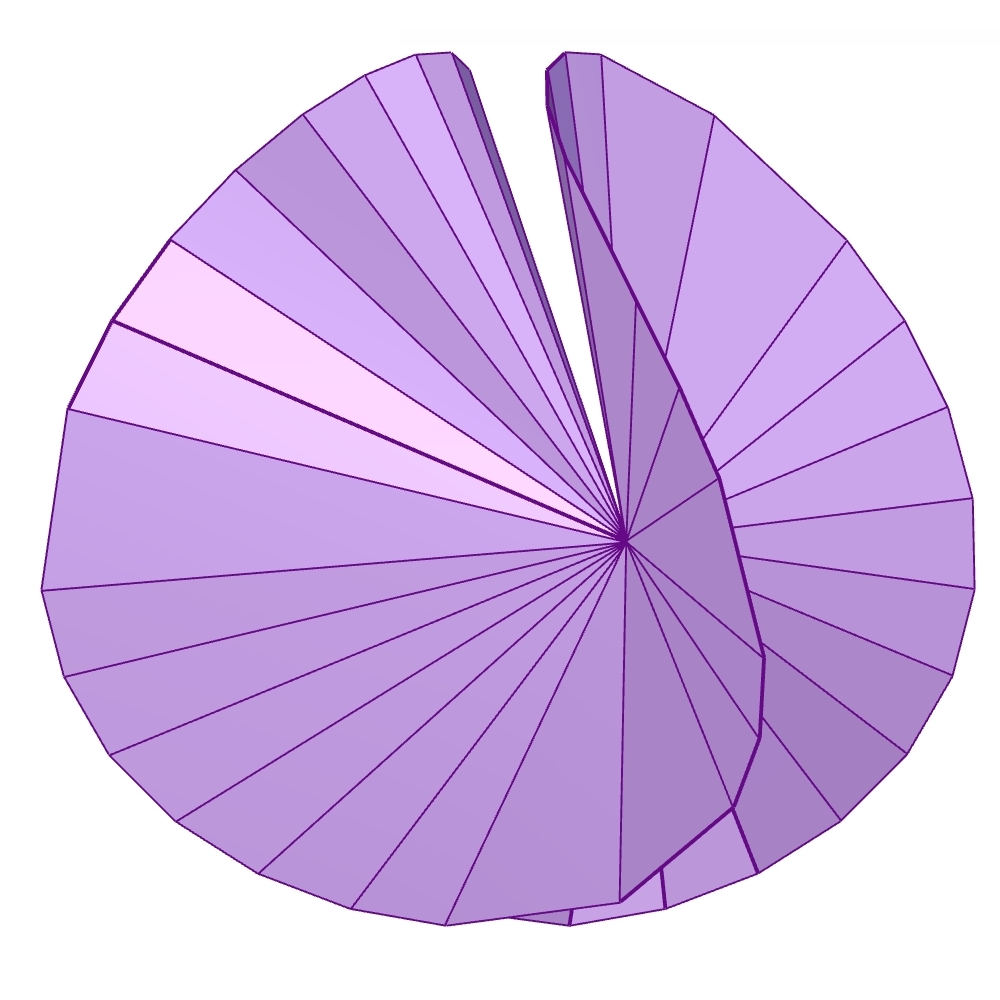}
		\caption{}
	\end{subfigure}
	\begin{subfigure}[b]{.19\linewidth}
		\centering
		\includegraphics[width=\textwidth]{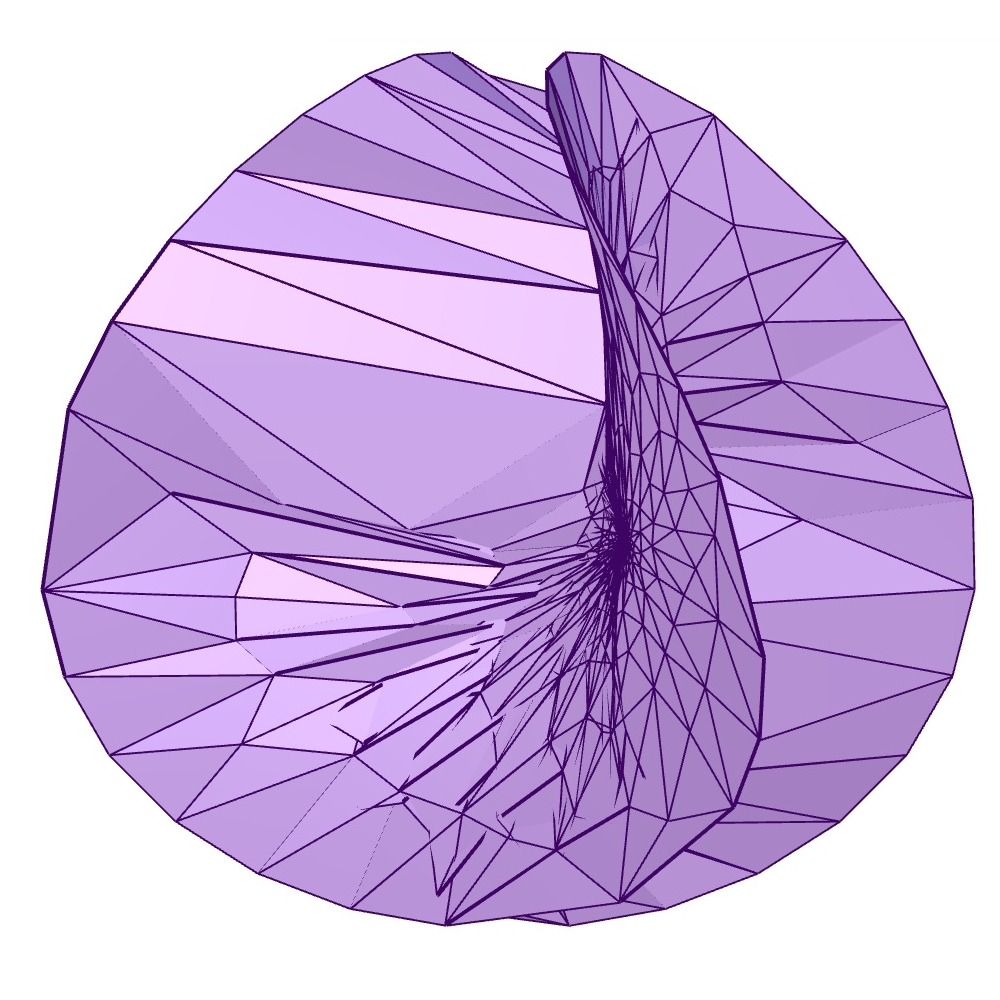}
		\caption{}
	\end{subfigure}
	\begin{subfigure}[b]{.19\linewidth}
		\centering
		\includegraphics[width=\textwidth]{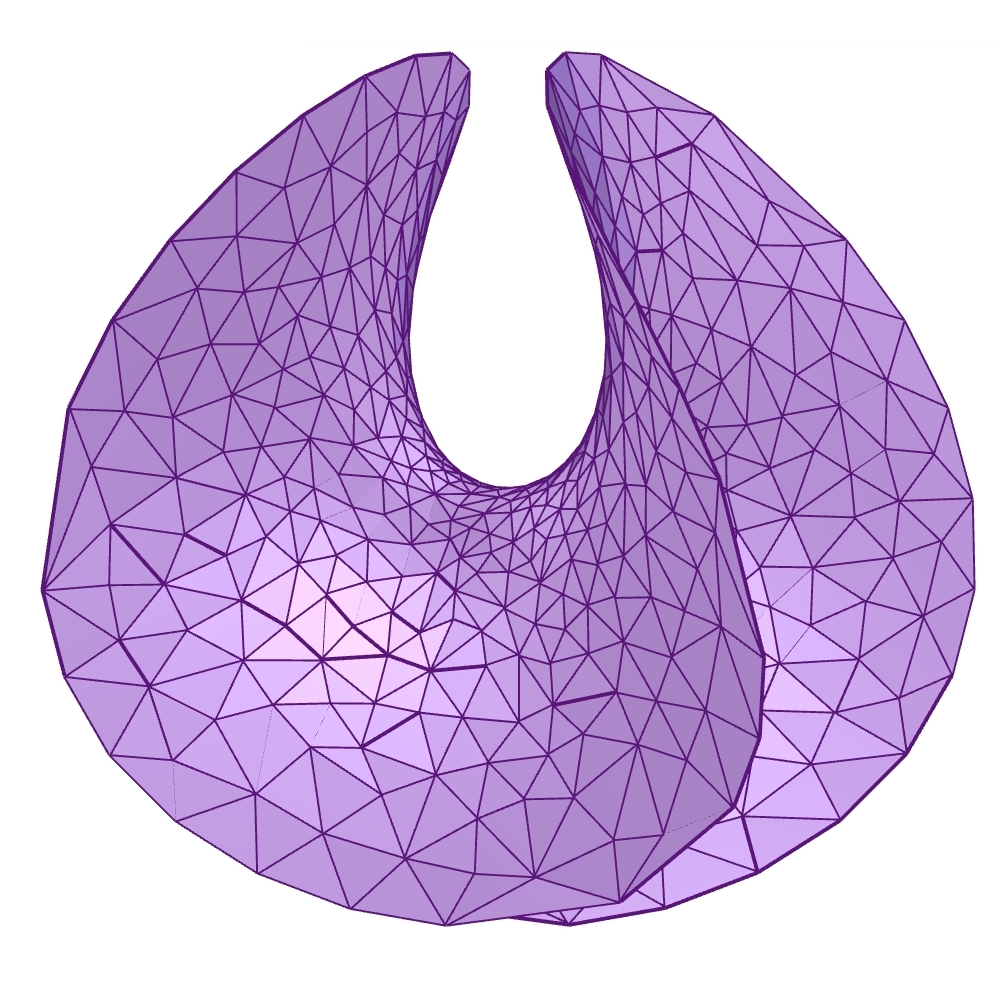}
		\caption{}
	\end{subfigure}
	\caption{Limitations of the piecewise linear discretization of the Dirichlet energy. (a) Source $\source$: a flat disk embedded in $\R^3$. (b) Target $\target$: enneper. (c) Initial piecewise linear map. (d,e) Final maps that minimize the energies in Eqs.~\eqref{eq:energy_eck}. and Eq.~\eqref{eq:energy_izeki}, respectively. See the text for details.} 
	\label{fig:enneper}
\end{figure}

In this section, we describe the basic construction of the Dirichlet energy and point out its advantages and flaws in the context of surface-to-surface correspondence; we also provide basic constructions for approximating the Dirichlet energy of a map between discrete surfaces.  In \S\ref{sec:construction}, we then propose a modified notion of harmonicity designed to avoid singularities and asymmetry in the surface-to-surface correspondence pipeline.

\subsection{Smooth Surfaces}
Following~\cite{urakawa-1993,nishikawa-2000}, harmonic maps between smooth surfaces are defined as the critical points of the \emph{Dirichlet energy}:
\begin{equation}
E\left[\phi_{12}\right]\eqdef \frac{1}{2}\int_\source |\drm\phi_{12}|^2\,dv_1,
\label{eq:energy_smooth}
\end{equation}
where $\drm\phi_{12}$ is the \emph{map differential} and $dv_1$ is the volume element of $\source$. 
$E[\phi_{12}]$ measures the total stretch of $\source$ after it is warped onto $\target$, as measured by the integrated norm of the Jacobian $\drm\phi_{12}$. 
Formally, given an orthonormal basis $\{e_1,e_2\}$ for $T_p \source$ at $p\!\in\!\source$, the integrand can be expanded as 
$$|\drm\phi_{12}|^2=\sum_{i=1}^2 \langle\drm\phi_{12}(e_i),\drm\phi_{12}(e_i)\rangle_{g_2(\phi_{12}(p))},$$
where $g_2$ is the metric of $\target$. 

Existence, uniqueness, and regularity of harmonic maps given assumptions on the geometry/topology of $\source$ and $\target$ as well as the homotopy class of $\phi_{12}$ is a key theme in the twentieth-century differential geometry literature.  A landmark paper by Eells and Sampson~\shortcite{eells1964harmonic} proves existence of a harmonic map in each homotopy class under the assumption that $\target$ has non-positive curvature.  The proof technique in this paper is attractive from a computational perspective:  Essentially they start with an arbitrary map in the prescribed homotopy class and use an analog of gradient descent to decrease the Dirichlet energy.

A key drawback of the Eells and Sampson proof technique from a computational perspective, however, highlights an issue with harmonic correspondence in the context of algorithmic mapping between surfaces.  In particular, their gradient descent procedure can fail when the target $\target$ has regions of positive curvature.  Roughly, this singular behavior is explained by the fact that the objective $|\drm\phi_{12}|^2$ is minimized globally by $\drm\phi_{12}\equiv0$, the constant map!  This observation highlights the difference between harmonic mapping and elastic models like the ones proposed in~\cite{sorkine2007rigid,chao2010simple}, which seek $\drm\phi_{12}$ to be close to a rotation matrix rather than to the zero matrix.  We will address this issue in our ``reversible harmonic'' formulation by adding the Dirichlet energy of $\phi_{12}^{-1}$ for the case of diffeomorphic correspondence; this has the added benefit of making forward maps $\phi_{12}$ and reverse maps $\phi_{21}$ critical points of the same objective function.

\subsection{Triangle Meshes}
For surfaces that are discretized as triangle meshes $M_i$, represented by their vertex, edge and face sets $(\mV_i,\mE_i,\mF_i)$, a pointwise vertex map $\phi_{12}$ assigns a point on a face of $\target$ to each vertex of $M_1$. The extension of the vertex map to the interior of faces of $\source$ determines the corresponding Dirichlet energy. 

If the map is assumed to be \emph{affine} on every face $f\in \mF _1$, then the Dirichlet energy is given by 
\begin{equation}
E\left[\phi_{12}\right] = \frac{1}{2}\sum_{f\in \mF _1} |\drm\phi_{12} \left( f \right)|^2 _2 a_f,
\end{equation} 
where $\drm \phi_{12}\left( f \right)\!\in\!\mathbb{R}^{2\times 2}$ is the unique linear transformation between $f$ and its image triangle $\phi_{12}(f)$, and $a_f$ is the area of $f$~\cite{pinkall1993computing}. 
Equivalently, the energy can also be written as:
\begin{equation}
E\left[\phi_{12}\right]\eqdef \frac{1}{4}\sum_{\left( u,v \right) \in \mE_1 }\! w_{uv} \, \| \phi_{12} \left( u \right) - \phi_{12} \left( v \right) \|_2 ^2,
\label{eq:energy_eck}
\end{equation}
where $w_{uv}$ is the \emph{cotangent weight} of the edge $\left( u,v \right)$. This energy is convex and quadratic in the images of the vertices of $\source$, and is therefore straightforward to minimize efficiently when $\phi_{12}$ is unrestricted, e.g.\ for planar parameterization~\cite{levy2002least}.

When $M_2$ is a non-Euclidean space, $\phi_{12}$ should be restricted to lie on $\target$, leading to a constrained optimization problem that is harder to solve. In addition, a more serious issue is the linearity assumption itself. When $\phi_{12}$ does not sample the target surface $\target$ well, the linear extension of $\phi_{12}(\mV_1)$ can be far from $\target$. In this case, minimizing the Dirichlet energy of the piecewise-affine map can lead to incorrect results.

Consider for example, as in Figure~\ref{fig:enneper}, mapping a disk $\source$ (a) to an enneper surface $\target$ (b) with Dirichlet boundary conditions; since the target has negative curvature, in the smooth case~\cite{eells1964harmonic} gradient flow will reach a harmonic map $\phi_{12}:\source\rightarrow\target$. 
An initial map (c) maps all the interior vertices of $\source$ to a single interior vertex on $\target$, and the boundary of the disk is mapped to the boundary of the enneper. 
Minimizing Eq.~\eqref{eq:energy_eck} using gradient descent, the analog of Eells \& Sampson's heat flow, with the side constraint that $\phi_{12}(\mV_1)$ is restricted to lie on $\target$, leads to a map (d) that is clearly not smooth. 
Effectively, Eq.~\eqref{eq:energy_eck} aims to place the image of every vertex of $\source$ in the Euclidean weighted average of the image of its neighbors. When the affine map samples the target poorly, this strategy fails to generate an approximation of a smooth map.

Alternatively,~\citet{izeki2005combinatorial} suggest an intrinsic formulation, the \emph{geodesic harmonic energy}, which replaces the Euclidean distances in Eq.~\eqref{eq:energy_eck} with the geodesic distances $d_{M_2}\!\left(\cdot,\cdot\right)$, as follows:
\begin{equation}
E_D\left[\phi_{12}\right]\eqdef \sum_{\left( u,v \right) \in \mE_1 }\! w_{uv}\, d_{M_2}^2\!\left( \phi_{12}\left( u \right), \phi_{12}\left( v \right) \right).
\label{eq:energy_izeki}
\end{equation}
As shown in Figure \ref{fig:enneper}(e), minimizing this energy instead of the Euclidean one yields a significantly better result at the cost of having to compute geodesic distances. 
Motivated by this idea, we propose to use this energy as the main building block in a shape mapping algorithm. We reformulate it to allow efficient optimization and combination with other terms that address the case of positively-curved target surfaces $\target$, as described in the following section.

\section{Reversible Harmonic Maps}\label{sec:construction}

\paragraph*{Notation.}
We represent a triangle mesh $M$ by its vertex, edge and face sets $(\mV,\mE,\mF)$, respectively,  where we denote $n = \nos{V}$, and its given embedding by $V\!\in \!\xR^{n \times 3}$.
We denote scalar functions $g\!:\!M \!\to \!\xR$ by a vector of coefficients of piecewise linear hat functions, with $g\! \in\! \xR^n$. The squared $l_2$ norm of a function on the surface is given by $\|g\|^2_M = g^T \massV g$, where $\massV\! \in\! \xR^{n \times n}$ is the diagonal (lumped) mass matrix of the vertices. The total area of the mesh is denoted by $s = \Tr(A)$. The squared gradient norm is given by $\|g\|^2_W = g^T W g$, where $W$ is the matrix of cotangent weights. Similarly, for matrices $G \in \xR^{n \times k}$ whose columns are scalar functions, we use the matrix trace: $\|G\|_M^2 = \Tr(G^T \massV G)$, $\|G\|_W^2 = \Tr(G^T W G)$. 
When two meshes are involved we use a subscript, e.g. $\massV_i$ is the mass matrix of $M_i$. 	

\begin{figure}[b!]
	\centering
	\begin{subfigure}[b]{.24\linewidth}
		\centering
		\includegraphics[width=\textwidth]{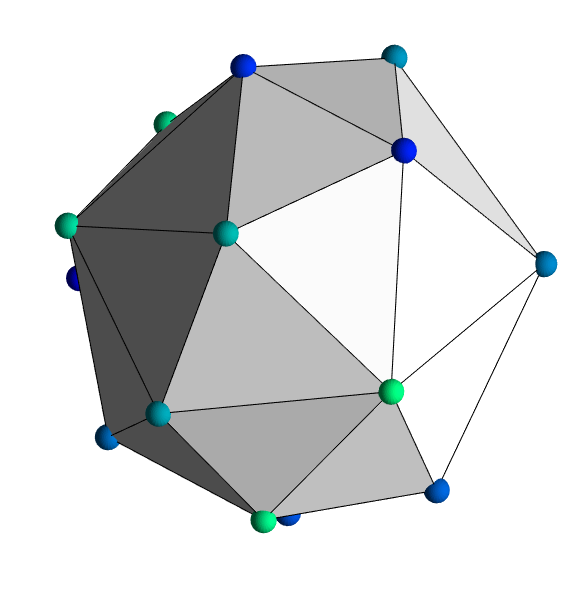}
		\caption{}
	\end{subfigure}	
	\begin{subfigure}[b]{.24\linewidth}
		\centering
		\includegraphics[width=\textwidth]{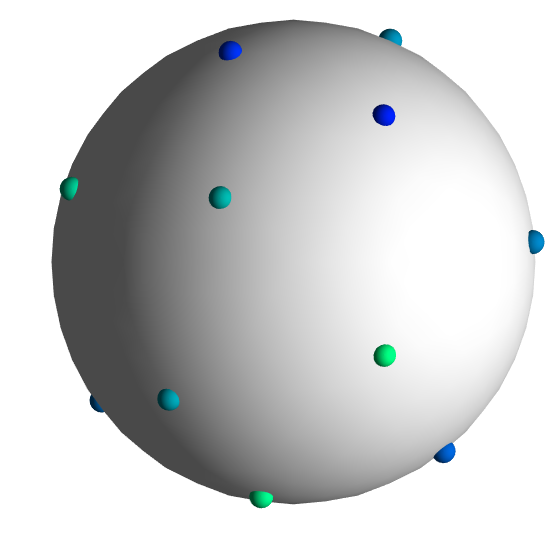}
		\caption{}
	\end{subfigure}
	\begin{subfigure}[b]{.24\linewidth}
		\centering
		\includegraphics[width=\textwidth]{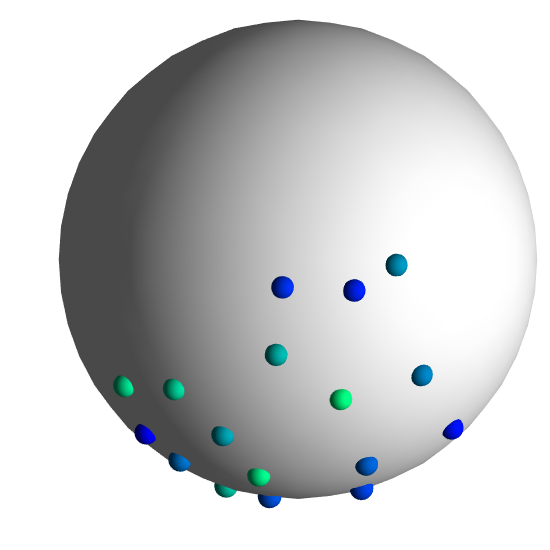}
		\caption{}
	\end{subfigure}		
	\begin{subfigure}[b]{.24\linewidth}
		\centering
		\includegraphics[width=\textwidth]{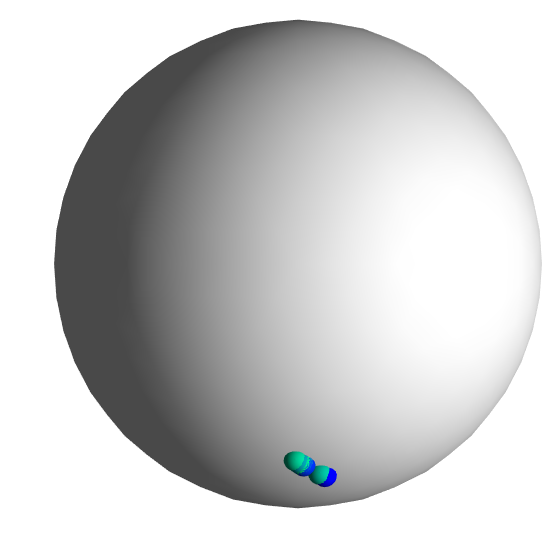}
		\caption{}
	\end{subfigure}	
	\caption{Collapse of a harmonic map. Mapping from a low resolution sphere (a) to a high resolution sphere, starting from the ground truth map (b). The map quickly ``slides'' to a single hemisphere (c) and then degenerates (d).} 
	\label{fig:collapse_sphere}
\end{figure}
\subsection{Energy}
While the geodesic harmonic energy can be effective, harmonic maps in general can become degenerate and map large regions to a single point. 
Indeed, the map taking all the points on $\source$ to a single point on $\target$ is harmonic. An example of such behavior is demonstrated in Figure~\ref{fig:collapse_sphere}. Here, the source $\source$ is a sphere with a small number of triangles (a), that is mapped to a high-resolution target sphere $\target$.
Even if the initial map is the ground-truth map (b) between the spheres, during the optimization the map quickly ``slides over'' to a single hemisphere of the target sphere (c), and then degenerates and collapses to a single point (d). Intuitively, in the discrete case, we can think of $M_1$ as an elastic fishnet instead of a continuous rubber sheet, stretched over $M_2$ and allowed to compress. Then, in addition to the usual degeneracies in the smooth case, the target surface can effectively ``slip through'' one of the holes in the net, allowing the map to degenerate to a single point.

To prevent this, we minimize the geodesic harmonic energies of both the \emph{forward} and \emph{backward} maps, together with a \emph{reversibility} constraint relating both maps. As we later demonstrate, this approach is highly effective in generating non-degenerate harmonic maps.

\paragraph{Smoothness.} 
Given two triangle meshes $M_1,M_2$, and maps $\phi_{12},\phi_{21}$, the total harmonic energy of both forward and backward maps is given by
\begin{equation}
E_D\left[\phi_{12},\phi_{21}\right] = \sum_{\substack{i,j\in\{1,2\}\\ i\neq j}}
\frac{1}{s_i}E_D\left[\phi_{ij}\right],
\label{eq:harmonic_energy}
\end{equation}
where $E_D$ is given in Equation~\eqref{eq:energy_izeki}.

\begin{figure}[t]
	\centering
	
	\begin{subfigure}[b]{.22\linewidth}
		\centering
		\includegraphics[width=\textwidth]{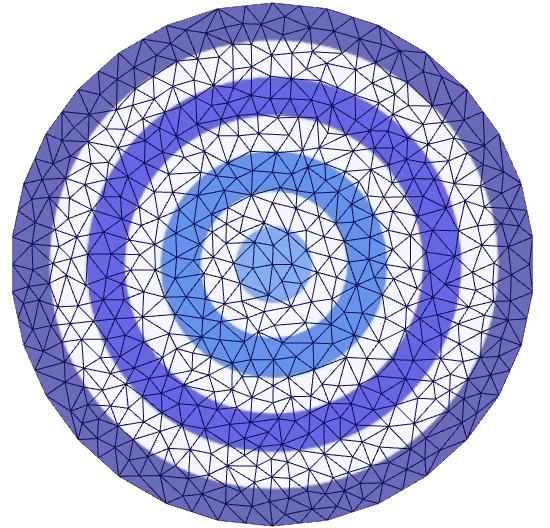}
		\caption{}
	\end{subfigure}
	\begin{subfigure}[b]{.24\linewidth}
		\centering
		\includegraphics[width=\textwidth]{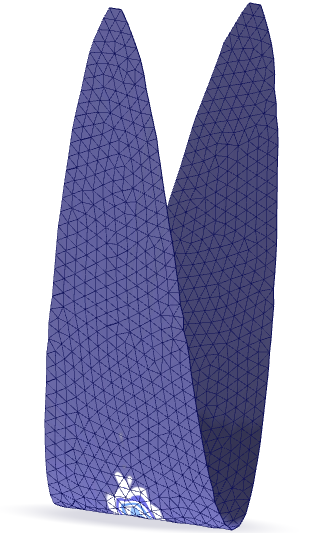}
		\caption{}
	\end{subfigure}	
	\begin{subfigure}[b]{.24\linewidth}
		\centering
		\includegraphics[width=\textwidth]{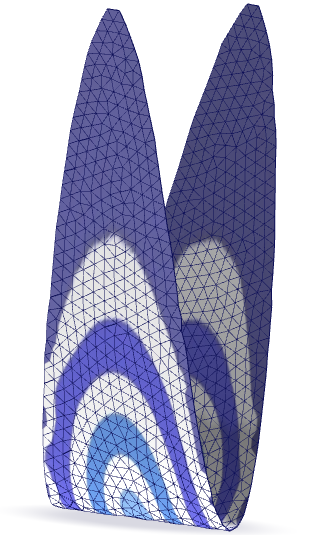}
		\caption{}
	\end{subfigure}			
	\begin{subfigure}[b]{.27\linewidth}
		\centering
		\includegraphics[width=\textwidth]{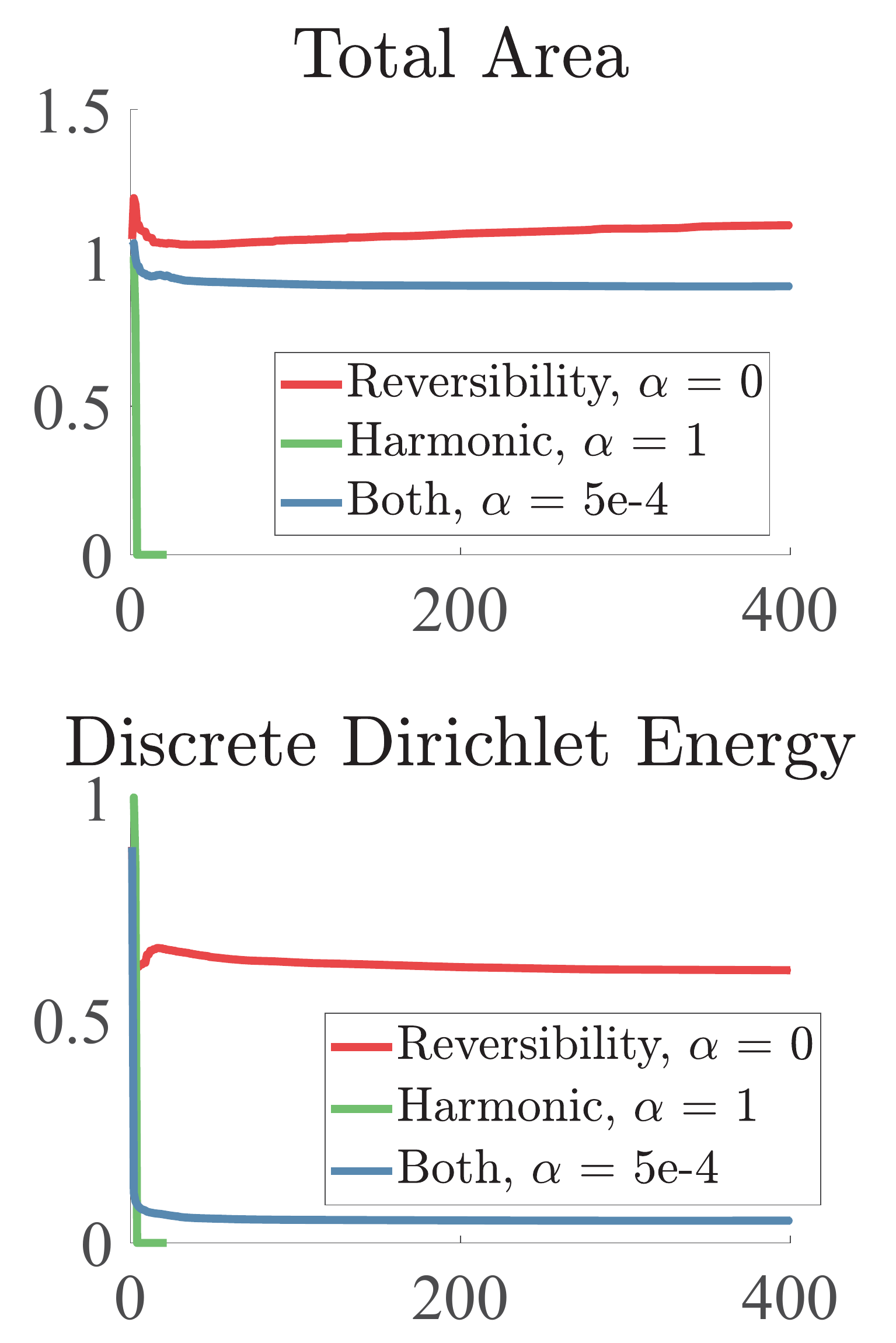}
		\caption{}
	\end{subfigure}
	
	\caption{Preventing collapse with reversibility. We use different $\alpha$ values, and visualize the result by mapping a texture from the target (a) to the source (b,c). In addition, we show the graphs of the total area of the image (d,top) and the discrete geodesic Dirichlet energy (d,bottom) during the optimization. Note that when $\alpha=0$, (b) and (d, red plots), the map is not collapsed but the Dirichlet energy is high. When $\alpha=1$, (d, green plots) the map collapses, as is evident by the zero total area. Finally, taking $\alpha=5 \cdot 10^{-4}$, (c, d blue plots) leads to a good balance between the energy components.} 
	
	\label{fig:geodesic_vs_euclidean_vs_embedding}
\end{figure}

\paragraph{Reversibility.}
We define the reversibility energy similarly to~\cite{kovnatsky2013coupled,ezuz2017deblurring} as:
\begin{equation}
E_R\left[\phi_{12},\phi_{21}\right] = \sum_{\substack{i,j\in\{1,2\}\\ i\neq j}}
\frac{1}{s_i^2}\sum_{p_i \in \mV_i} d_{M_i}^2 \left(\phi_{ji}\left(\phi_{ij}(p_i)\right),p_i\right)A_i(p_i). 
\label{eq:consistency_energy}
\end{equation}
The reversibility energy prevents the maps from collapsing. In the smooth case, if the reversibility energy is bounded pointwise, it easily follows that the maps are close to being injective and surjective, as we show in the Appendix. 
For both energies, care is required to handle correctly meshes of different scales, hence the normalization by $s_i$, the total area of $M_i$.

Finally, the full energy is given by:
\begin{equation}
E\left[\phi_{12},\phi_{21}\right] = \alpha E_D\left[\phi_{12},\phi_{21}\right] +  \left(1 - \alpha\right)E_R\left[\phi_{12},\phi_{21}\right], 
\label{eq:full_energy}
\end{equation}
where the parameter $\alpha\!\in\!\left[0,1\right]$ controls the trade-off between smoothness and reversibility. 

To demonstrate the effect of the different components of the energy we set the parameter $\alpha$ to different values. Figure~\ref{fig:geodesic_vs_euclidean_vs_embedding} shows the results of this experiment, computing a map between a disk (a) and a folded disk. An initial map maps all the interior vertices of $\source$ to a single interior vertex on $\target$, and the boundary of the disk is mapped to the boundary of the folded disk. In (b,c) we visualize the resulting map by mapping a texture from (a) to the source, and in (d) we show the graphs of the total area of the image (top) and discrete geodesic Dirichlet energy, computed using Eq.~\eqref{eq:energy_izeki} and normalized with respect to its value in the first iteration (bottom), during the optimization. Taking $\alpha=0$ models reversibility only, and leads to a non-collapsed map, yet with a high Dirichlet energy (b). On the other hand, taking $\alpha=1$ models harmonicity only, and leads to a map that collapses the image to a single point (the total area is zero in the graph). Finally, taking $\alpha\!=\!5\!\cdot\! 10^{-4}$ leads to a balance between the harmonic energy and reversibility, and a good overall map (c). 
Note that the mapped boundary vertices were not constrained to lie on the boundary of the target shape during the optimization. For all our experiments, we use $\alpha\!=\!5\!\cdot\! 10^{-4}$.

Minimizing the energy in Eq.~\eqref{eq:full_energy} requires computing the gradient of the geodesic distances with respect to the forward and backward maps, as well as tracing vector fields on the surface, which are both computationally heavy.
We therefore apply two approximations to address these issues.

\subsection{Energy approximation}

\subsubsection{Notation} 
Any point $p \!\in\! M$ can be represented uniquely using its barycentric coordinates $\omega_l(p), l\!\in\!\{1,..,3\}$ with respect to the face $f(p) = (v_1, v_2, v_3) \!\in\! \mF$ it lies on. We denote by $\lambda(p) \!\in\! \xR^{1\times n}$ the row vector that is zero everywhere except at the vertices of $f(p)$, where we have $\lambda(p)[v_l] = \omega_l(p)$. In addition, we denote the \emph{feasible row set} of $M$, i.e. the set of all possible such vectors, by $\mP\! = \!\{\lambda(p)  \,| \,p \!\in\! M\}$. Finally, the \emph{feasible set} of all possible precise maps from $M_1$ to $M_2$ is given by $\mP_{12}\! =\! \{P_{12} \!\in\! \xR^{n_1 \times n_2} \, | \, P_{12}(l,:) \!\in\! \mP_2 ,\,\, \forall l\!\in\!\{1..n_1\}\}$, where $P(l,:)$ denotes the $l$-th row of the matrix $P$.
Thus, any map $\phi_{12}$ can be represented using a matrix $P_{12}$, by setting $P_{12}(l,:) = \lambda_2(\phi_{12}(v_l)), \,\, \forall l\in\{1..n_1\}$, which, by definition, is in the feasible set $\mP_{12}$. Furthermore, the matrix $P_{12} V_2 \in \mathbb{R}^{n_1\times 3}$ represents the images of the vertices $\mV_1$ under the map $\phi_{12}$.

\subsubsection{High-dimensional embedding}
As we have seen, if the target space is \emph{Euclidean} then the geodesic distances are Euclidean distances, and the optimization is simple and efficient. 
Following similar ideas in the literature~\cite{bronstein2006generalized}, we therefore suggest to use a \emph{high-dimensional Euclidean embedding} as a proxy for fast geodesic distance computation.  

Given a mesh $M$, we seek an embedding $x : \mV \to \xR^m$, for $m \ll n$ such that the Euclidean distance $\| x \left( u \right) - x \left( v \right) \|$ approximates well the geodesic distance $d_M(u,v)$, for all $u,v \!\in\! \mV$. The literature on such embeddings is quite vast, and we chose to leverage the method suggested by~\citet{panozzo2013weighted} that relies on multidimensional scaling~\cite{cox2000multidimensional}. Any other embedding method could be used as well, as long as the geodesic distances are well approximated. We took $m = 8$ in all our experiments, following~\citet{panozzo2013weighted}.

Our goal now is to compute a harmonic map between the high-dimensional Euclidean embeddings. We denote by $X \!\in\! \xR^{n \times m}$ the matrix whose rows are the embeddings $x(v), \forall v \!\in\! \mV$. 
Rewriting the harmonic and reversibility energies in terms of the high-dimensional embeddings, and in matrix form, leads to:
\begin{equation}
E\left(P_{12},P_{21}\right) = \sum_{\substack{i,j\!\in\!\{1,2\}\\ i\neq j}}\alpha \frac{1}{s_i}\| P_{ij} X_j \|_{W_i}^2 + (1-\alpha)\frac{1}{s_i^2}\|P_{ij} P_{ji} X_i - X_i\|_{M_i}^2.
\label{eq:full_energy_relaxed}
\end{equation}
Note that the weights in the harmonic energy, given now in matrix form in $W_i$, remain the same for the high-dimensional embedding, since the embedding is nearly isometric. 
\subsubsection{Half quadratic splitting}
While Equation~\eqref{eq:full_energy_relaxed} can be minimized using gradient descent, we found, similarly to~\cite{ezuz2017deblurring}, that it is more efficient to use the \emph{half quadratic splitting} optimization method~\cite{geman1995nonlinear}  (see also e.g.~\cite{wang2008new,zoran2011learning}). We introduce auxiliary variables $X_{ij} \in \xR^{n_i \times m}$, which estimate the images of the vertices $\mV_i$ given by $P_{ij}X_j$. Substituting, the energies are:
\begin{equation}
\bar{E}_D(X_{ij}) = \frac{1}{s_i}\| X_{ij} \|_{W_i}^2, \quad
\bar{E}_R(P_{ij},X_{ji}) = \frac{1}{s_i^2}\|P_{ij} X_{ji} - X_i\|_{M_i}^2,
\end{equation}
for $i,j\!\in\!\{1,2\},i\neq j$. In addition, we need soft constraints for the auxiliary variables:
\begin{equation}
\bar{E}_Q(P_{ij},X_{ij}) = \frac{1}{s_i s_j}\|X_{ij} - P_{ij}X_j\|_{M_i}^2.
\end{equation}
The full energy is now: 
\begin{align}
\begin{split}
&\bar{E}\left(P_{12},P_{21},X_{12},X_{21}\right) = \\	&\sum_{\substack{i,j\in\{1,2\}\\ i\neq j}}\alpha \bar{E}_D(X_{ij}) + (1-\alpha)\bar{E}_R(P_{ij},X_{ji}) + \beta \bar{E}_Q(P_{ij},X_{ij}),
\label{eq:full_energy_hqs}
\end{split}
\end{align}
where $\beta$ controls the accuracy of the auxiliary variables. 
In practice, an update scheme for $\beta$ is often tailored per application, with the general guideline of increasing $\beta$ as the iterations advance~\cite{wang2008new}. In practice, in all of our experiments, we took $\beta = 5\cdot 10 ^{-3} k$ where $k$ is the optimization iteration number for the first 100 iterations, and then kept the value of $\beta$ constant until convergence. 

\subsection{The optimization problem}
Our optimization problem is now given by:
\begin{equation}
\label{eq:energy_for_opt}
\begin{aligned}
& \underset{P_{12}, P_{21},X_{12},X_{21}}{\text{minimize}}
& & \bar E(P_{12}, P_{21},X_{12},X_{21}) \\
& \,\,\,\,\, \text{subject to}
& & P_{12} \!\in\! \mP_{12},\,\, P_{21} \!\in\! \mP_{21}, 
\end{aligned}
\end{equation}
where $\mP_{ij}$ is the feasible set of precise maps from $M_i$ to $M_j$.
Despite the two approximations that we used, solving this optimization problem succeeds in decreasing the total discrete Dirichlet energy from Eq.~\eqref{eq:energy_izeki} while preventing the map from collapsing, as is illustrated in Figure~\ref{fig:iterations}. In addition to the energy, we show the initial map that was created from landmarks as described in Section~\ref{sec:initialization}, and the intermediate map at a few iterations. 

\section{Optimization}
\label{sec:opt}

Our optimization problem has block structure, in the sense that if some of the variables are kept fixed it becomes a linear least squares problem. We therefore chose to use block coordinate descent (see e.g.~\cite{xu2013block}) as the optimization algorithm. 

\subsection{Block coordinate descent}
In each sub-iteration we solve for one of the matrices $X_{ij}, P_{ij}$, while keeping the others fixed. Since the energy is quadratic in all the variables, every sub-iteration involves a relatively simple optimization problem, with the only complication arising because of the non-convex feasible sets $\mP_{ij}$. 

\begin{algorithm}[b]
	\fbox{\hspace{-.1in}\parbox{\columnwidth}{%
			\begin{algorithmic}
				\State\textbf{Input:} Two triangles meshes $M_1$, $M_2$, initial $P_{12}, P_{21}, X_{12}, X_{21}$
				\State\textbf{Output:} $P_{12}, P_{21}$
				\algspace
				\State \textbf{For} $k= 1\ldots N$ 
				\State \quad \quad \textbf{For} $i= 1,2$
				\State \quad \quad \quad \quad $j = 3-i$
				\State	\quad \quad\quad  \quad $P_{ij} \leftarrow \,\,\, \,\underset{P\in {\mP}_{ij}}{\text{argmin}} \,\, \bar{E}_R(P,X_{ji}) +		\bar{E}_Q(P,X_{ij}) $
				\State  \quad\quad \quad  \quad $X_{ij} \leftarrow \underset{X\in \xR^{n_i \times m}}{\text{argmin}} \bar{E}_D(X) + \bar{E}_R(P_{ji},X) +		\bar{E}_Q(P_{ij},X)$
				\State\quad\quad  \textbf{end}
				\State\textbf{end}
				
			\end{algorithmic}
		}}
		\caption{Alternating minimization.\vspace{-.2in}}\label{alg:flow}
	\end{algorithm}
	
	\paragraph{Optimizing for $X_{ij}$.} When $P_{12}, P_{21}$ are fixed, the optimization problem is a linear least squares minimization of the form $\|A X_{ij} - B\|^2_2$, with known matrices $A$ and $B$, where $A$ is sparse, which we solve using a direct method. The system is highly over-constrained, as even if $P_{ji}$ degenerates, the system is well-conditioned due to the term $\bar{E}_Q$, as long as the vertex areas of the input mesh $M_i$ do not vanish. 
	\paragraph{Optimizing for $P_{ij}$.} When $X_{12},X_{21}$ are fixed, the energy has the form $\|P_{ij} A - B\|_2^2$, where $A,B$ are known, with the constraints that $P_{ij}\!\in\!\mP_{ij}$. Following~\cite{ezuz2017deblurring}, the optimization is done by solving for every row of $P_{ij}$ separately. Intuitively, we can think of $B$ as a high-dimensional embedding of the faces of $M_j$, and of $A$ as a high-dimensional point cloud. The optimal $P_{ij}$ projects each point in $A$ to its closest point on the faces given by $B$.  
	As shown in~\cite{ezuz2017deblurring}, this process is guaranteed to find $P_{ij}$ which are globally optimal when $X_{ij}$ are kept fixed.
	
	\paragraph{Stopping criterion.} The alternating descent guarantees that the energy is reduced at every iteration, since the sub-iterations find a global optimum of the reduced optimization problems. 
	In practice, we stopped the optimization when the change of energy was less than $10^{-9}$, or after a maximum of $N = 200$ iterations. In most cases, we have observed convergence of the energy to high precision even when early stopping after $N$ iterations was used.
	
	We provide the details of the alternating descent in Algorithm 1. 
	
	\subsection{Initialization} Our method is general and can receive as input various initial data. Depending on the input, we describe the initialization of the variables $P_{ij}, X_{ij}$. Given a pointwise map $\phi_{12}: \mV_1 \to M_2$ its corresponding matrix representation is given by $P_{12}(l,:) = \lambda_2(\phi_{12}(v_l))$ for $l \!\in\!\{1,..,n_1\}$.
	Similarly, given a matrix representation $P_{ij} \in \mP_{ij}$, we have that $\phi_{ij}(v_l) = (P_{ij}V_j)(l,:) \!\in\! M_j$. Therefore, in the following refer to either $P_{ij}$ or $\phi_{ij}$ according to which notation is more convenient.
	\begin{figure}[t]
		\centering
		\includegraphics[width=1\linewidth]{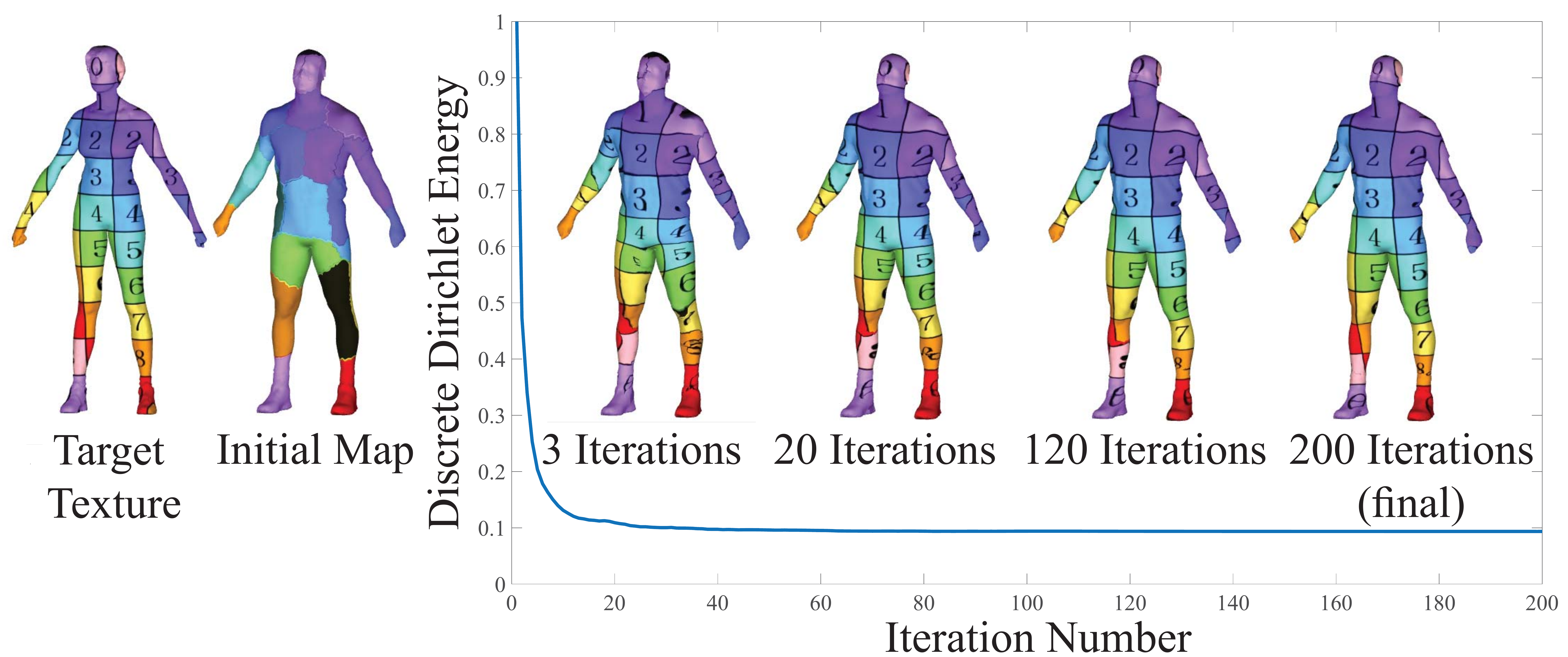}
		\caption{Optimization of Eq.~\eqref{eq:energy_for_opt}, starting from landmarks. The discrete Dirichlet energy in Eq.~\eqref{eq:energy_izeki} decreases, and the final map, visualized by texture transfer, is semantic and not distorted locally.}
		
		\label{fig:iterations}
	\end{figure}
	\label{sec:initialization}
	\paragraph*{Pointwise map.}
	Given a pointwise map $P_{12}$ we approximate an inverse map $P_{21}$  
	by taking $\phi_{21}(v_2) = \text{argmin}_{v \in V_1} \| \phi_{12}(v) - v_2 \|$.
	
	Then, the initialization of $X_{ij}$ is $P_{ij} X_j$. 
	\paragraph*{Functional map.}
	The term \emph{functional map}~\cite{ovsjanikov2012functional} denotes a map between scalar functions. It is a linear operator that can be represented using a matrix when scalar functions are represented in a linear basis. Let $\Psi_i \in \R ^{n _i \times k_i}$ be a matrix whose columns are basis functions of a subspace of scalar functions on $M_i$, where each function is piecewise linear and is defined by values assigned to vertices.
	Given a pointwise map that maps vertices of $M_1$ to points on $M_2$, the corresponding functional map $C_{12}\in \R ^{k_1 \times k_2}$ maps functions on $M_2$ to functions on $M_1$, represented in the reduced basis. Therefore, given two functional maps $C_{12}$ and $C_{21}$, we initialize $X_{ij} = \Psi_i C_{ij} \Psi_j ^\dagger X_j$. 
	Optimizing $P_{ij}$ does not require initialization. Figure~\ref{fig:fmaps_texture} shows results where functional maps were used for initialization.
	
	\begin{figure*}[t]
		\centering
		\includegraphics[width=.3\linewidth]{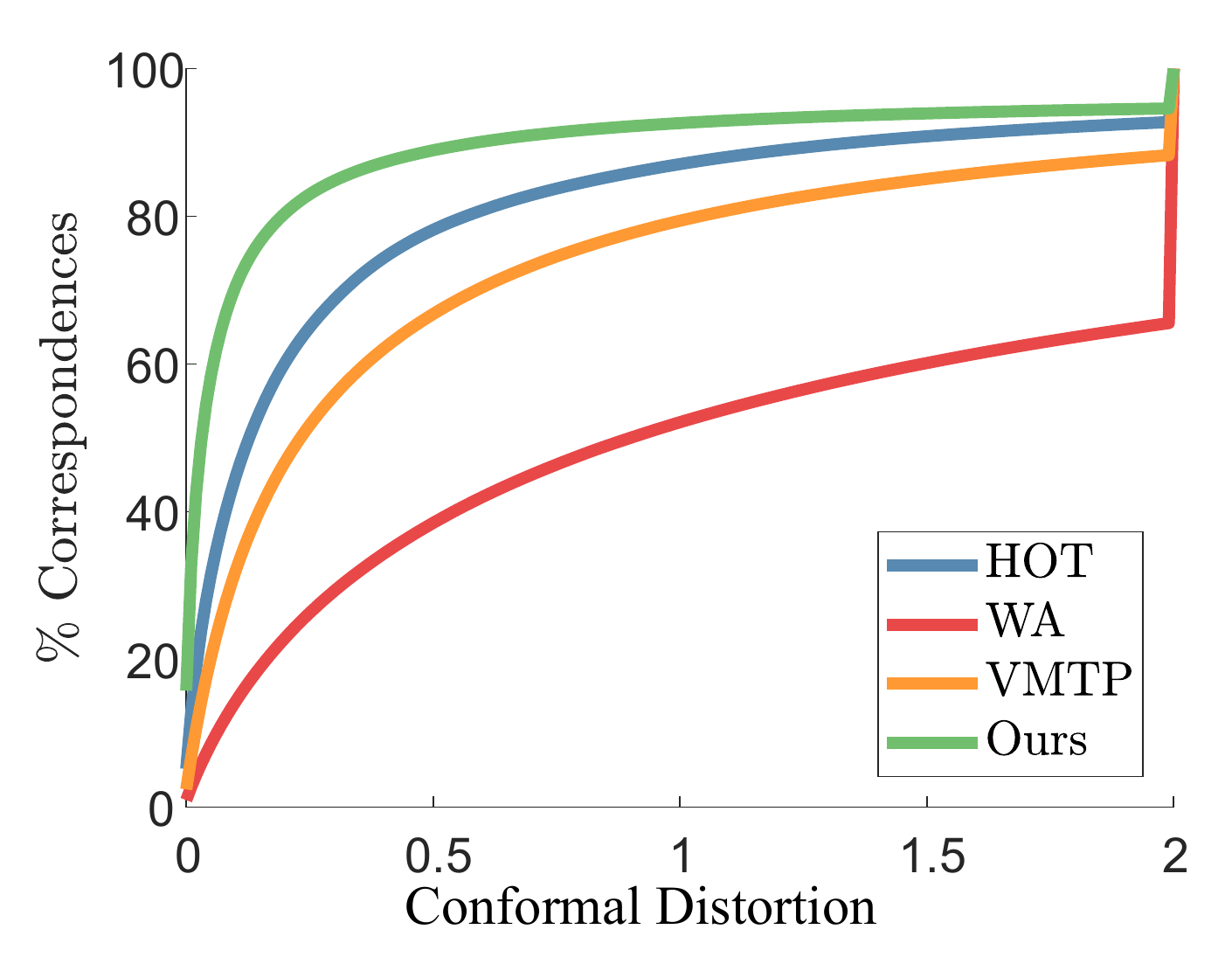}
		\hspace{.03\linewidth}
		\includegraphics[width=.3\linewidth]{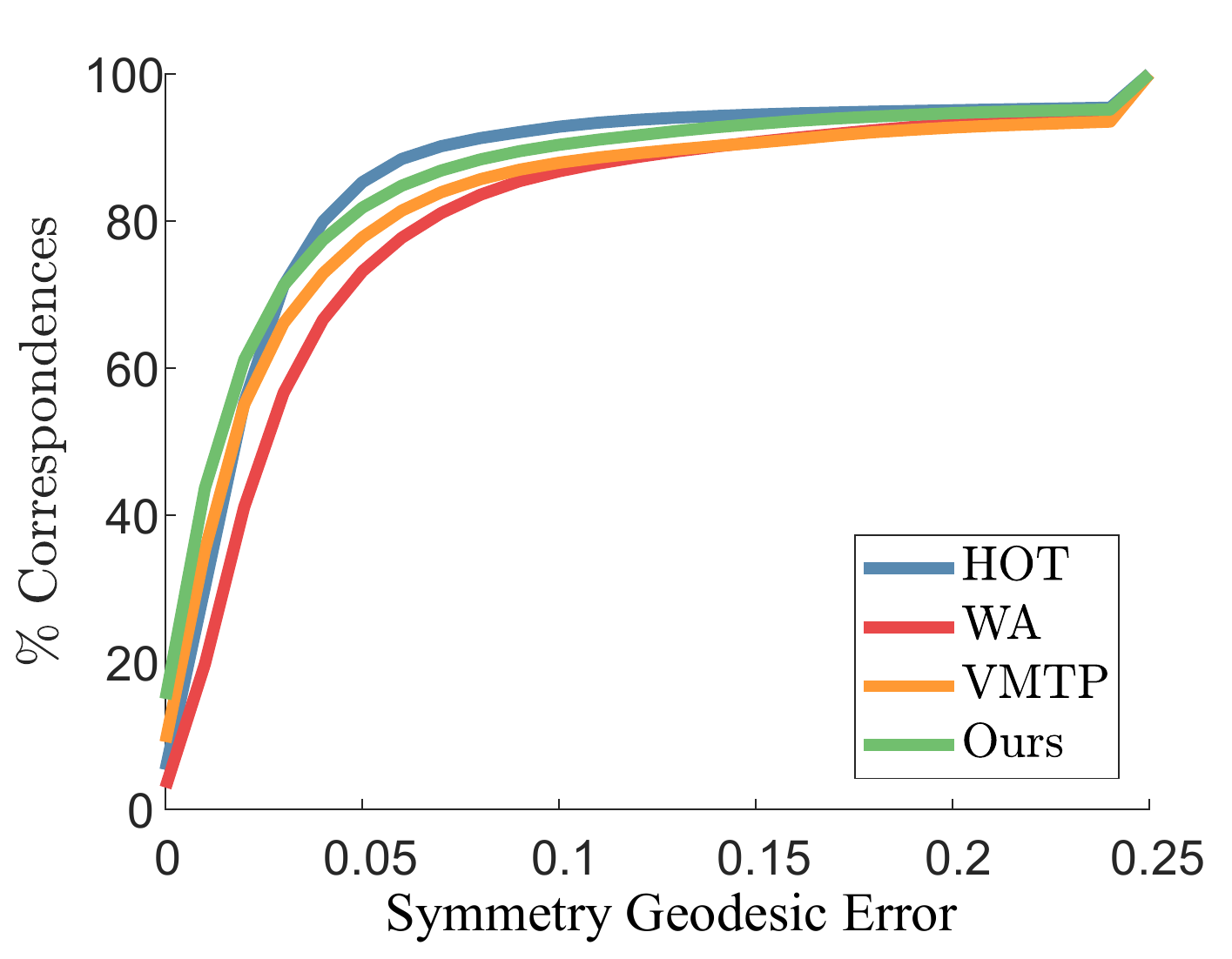}	
		\hspace{.03\linewidth}
		\includegraphics[width=.3\linewidth]{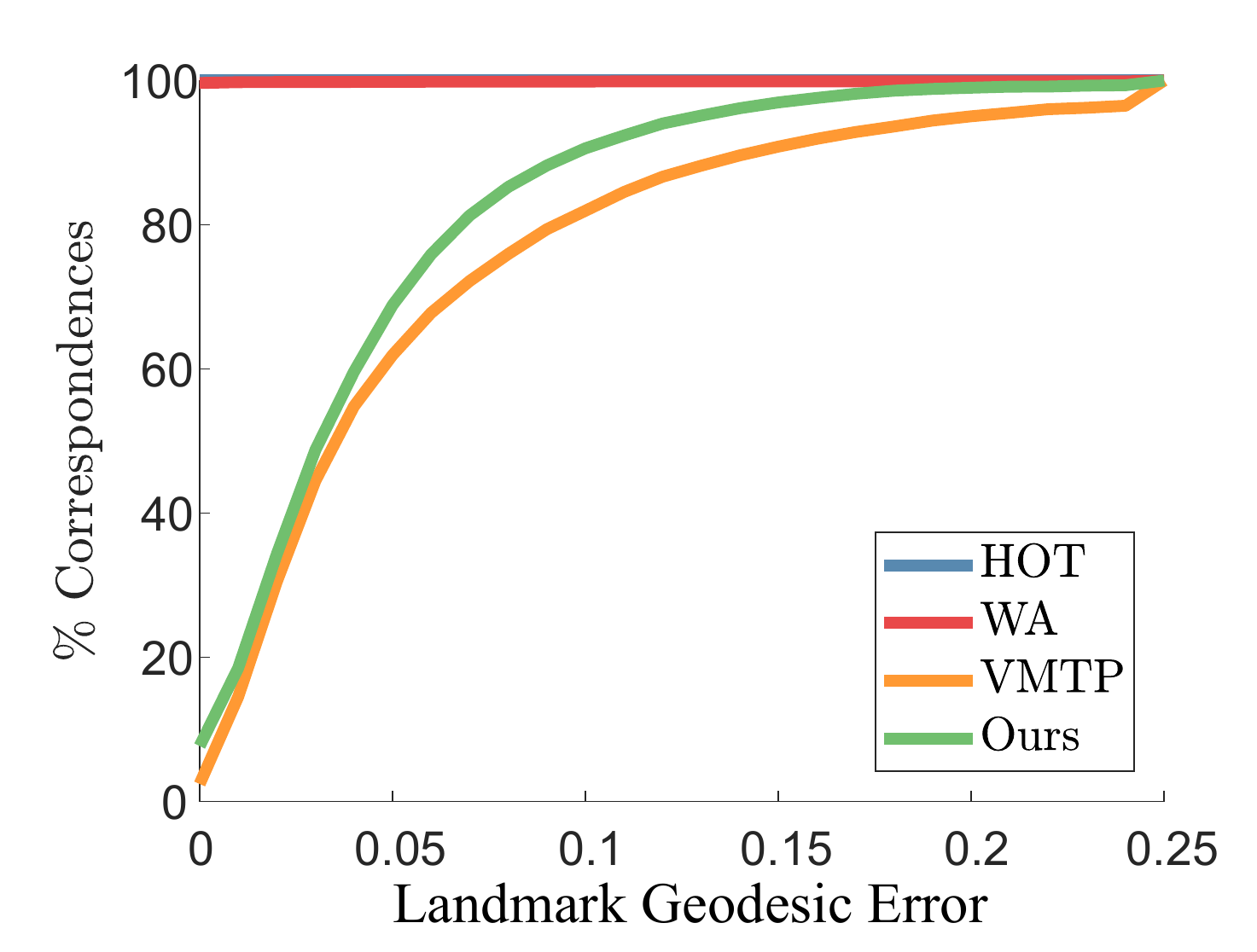}
		\caption{Quantitative comparison on the SHREC dataset, measuring, from left to right: conformal distortion, compatibility with symmetries and distance from ground truth landmarks. Note that we achieve a better conformal distortion, and comparable symmetry geodesic error. Furthermore, note that WA and HOT do not modify their input landmarks, while our method and VMTP do. Compared to VMTP we achieve a better landmark geodesic error.} 	
		\label{fig:shrec_graphs}
	\end{figure*}
	
	\begin{figure}[b!]
		\centering
		\includegraphics[width=.19\linewidth]{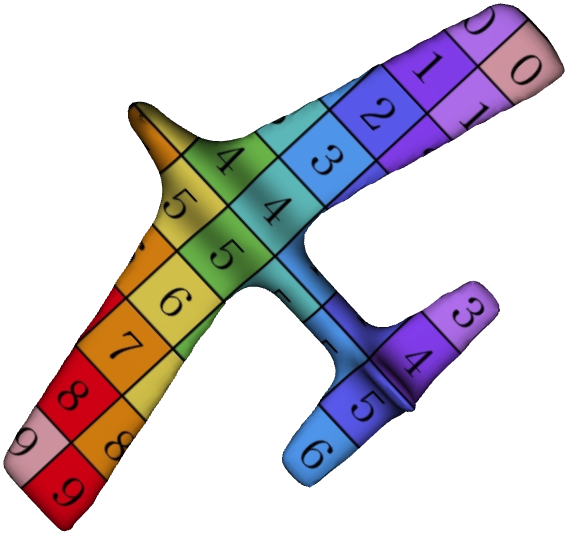}
		\includegraphics[width=.19\linewidth]{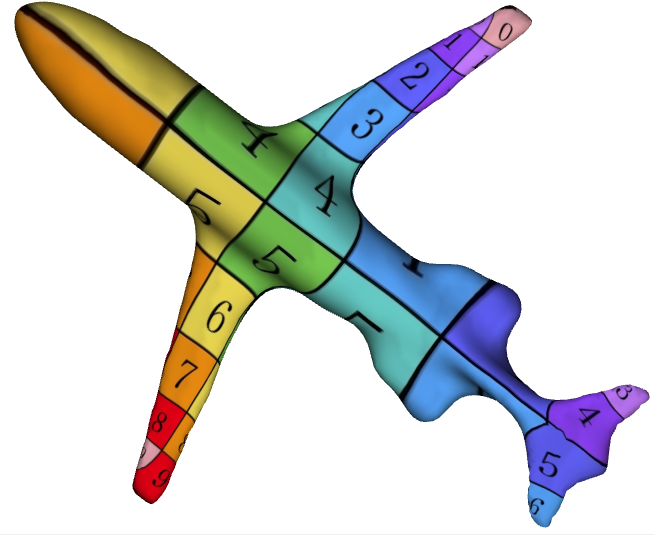}
		\includegraphics[width=.19\linewidth]{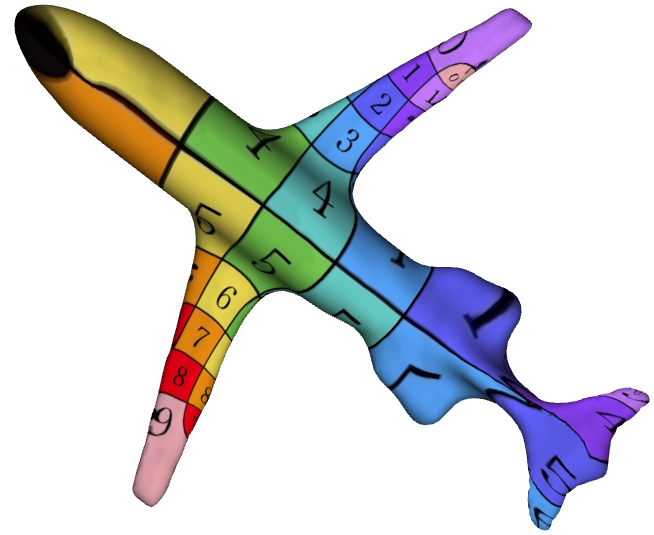}
		\includegraphics[width=.19\linewidth]{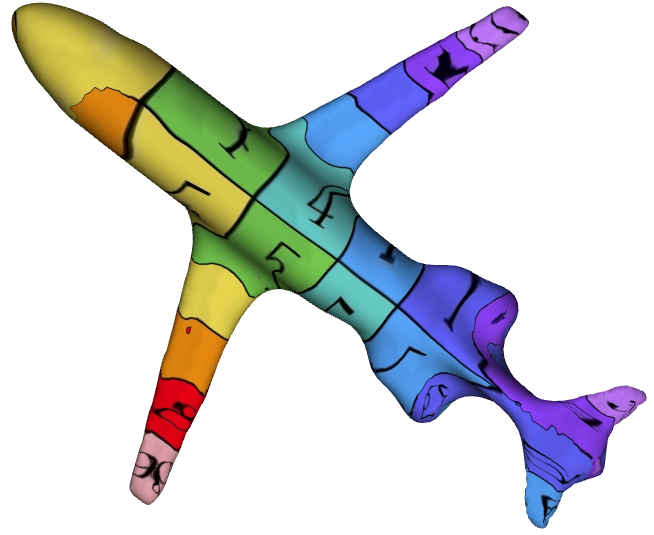}
		\includegraphics[width=.19\linewidth]{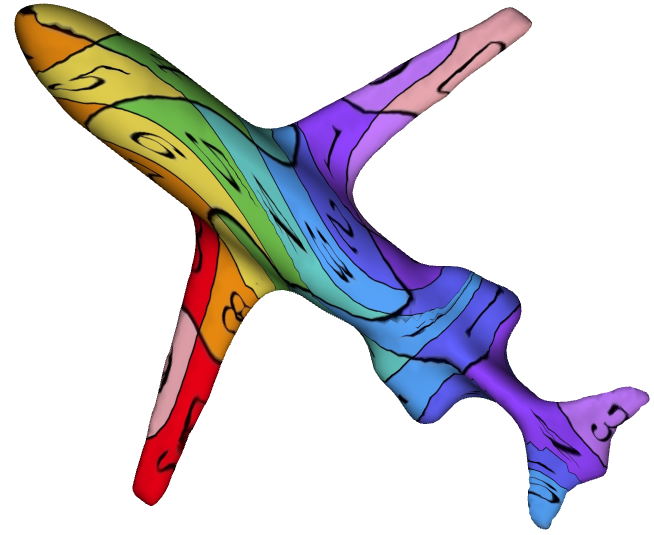}
		\vspace{0.03\linewidth}
		\includegraphics[width=.19\linewidth]{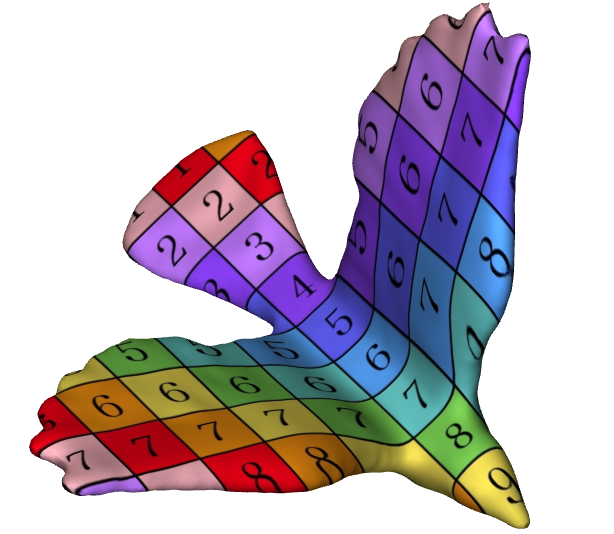}
		\includegraphics[width=.19\linewidth]{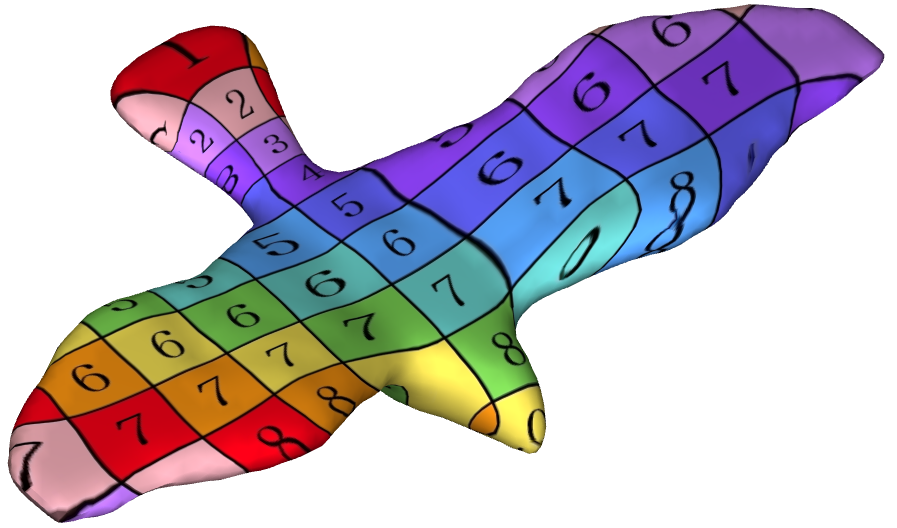}
		\includegraphics[width=.19\linewidth]{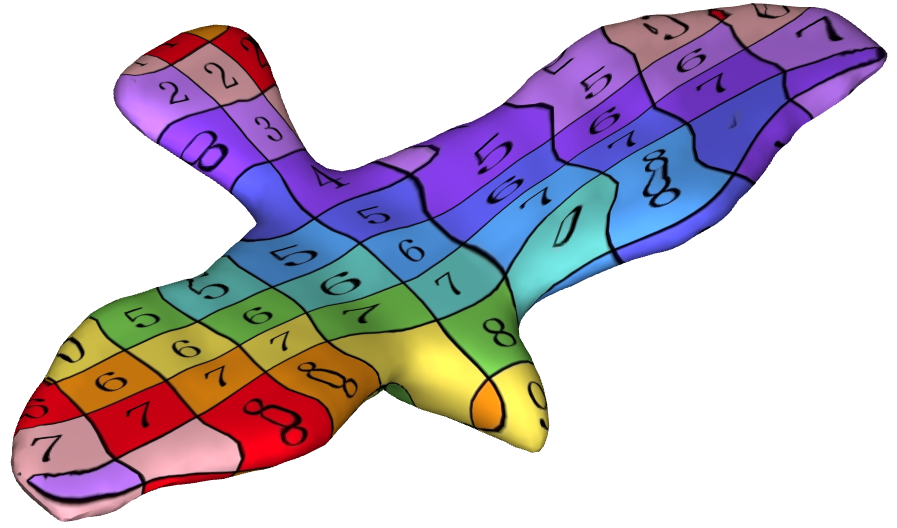}
		\includegraphics[width=.19\linewidth]{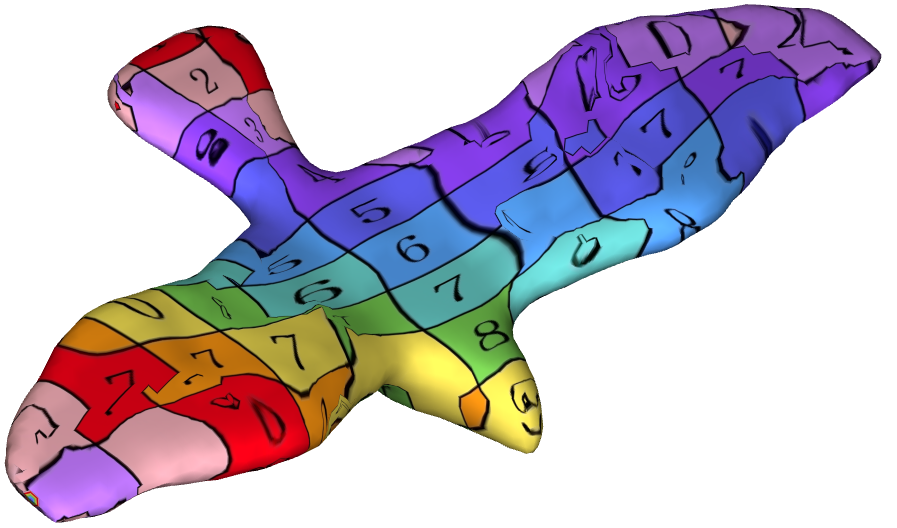}
		\includegraphics[width=.19\linewidth]{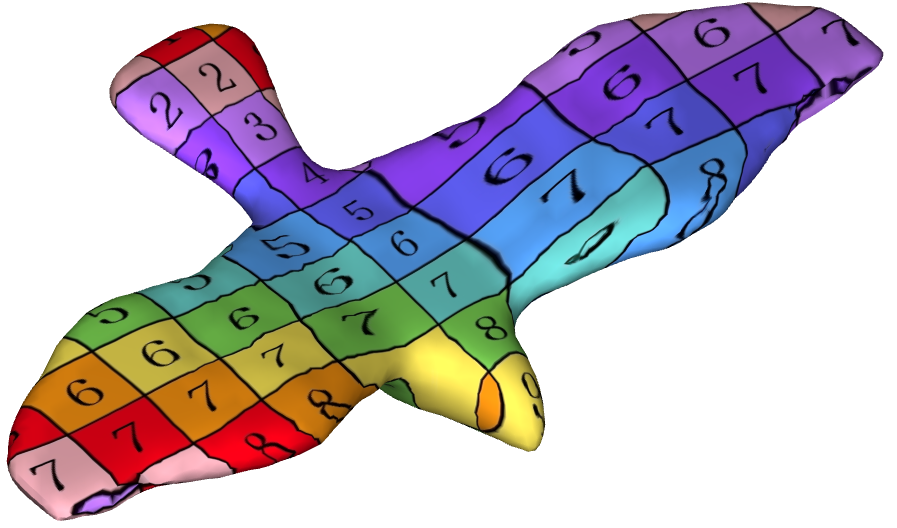}

		\includegraphics[width=.19\linewidth]{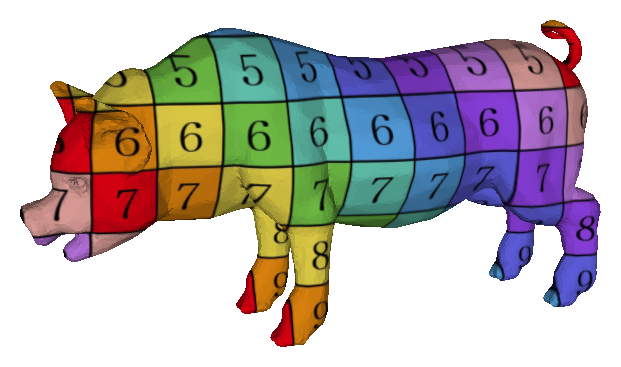}
		\includegraphics[width=.19\linewidth]{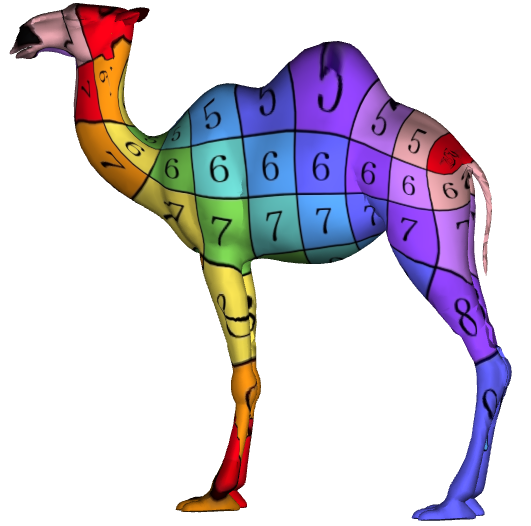}
		\includegraphics[width=.19\linewidth]{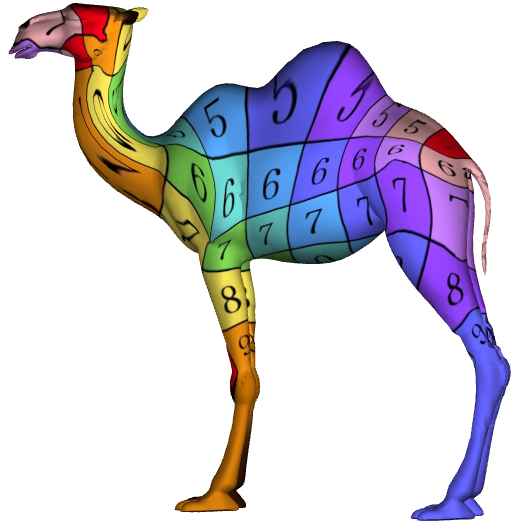}
		\includegraphics[width=.19\linewidth]{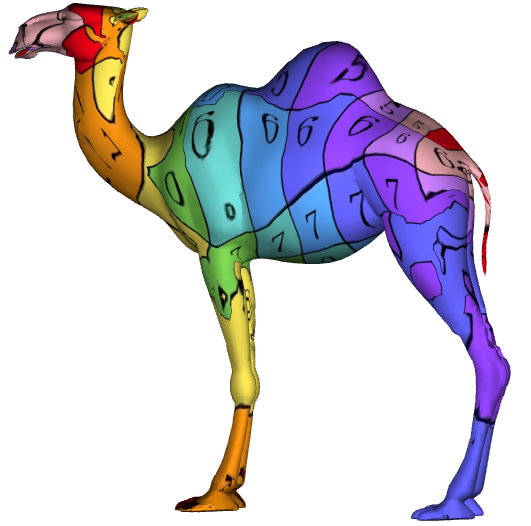}
		\includegraphics[width=.19\linewidth]{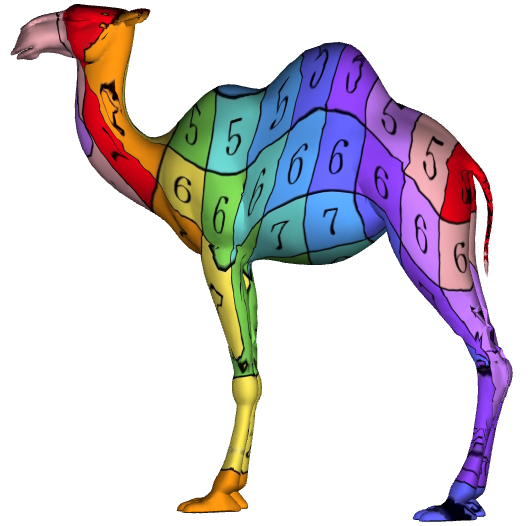}
		\vspace{0.02\linewidth}
		
		\includegraphics[width=.19\linewidth]{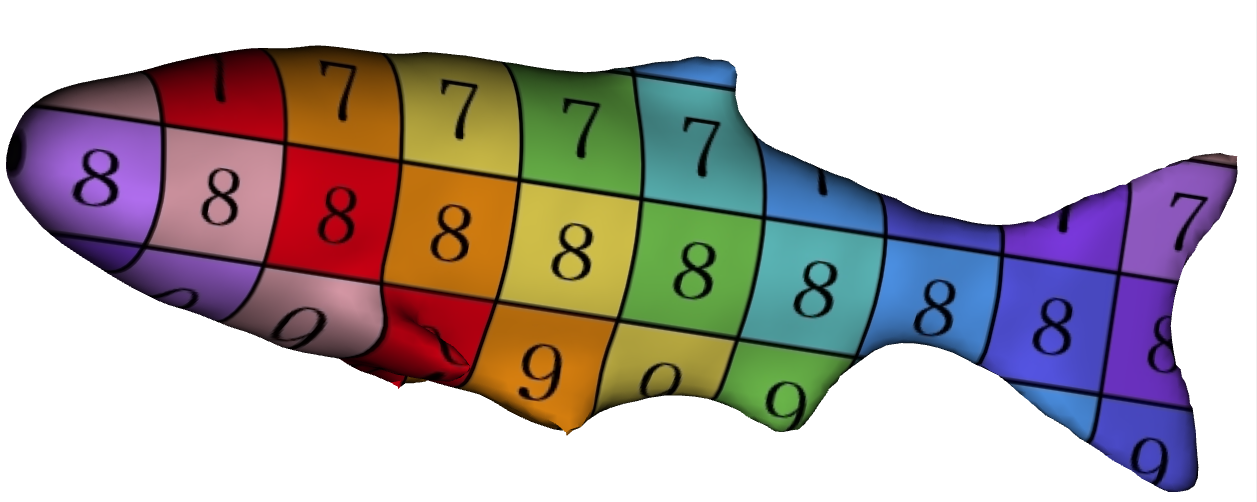}
		\includegraphics[width=.19\linewidth]{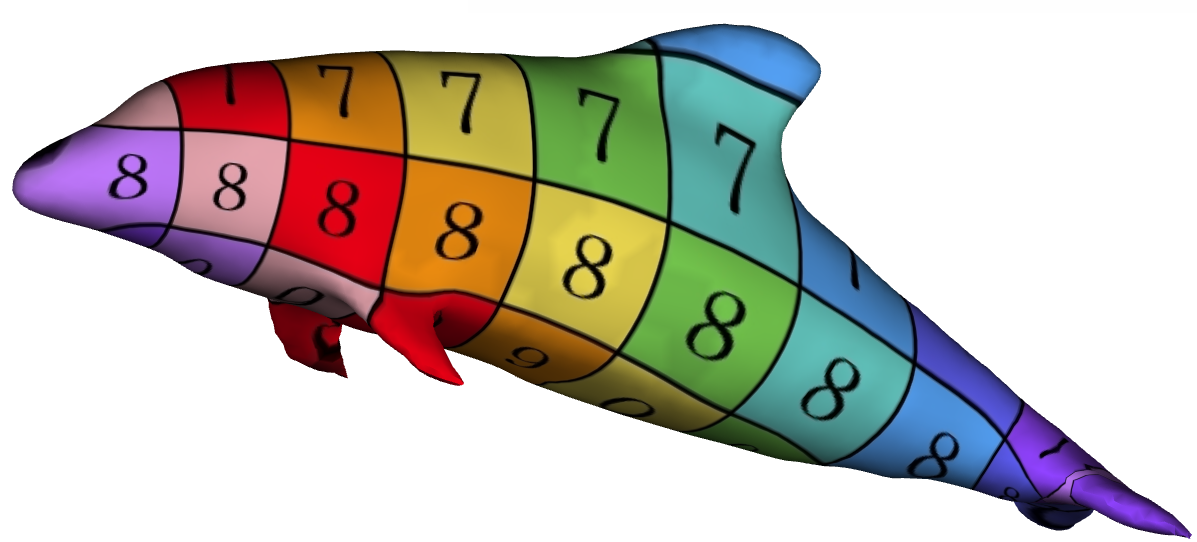}
		\includegraphics[width=.19\linewidth]{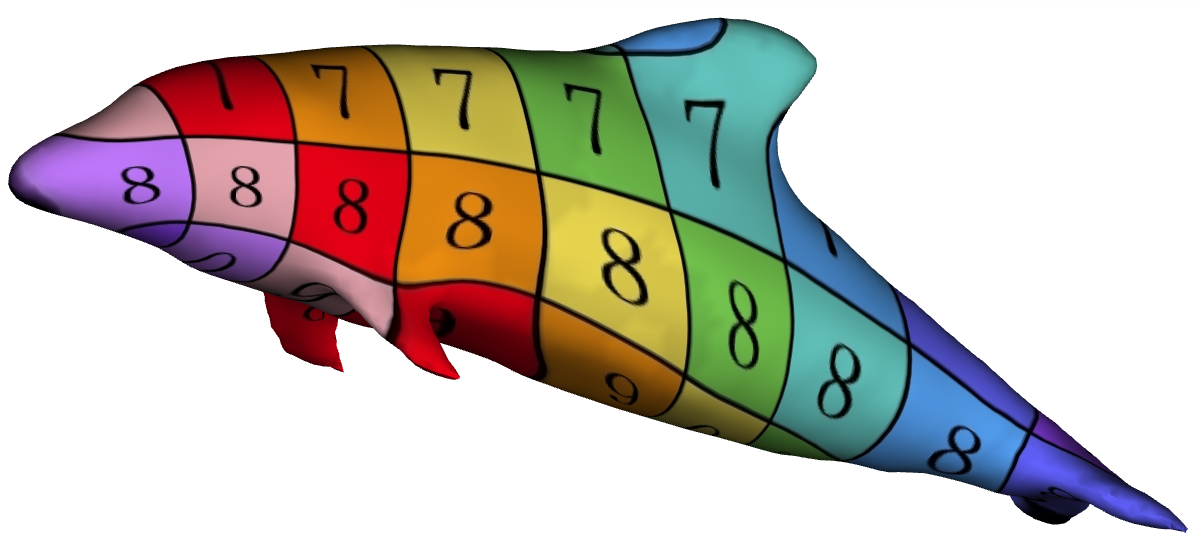}
		\includegraphics[width=.19\linewidth]{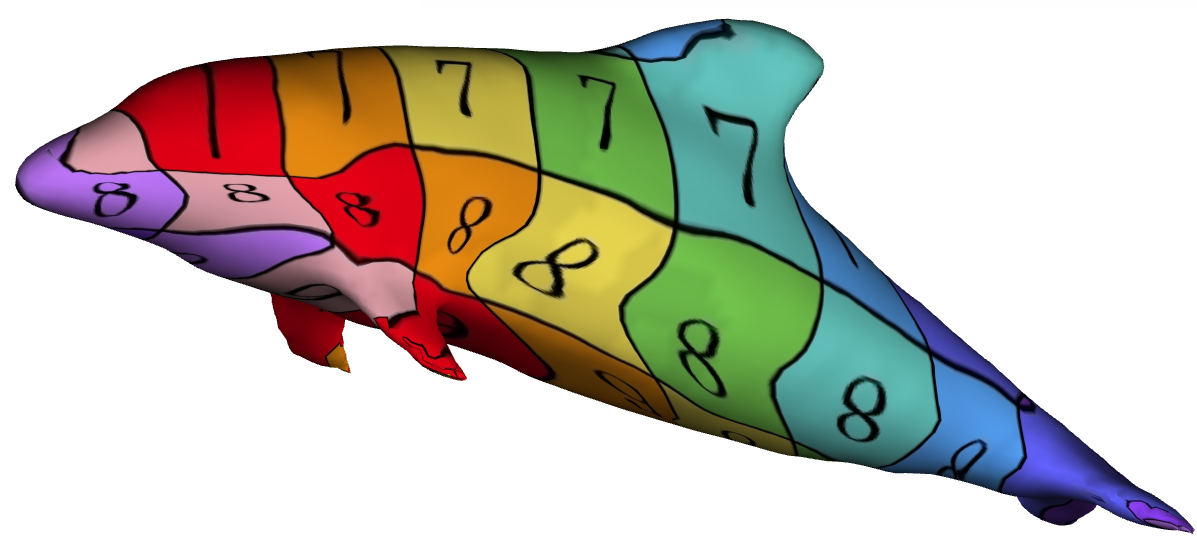}
		\includegraphics[width=.19\linewidth]{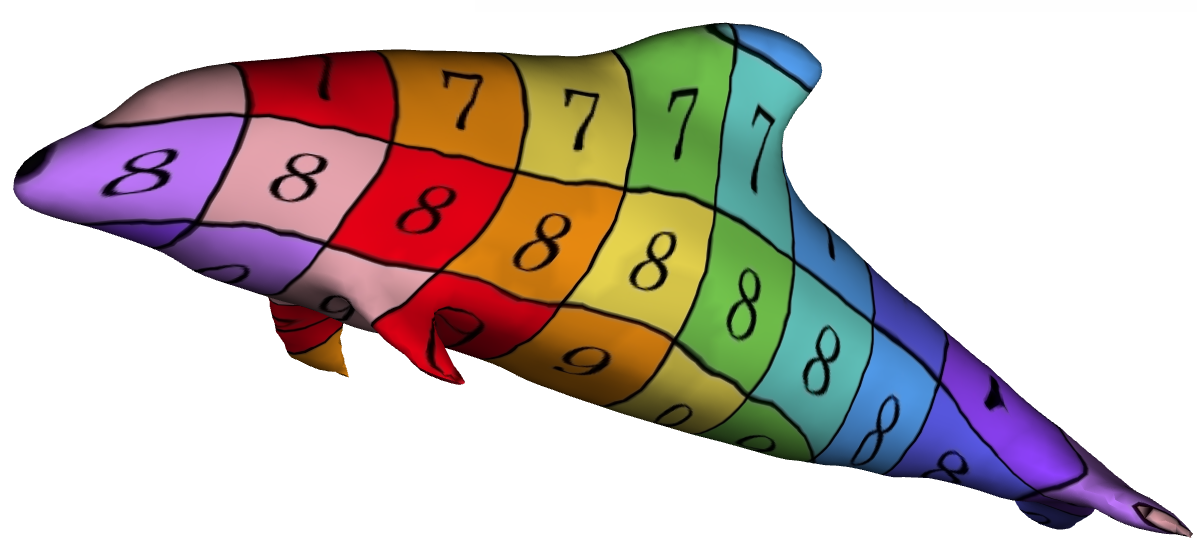}
		\includegraphics[width=.19\linewidth]{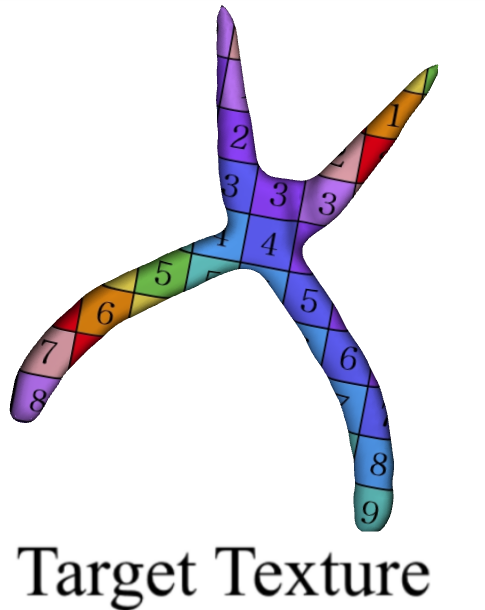}
		\includegraphics[width=.19\linewidth]{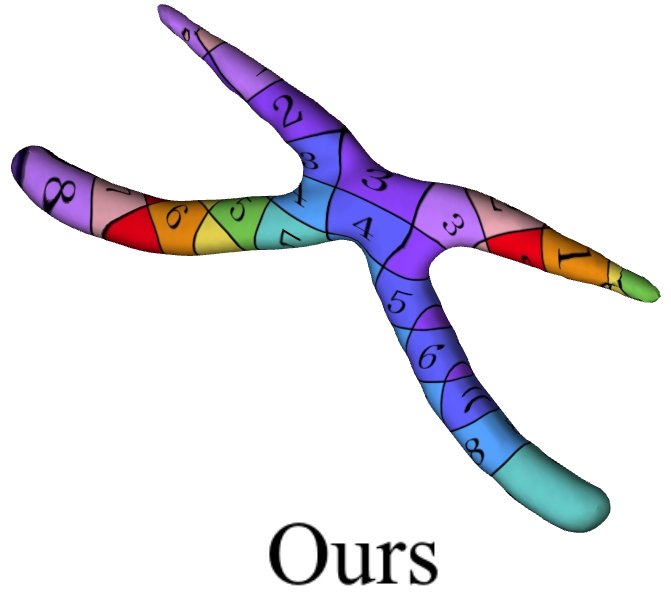}
		\includegraphics[width=.19\linewidth]{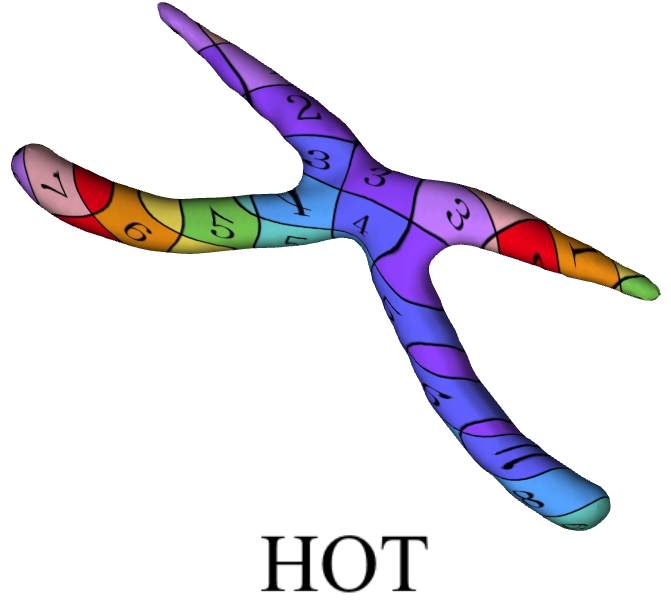}
		\includegraphics[width=.19\linewidth]{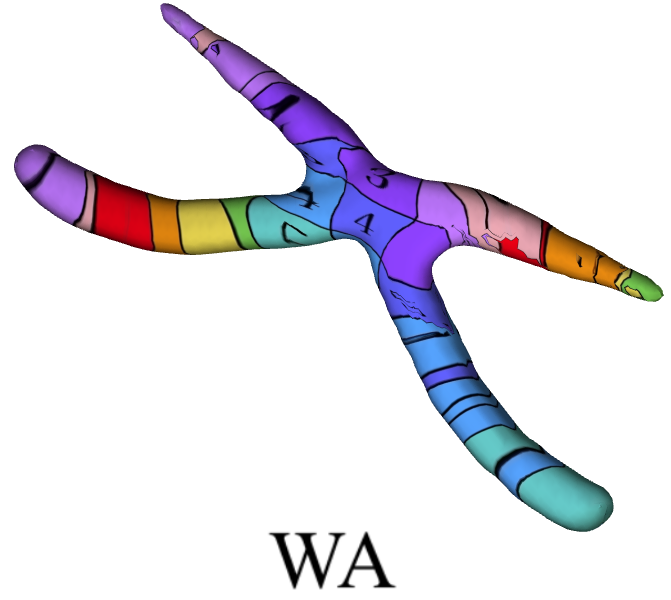}
		\includegraphics[width=.19\linewidth]{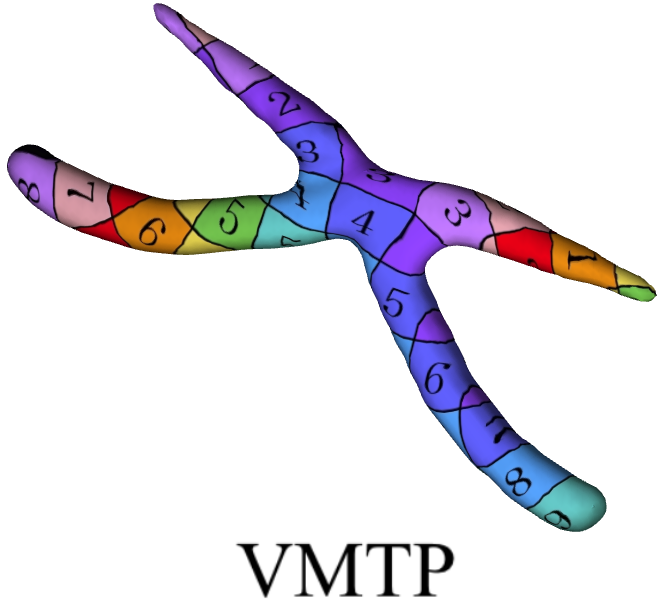}
		
		\caption{Qualitative results, from input landmarks. From left to right: target texture, our method, HOT~\cite{aigerman2016hyperbolic}, WA~\cite{panozzo2013weighted} and VMTP~\cite{mandad2017variance}. See the text for details.} 
		\label{fig:shrec_vis}
	\end{figure}
	\paragraph*{Landmarks.}
	Given $r$ input landmark pairs $ \left\lbrace (p_i, q_i) \right\rbrace $ where $p_i \!\in\! \mV_1, q_i \!\in\! \mV_2$ and $i\!\in\!\{1,..,r\}$, we first construct a rough initial pointwise map $P_{12}$ and then use it to initialize the rest of the variables, as previously described. 
	We first compute the geodesic Voronoi diagram on $M_1$ with centers $p_i$, and then set $\phi_{12}(v) = q_i, \, \forall v \in C_i$, where $C_i$ is the geodesic cell corresponding to the center $p_i$.
	Note that this initialization is highly degenerate, as all the points on $M_1$ are mapped to the landmarks $q_i$ on $M_2$, yet it is enough for our needs. 
	In Figure~\ref{fig:iterations} initialization using input landmarks is visualized.

	\subsection{Limitations}
	Using the projection step when optimizing for $P_{ij}$ has some limitations. First, as discussed in~\cite{panozzo2013weighted}, such a projection is not smooth. Furthermore, the closest point on the high-dimensional embedding of the triangle mesh might not be unique, therefore the solution of the optimization for $P_{ij}$ might alternate between two configurations with the same energy. Thus, while the energy is guaranteed to converge, we do not have a similar guarantee for the convergence of the solution. In practice, we have not encountered a case where these limitations posed a practical problem. In future work it could be possible to handle the first issue using a Phong projection, as in~\cite{panozzo2013weighted}, and the second issue using an additional regularization that penalizes diverting from the current solution. 
	
	\subsection{Timing}
	
	The most expensive step in the optimization process is the projection on a triangle mesh for optimizing $P_{ij}$. However, this procedure is highly parallelizable since the projection of each point is independent of the other points. We used CUDA 8 to implement the projection in parallel, while the rest of the optimization method was written in MATLAB. On a desktop machine with a TITANX GPU and an Intel Core i7 processor, 200 optimization iterations of our method, for shapes with 5K vertices, took around $115$ seconds.
	
	\section{Results}
	\label{sec:results}
	To validate our method we have compared with a variety of state-of-the-art mapping techniques, in accordance with the type of input they can accept. In addition, we show applications to shape interpolation and quad mesh transfer.

	\subsection{Quality metrics}
	To evaluate the quality of a map we measure its smoothness through its conformal distortion and its semantic accuracy, using the distance to the ground truth, when given. We also use alternative measures, such as symmetry and compatibility with ground truth segmentations, when no dense ground truth map is available. 
	
	\paragraph{Conformal distortion.} We use the definition by Hormann and Greiner~\shortcite[Eq. (3)]{hormann2000mips} for the conformal distortion of a single triangle $f\!\in\!\mF_1$: $\kappa(f)\!=\!\frac{\sigma_1}{\sigma_2}\!+\!\frac{\sigma_2}{\sigma_1}$, where $\sigma_1\!\geq\! \sigma_2$ are the singular values of the linear transformation which maps $f$ from $M_1$ to $M_2$. We subtract $2$ so that the minimal conformal distortion is zero and visualize the result as a cumulative graph showing the percentage of triangles with less than a certain distortion value. 
	\begin{figure}[t!]
		\centering
		\includegraphics[width=.19\linewidth]{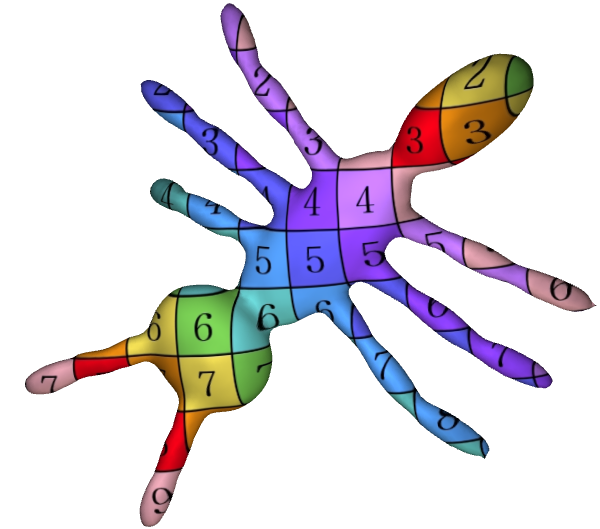}
		\includegraphics[width=.19\linewidth]{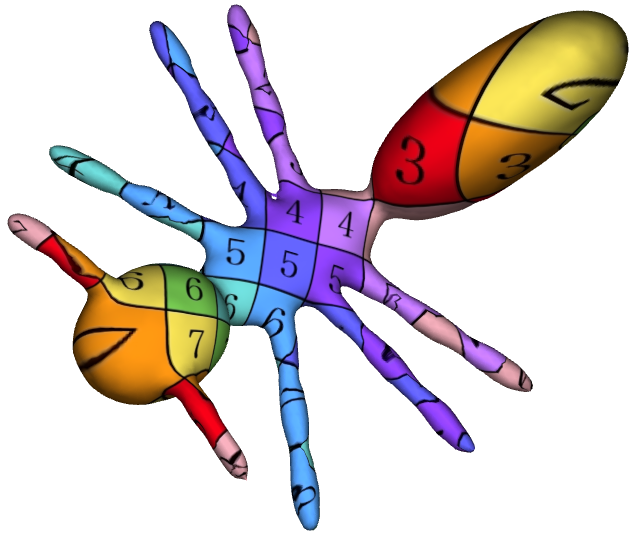}
		\includegraphics[width=.19\linewidth]{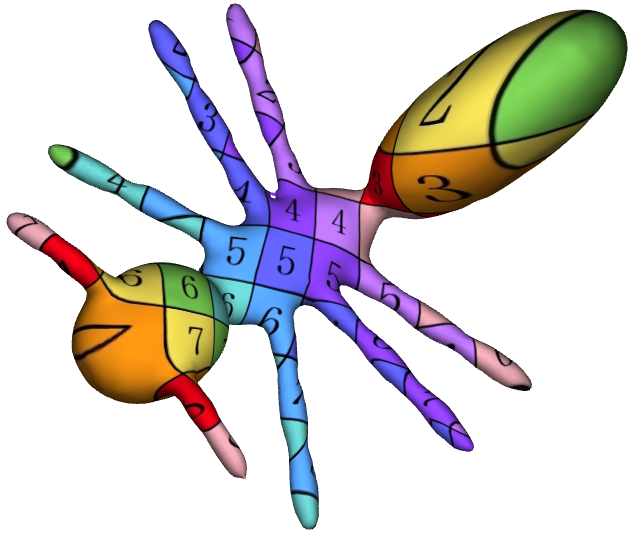}
		\includegraphics[width=.19\linewidth]{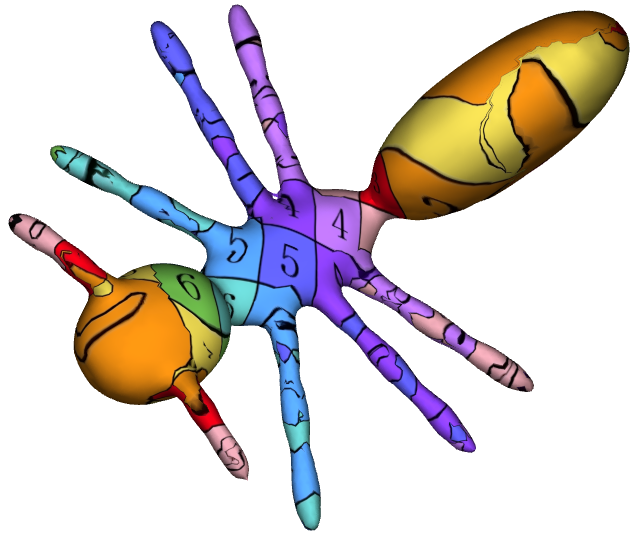}
		\includegraphics[width=.19\linewidth]{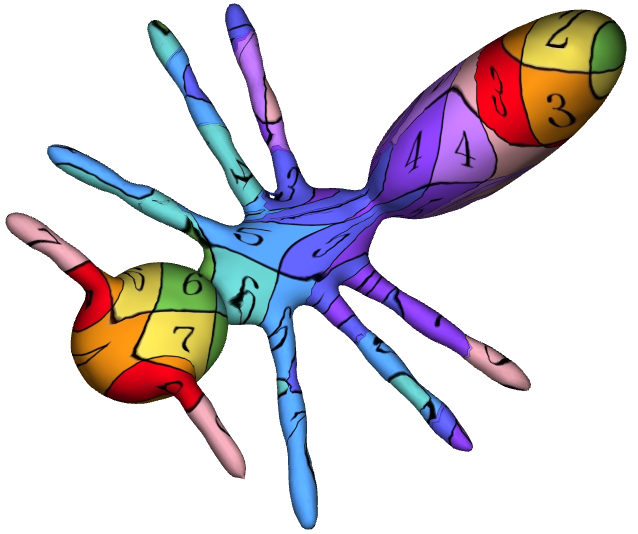}
		
		\includegraphics[width=.19\linewidth]{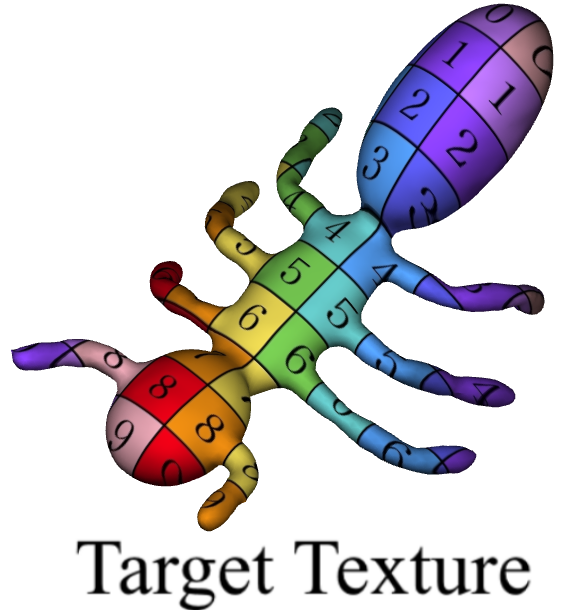}
		\includegraphics[width=.19\linewidth]{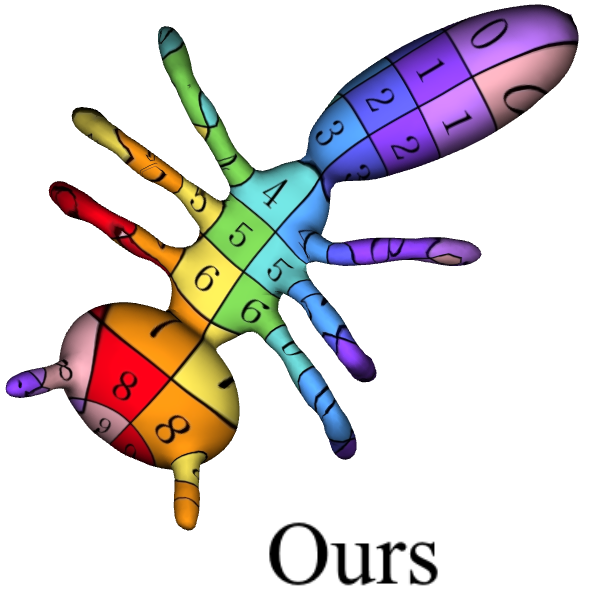}
		\includegraphics[width=.19\linewidth]{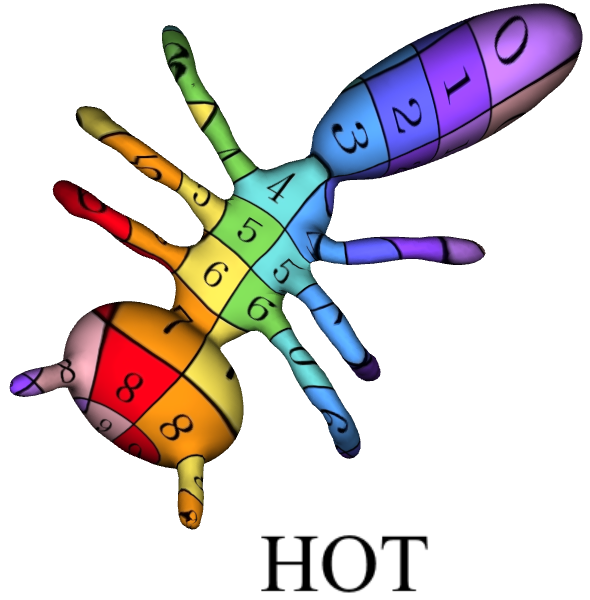}
		\includegraphics[width=.19\linewidth]{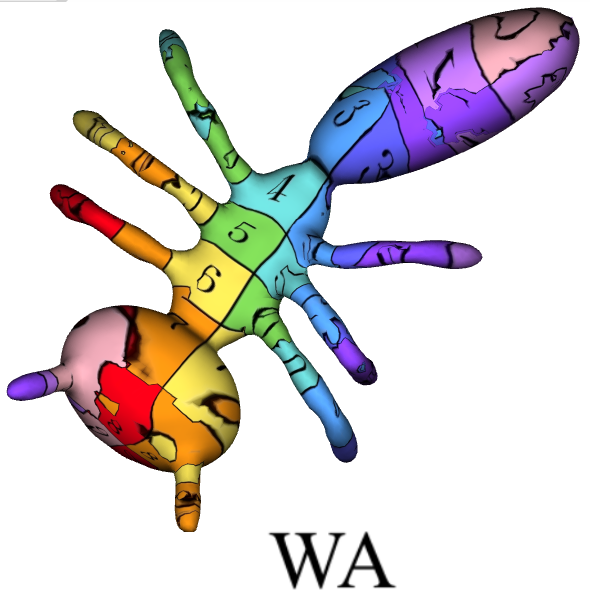}
		\includegraphics[width=.19\linewidth]{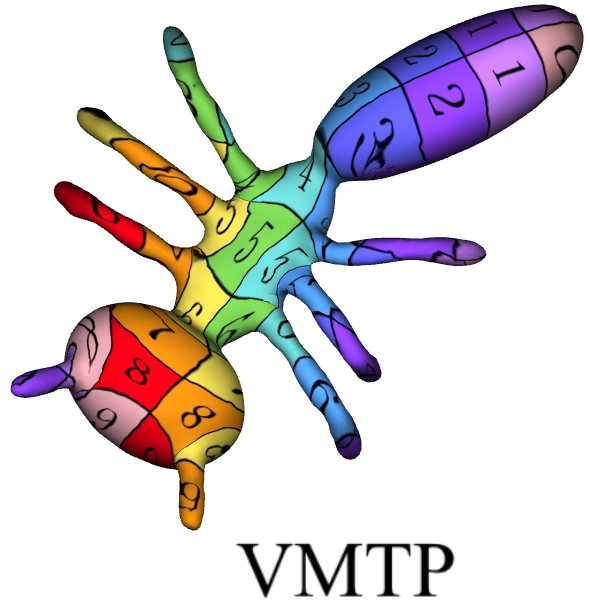}
		\caption{Qualitative results, from input landmarks. $M_1$ in the top row is genus 1 (there is a small hole between the right rear limbs), while $M_2$ is genus 0. From left to right: target texture, our method, HOT~\cite{aigerman2016hyperbolic}, WA~\cite{panozzo2013weighted} and VMTP~\cite{mandad2017variance}. See the text for details.} 
		\label{fig:shrec_ants}
	\end{figure}
	\begin{figure*}[t!]
		\centering
		\includegraphics[width=.31\linewidth]{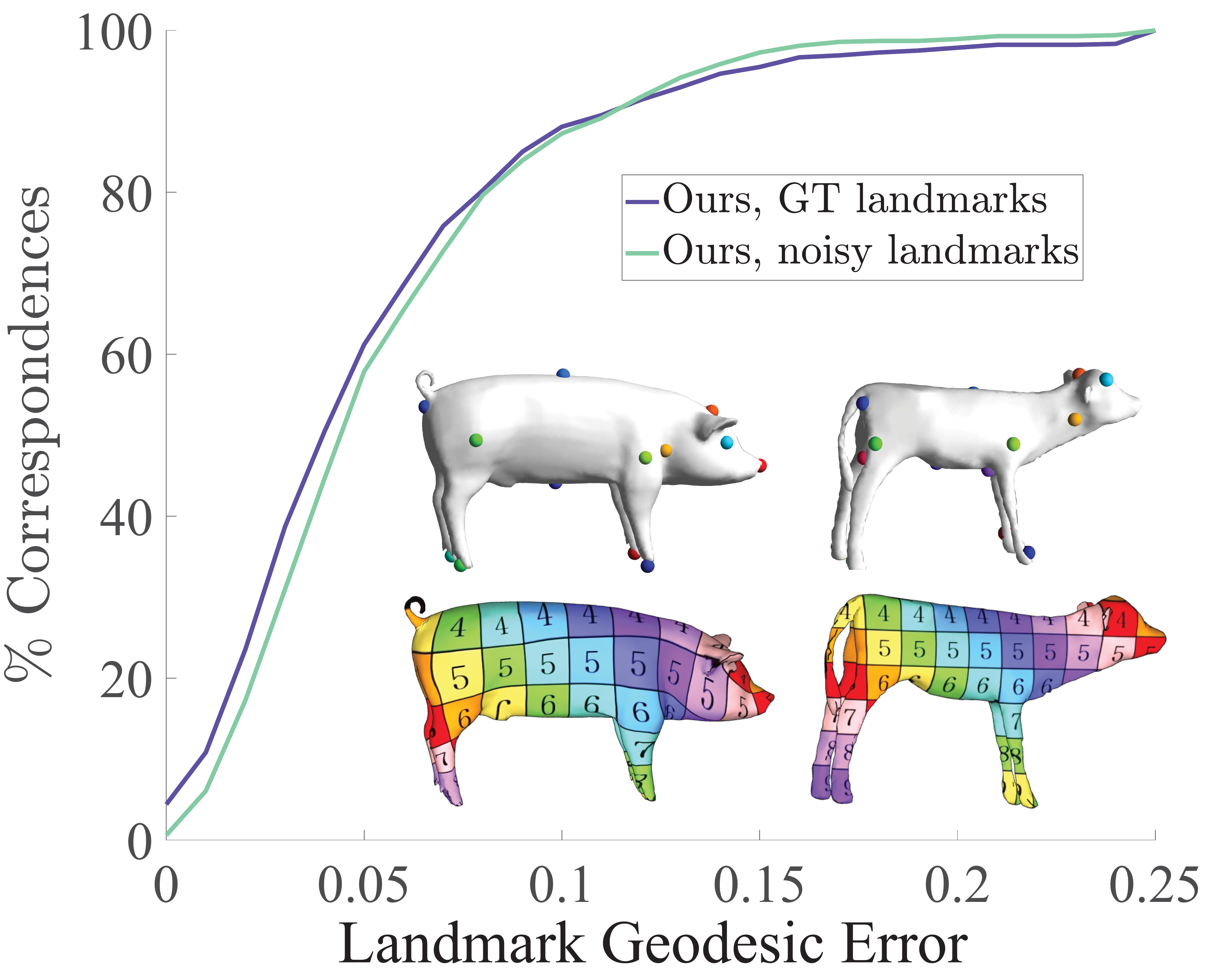}
		\includegraphics[width=.33\linewidth]{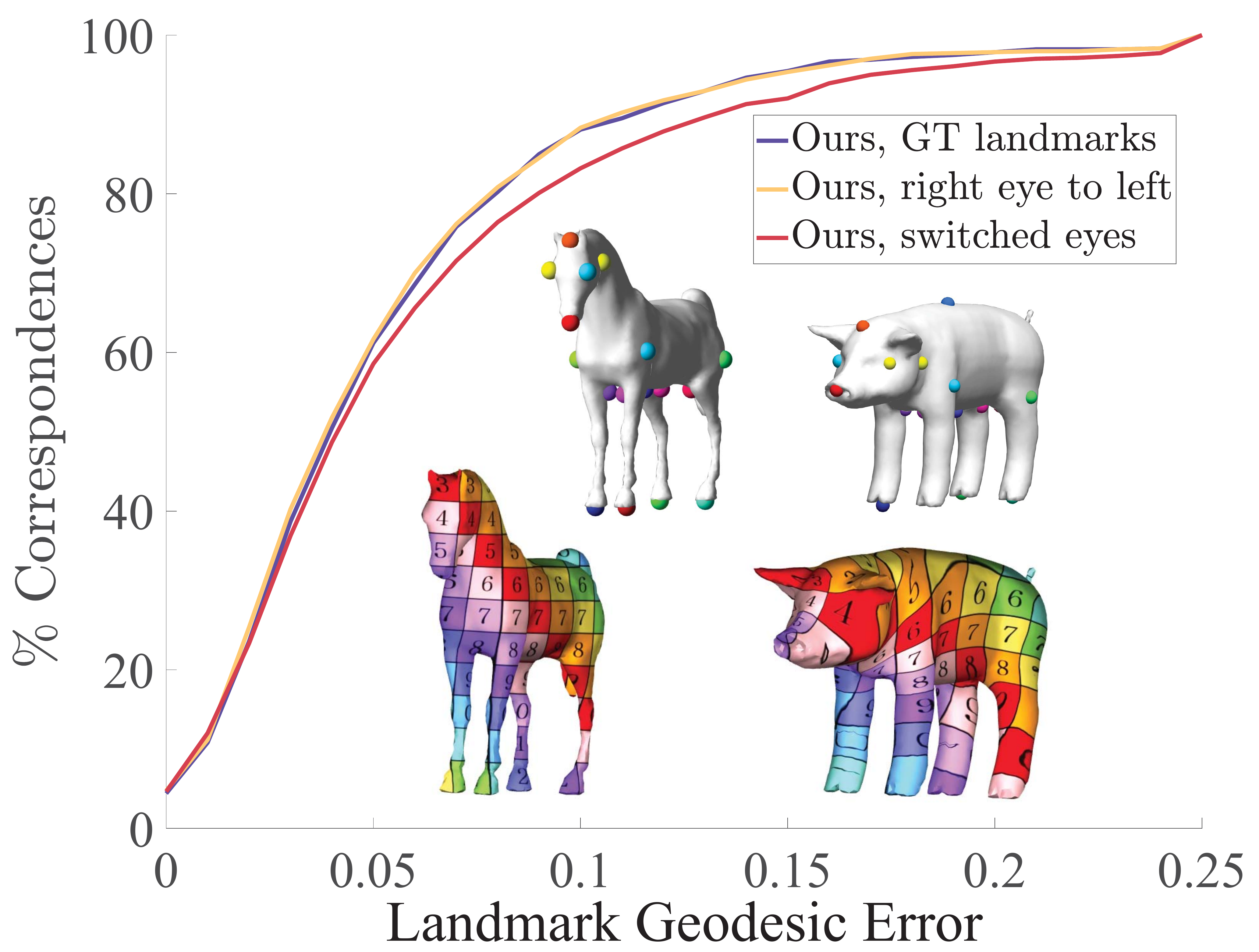}
		\includegraphics[width=.31\linewidth]{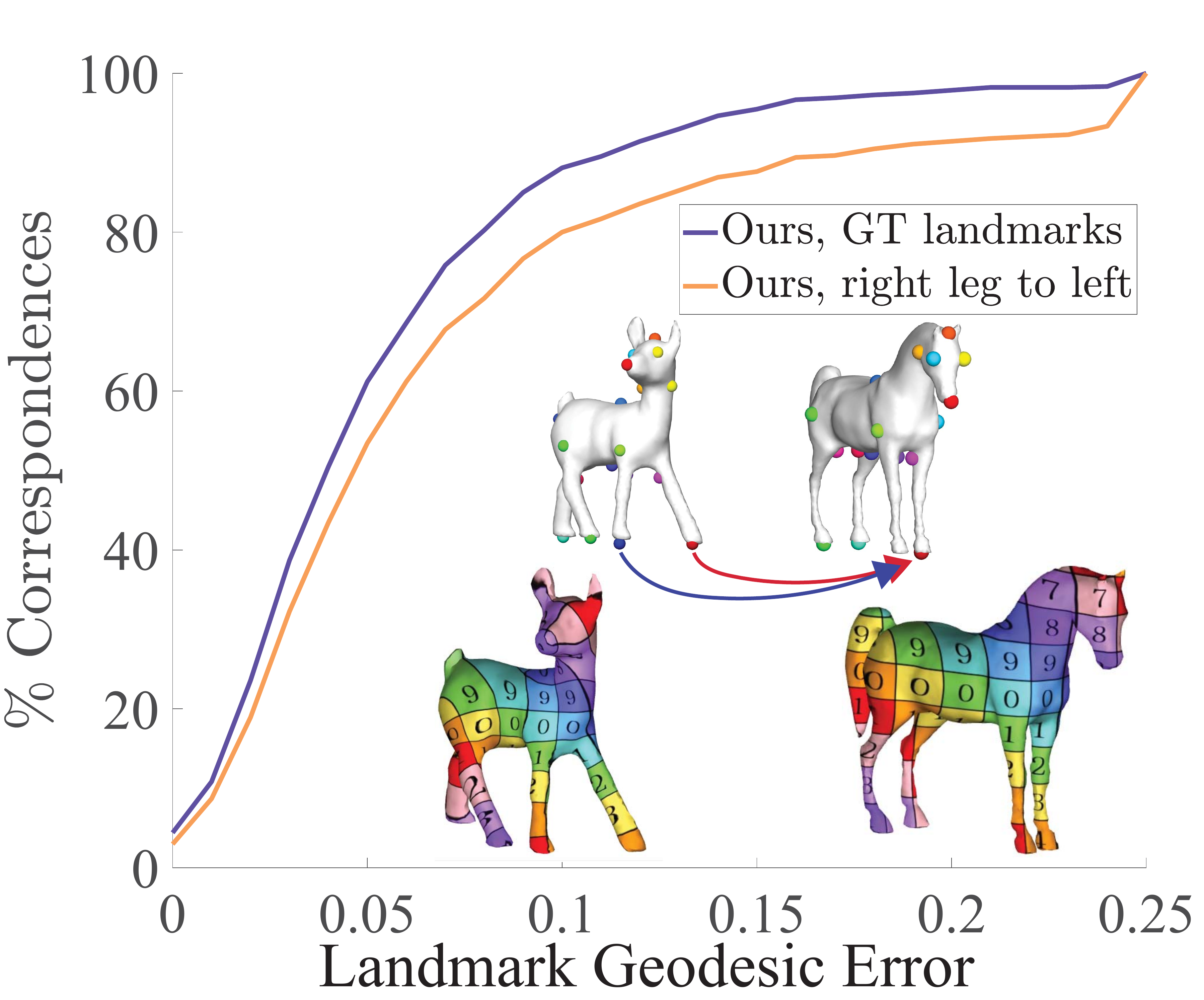}
		\caption{Mapping SHREC quadrupeds by our method, starting from noisy landmarks. Our method is only slightly affected by the noise even when the landmark modification is severe.} 
		\label{fig:noisy_landmarks}
	\end{figure*}  
	\paragraph{Distance from ground truth.} When a ground truth map is given, we measure the distance from the ground truth using the protocol suggested by~\citet[Section 8.2]{kim2011blended}. For every mapped vertex, we measure its geodesic distance from the ground truth location, relative to the square root of the total area of $M_2$, and visualize the percent of vertices whose distortion is less than a given value.

	\paragraph{Compatibility with segmentations.} 
	For some datasets ground truth labeled segmentations are available. In this case, for every pair of shapes and a given map we measure the consistency of the segmentation with respect to the map. This is done by computing the relative vertex area of vertices that are mapped to a face that belongs to the same segment as the source vertex.
	
	\paragraph{Compatibility with symmetry.} For some datasets a ground truth map is only known for a subset of the points, yet a full intrinsic symmetry can be computed for every shape separately. 
	We assume that a good map should respect the intrinsic symmetries of the source and target shapes, given by $S_1,S_2$, respectively. We therefore use these symmetries as input, and measure the compatibility of the map $\phi_{12}$ with the symmetries, given by the geodesic distance $d_{M_2}(S_2(\phi_{12}(v_1)), \phi_{12} (S_1 (v_1))), \forall\!v_1\!\in\!\mV_1$. We visualize the result using a cumulative graph, similarly to the ground truth error. To compute the symmetries we use the method by Kim et al.~\shortcite{kim2011blended}. We also manually filtered the results to use only the accurate symmetries.

	\subsection{Dataset: SHREC, input: landmarks}
	\label{sec:shrec_landmarks}
	We use the BIM benchmark~\cite{kim2011blended} that provides more than 200 pairs of highly non-isometric shapes from the SHREC dataset~\cite{shrec07} with user-verified landmarks.
	We compare our method with a state of the art parameterization based method~\cite{aigerman2016hyperbolic} (HOT) and the weighted averages method~\cite{panozzo2013weighted} (WA). 
	Both receive as input landmark points, which are not modified during the optimization, and generate precise maps.
	In addition we compare to the recent method by~\citet{mandad2017variance} (VMTP), that similarly gets as input landmark points, yet can modify them during the optimization. Since VMTP requires uniform isotropic meshes, we recursively add edges using the longest edge bisection method to meshes with less than 10K vertices, before applying VMTP.
	All the methods we compare with produce \emph{precise} maps, as vertex-to-vertex maps induce high local distortion.
	As input to our method we also use the user defined landmarks, and extend them to a full initial map as described in Section~\ref{sec:initialization}. The landmarks are not used after the initialization.
	
	\begin{figure}[b!]
		\centering
		\includegraphics[width=.3\linewidth]{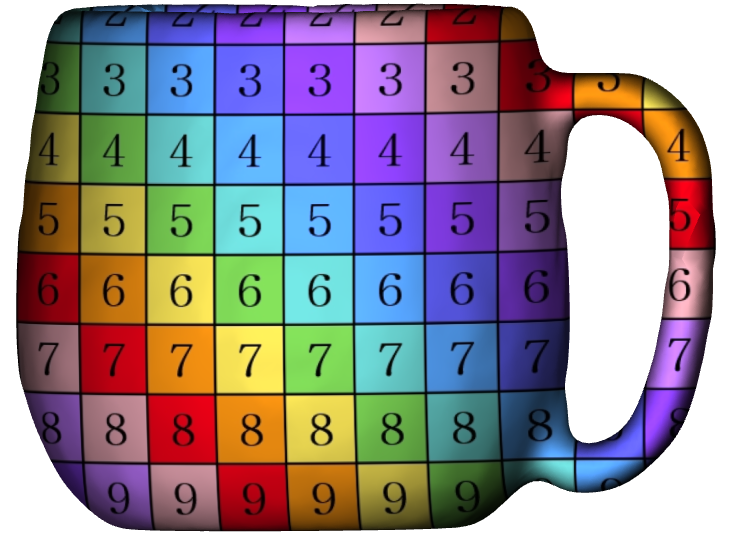}
		\includegraphics[width=.3\linewidth]{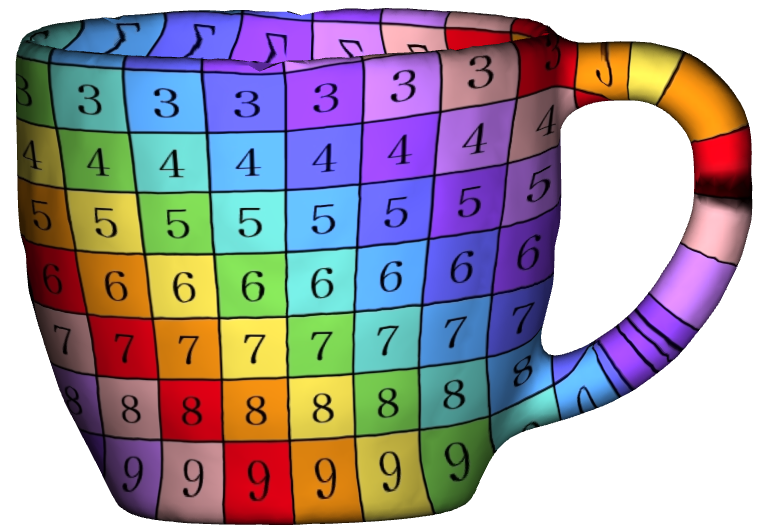}
		\includegraphics[width=.3\linewidth]{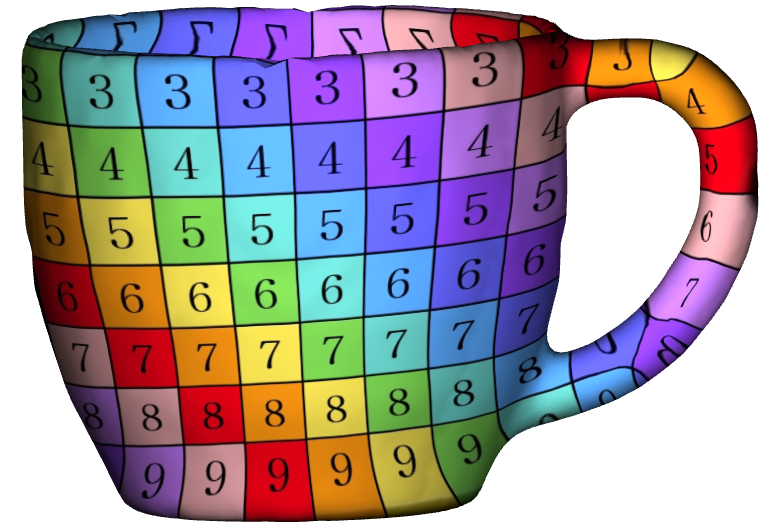}

		\includegraphics[width=.3\linewidth]{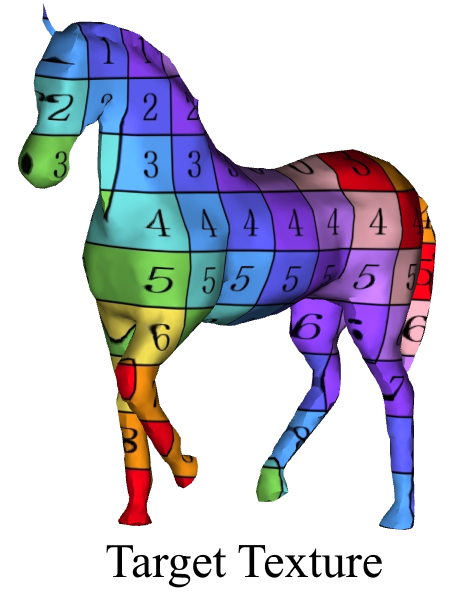}
		\includegraphics[width=.3\linewidth]{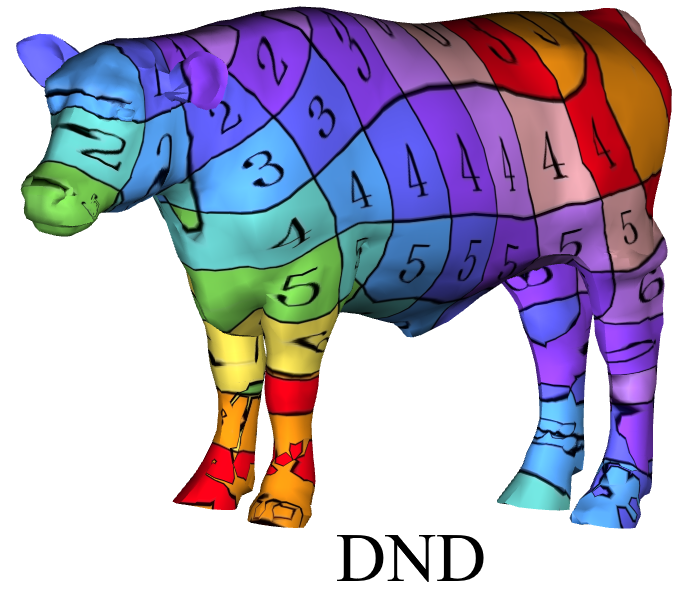}
		\includegraphics[width=.3\linewidth]{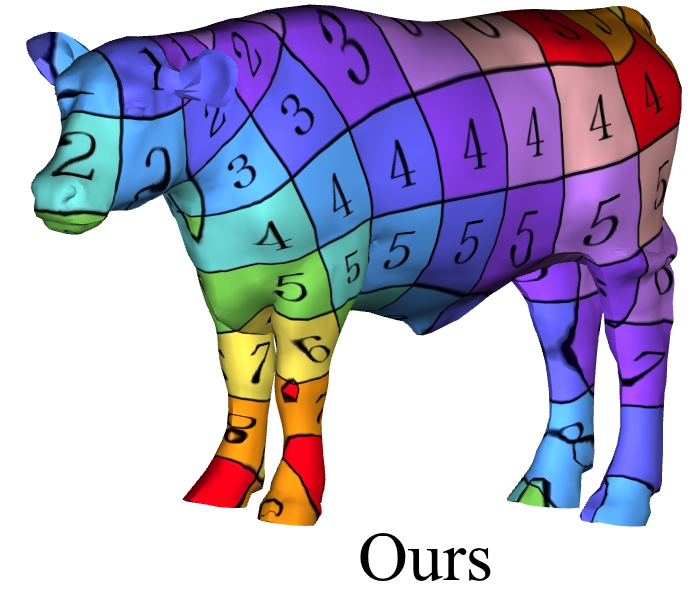}
		
		\includegraphics[width=.49\linewidth]{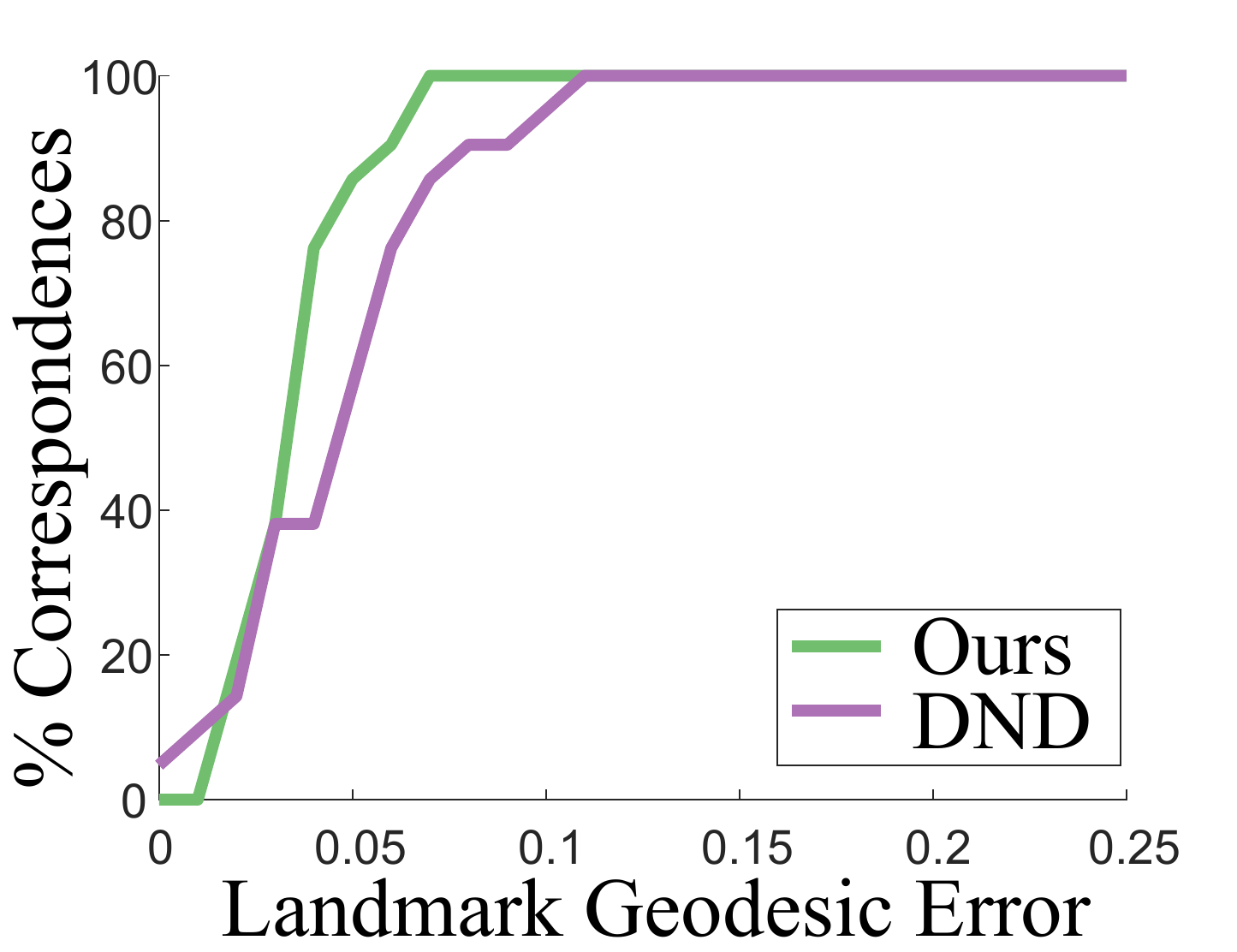}
		\includegraphics[width=.49\linewidth]{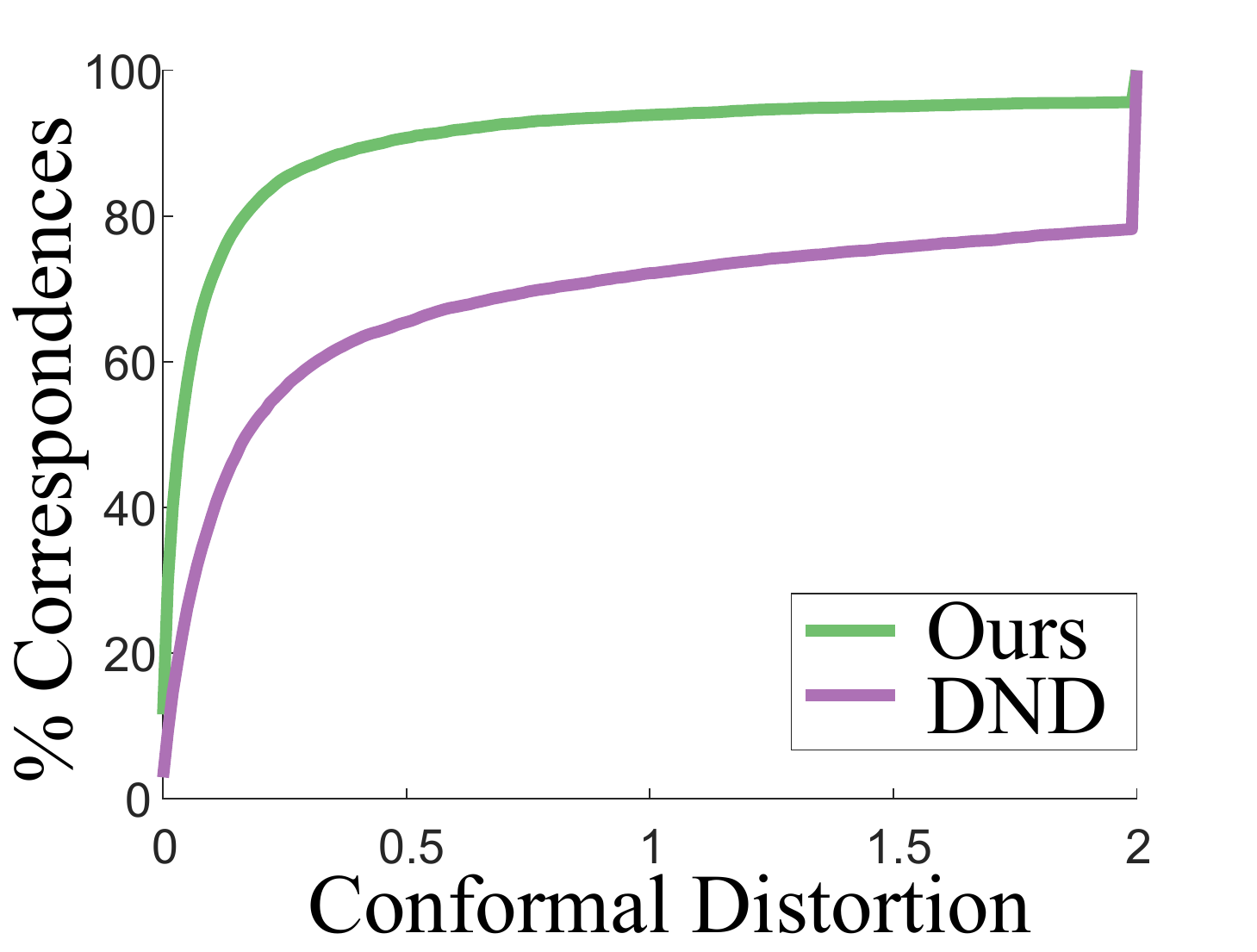}
		\caption{Qualitative and quantitative comparison starting with a functional map computed from landmarks. From left to right: target texture, \cite{ezuz2017deblurring} (DND), our method. Notice the difference at the cup handle and the legs. } 
		\label{fig:fmaps_texture}
	\end{figure}
	\begin{figure*}[t!]
		\centering
		
		\includegraphics[width=.23\linewidth]{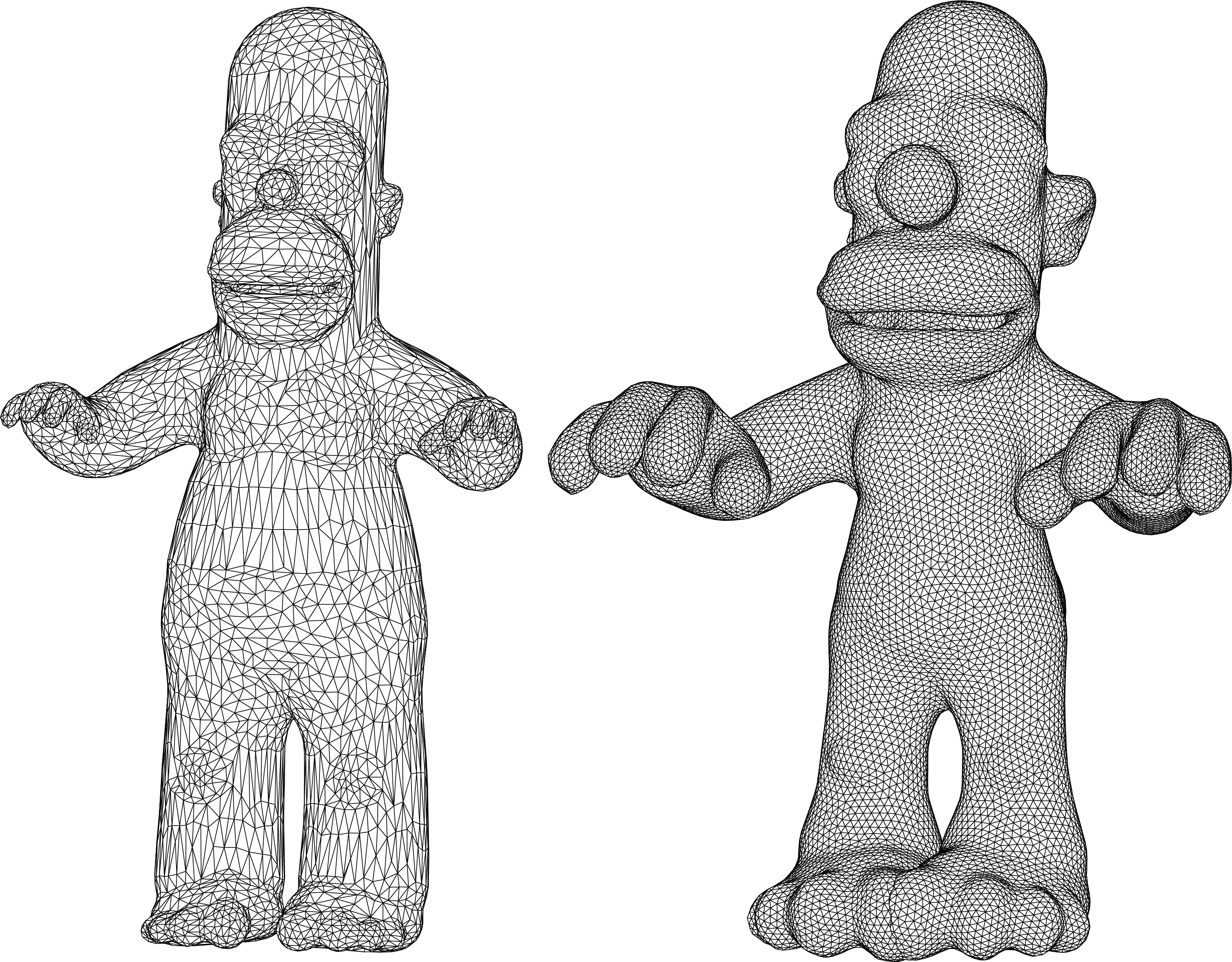}
		\includegraphics[width=.13\linewidth]{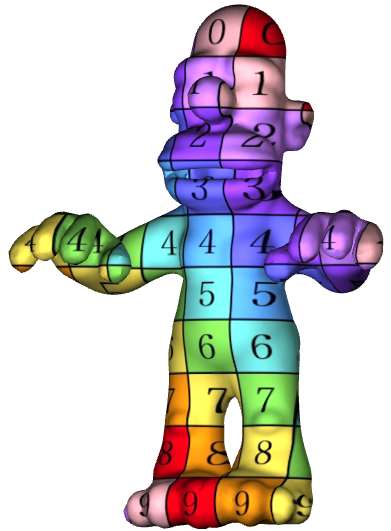}
		\includegraphics[width=.10\linewidth]{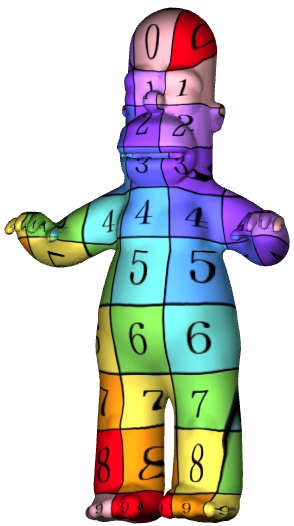}
		\includegraphics[width=.10\linewidth]{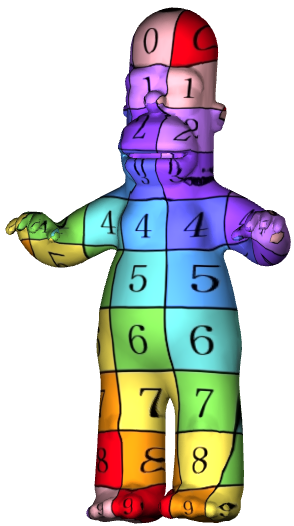}
		\includegraphics[width=.10\linewidth]{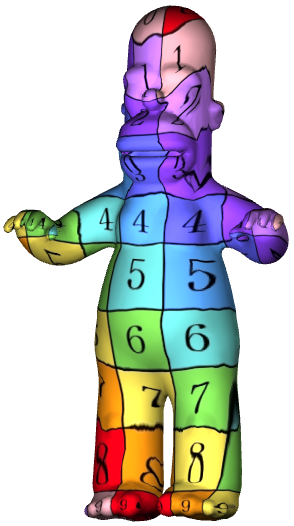}
		\includegraphics[width=.10\linewidth]{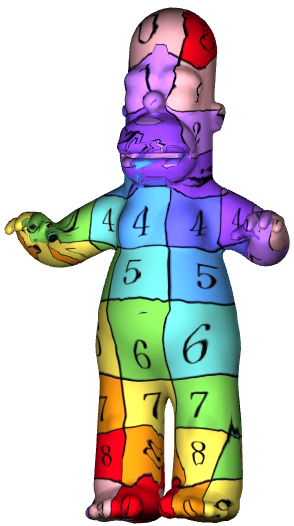}
		\includegraphics[width=.10\linewidth]{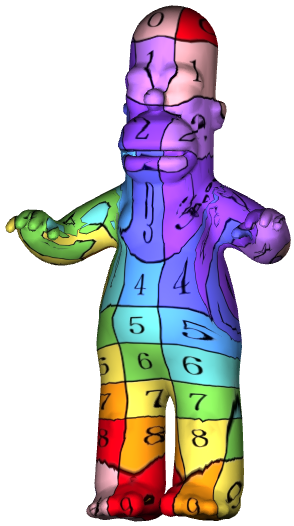}
		\vspace{0.01\linewidth}
		
		\includegraphics[width=.22\linewidth]{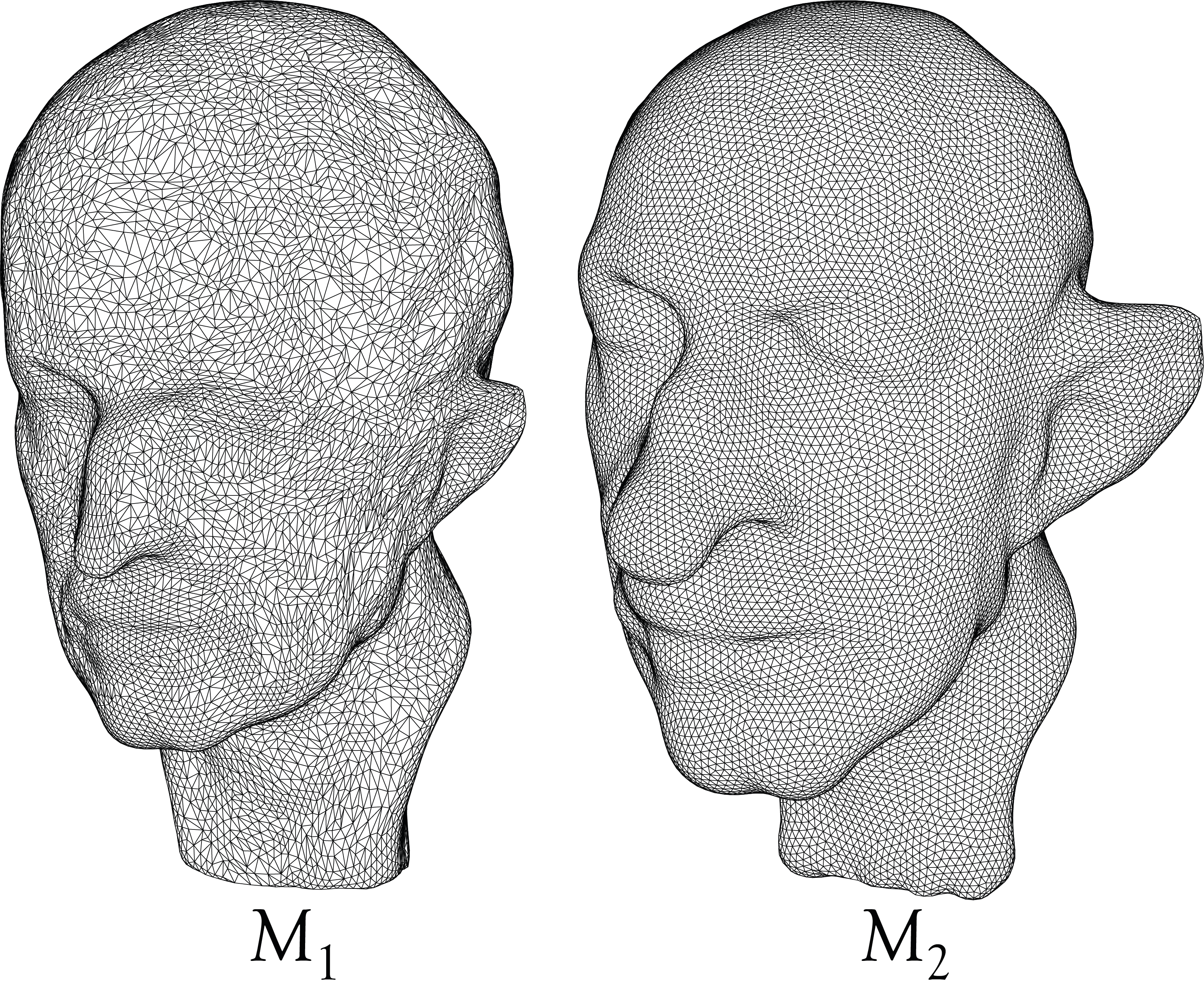}
		\includegraphics[width=.13\linewidth]{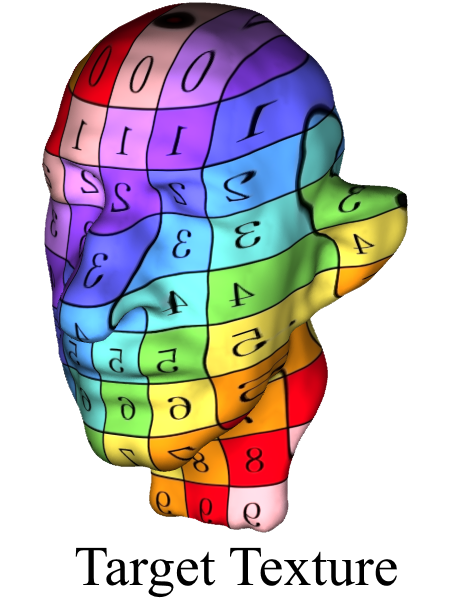}
		\includegraphics[width=.10\linewidth]{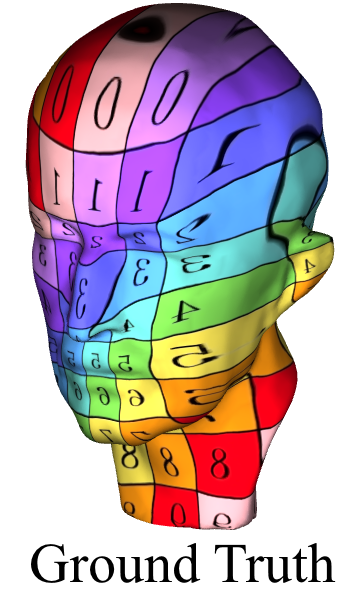}
		\includegraphics[width=.10\linewidth]{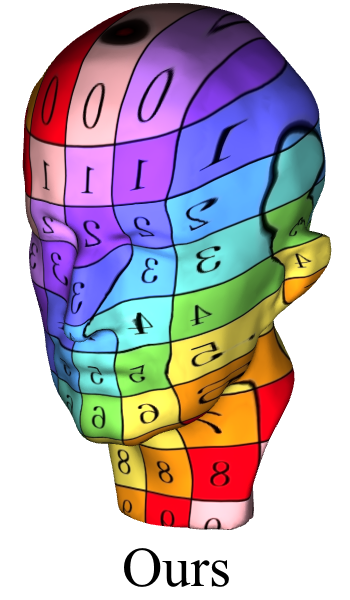}
		\includegraphics[width=.10\linewidth]{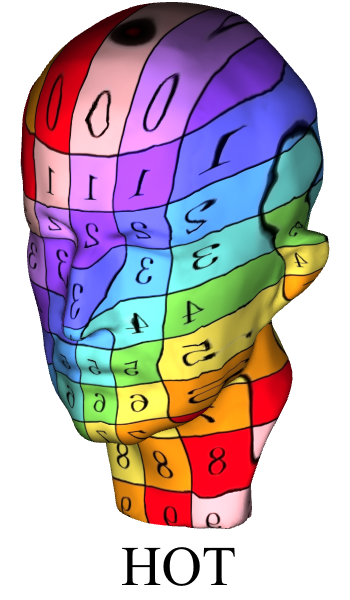}
		\includegraphics[width=.10\linewidth]{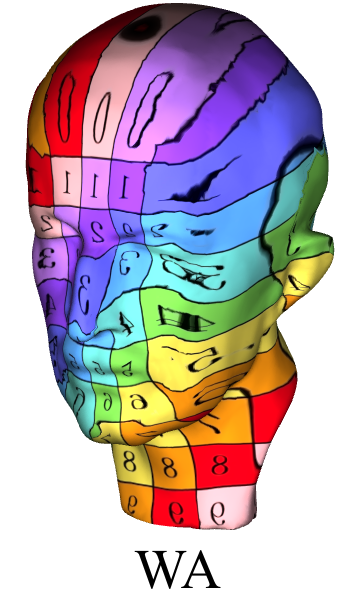}
		\includegraphics[width=.10\linewidth]{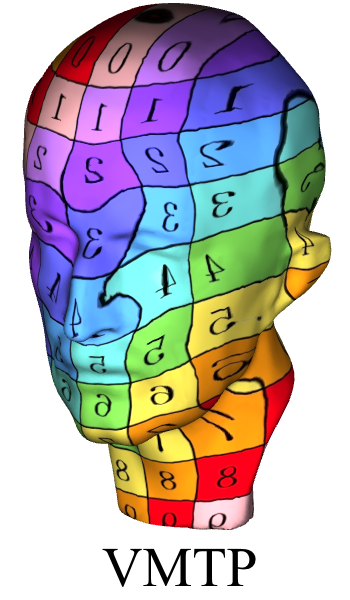}
		\caption{Qualitative result, caricatures. $M_2$ was generated by deforming $M_1$~\cite{Sela20151}, and the deformation defines a ground truth map. $M_2$ was remeshed so that the source and target shapes do not have the same connectivity.}
		\label{fig:homers_texture}
	\end{figure*}
	Quantitative results are shown in Figure~\ref{fig:shrec_graphs}, where we measure conformal distortion, compatibility with symmetries and distance from the ground truth landmarks. Note that our method achieves the best conformal distortion. In addition, the distance from the ground truth landmarks is also improved when compared to the other method which modifies them (VMTP). As shown in section~\ref{sec:noisy_landmarks}, the option to modify the input landmarks is valuable when the input is not completely reliable. 
	In terms of compatibility with symmetry, our method is comparable with existing techniques, notably achieving a better ratio of perfect matches with about 15\% of the vertices exactly symmetric for  our method, where the next best method has less than 10\% exactly symmetric vertices.
	On this dataset ground truth segmentations are also available~\cite{Kalogerakis:2010:labelMeshes,Chen:2009:ABF}, and measuring the relative mapped area which is compatible with the segmentations we have HOT: 90.39\%, WA: 90.35\%, VMTP: 81.8\%, our method: 89.62\%. Therefore, this measure also demonstrates that our maps are as compatible semantically as existing techniques, while being considerably more conformal.
	
	Qualitative results are shown in Figure~\ref{fig:shrec_vis}, where we have selected a subset of pairs to visually show the differences between the maps. In every row we show, from left to right, the target texture, and the results of our method, HOT, WA and VMTP. 
	Figure~\ref{fig:shrec_ants} shows qualitative results of the mapping methods for two pairs of ants, where the pair on the top row is of different genus. Despite the topological difference, our method generates a semantically accurate map.

	\subsection{Dataset: SHREC quadrupeds, input: noisy landmarks}
	\label{sec:noisy_landmarks}
	In many cases, the selection of the landmarks by the user has some variability (see, e.g.~\cite{chen2012schelling}), and it might be better to treat these landmarks as guidelines rather than exact ground truth. Our approach is compatible with this notion, since our method only uses the landmarks for initialization, and their final location will, in most cases, vary from their initial one. To check the sensitivity of our approach to the landmark locations, we repeated the experiment from Section~\ref{sec:shrec_landmarks} with various landmark modifications. Figure~\ref{fig:noisy_landmarks} shows the landmark geodesic error of the output maps when starting from the noisy landmarks compared to starting from the original landmarks. We ran the experiment on the ``quadrupeds'' class (20 pairs) from the SHREC dataset and did the following modifications: (a) moved every landmark randomly to a vertex in its $5$-ring neighborhood, (b) switched between the two eyes, or mapped both eyes on $M_1$ to a single eye on $M_2$, and (c) mapped both feet on $M_1$ to the same foot on $M_2$. In addition to the error graph we show some example maps, as well as the input noisy landmarks. As the figure shows, our results are not sensitive to landmark noise, and even a relatively severe modification, such as mapping both feet to the same foot, yields good qualitative and quantitative results. 
	\begin{figure}[b!]
		\centering
		\includegraphics[width=.49\linewidth]{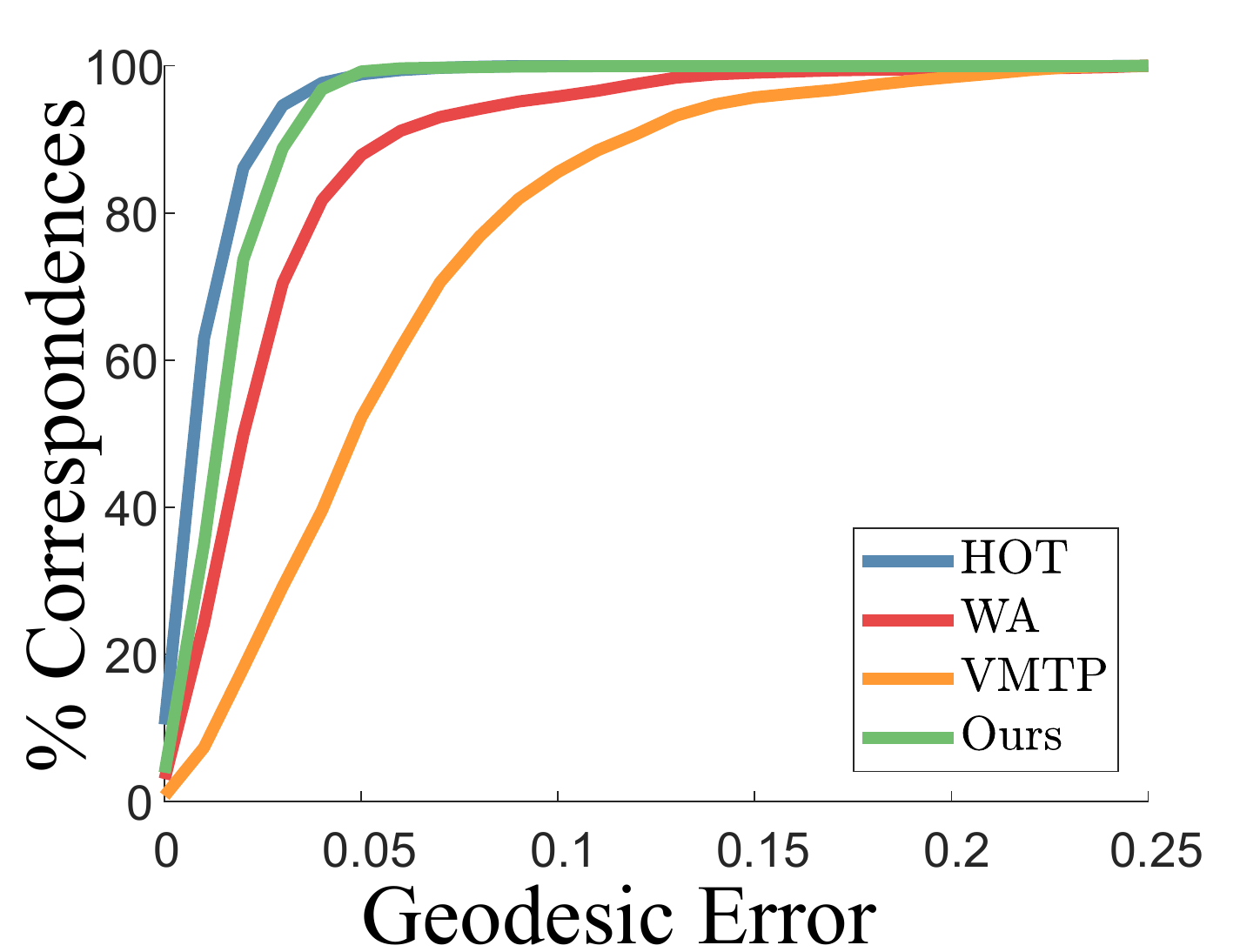}
		\includegraphics[width=.49\linewidth]{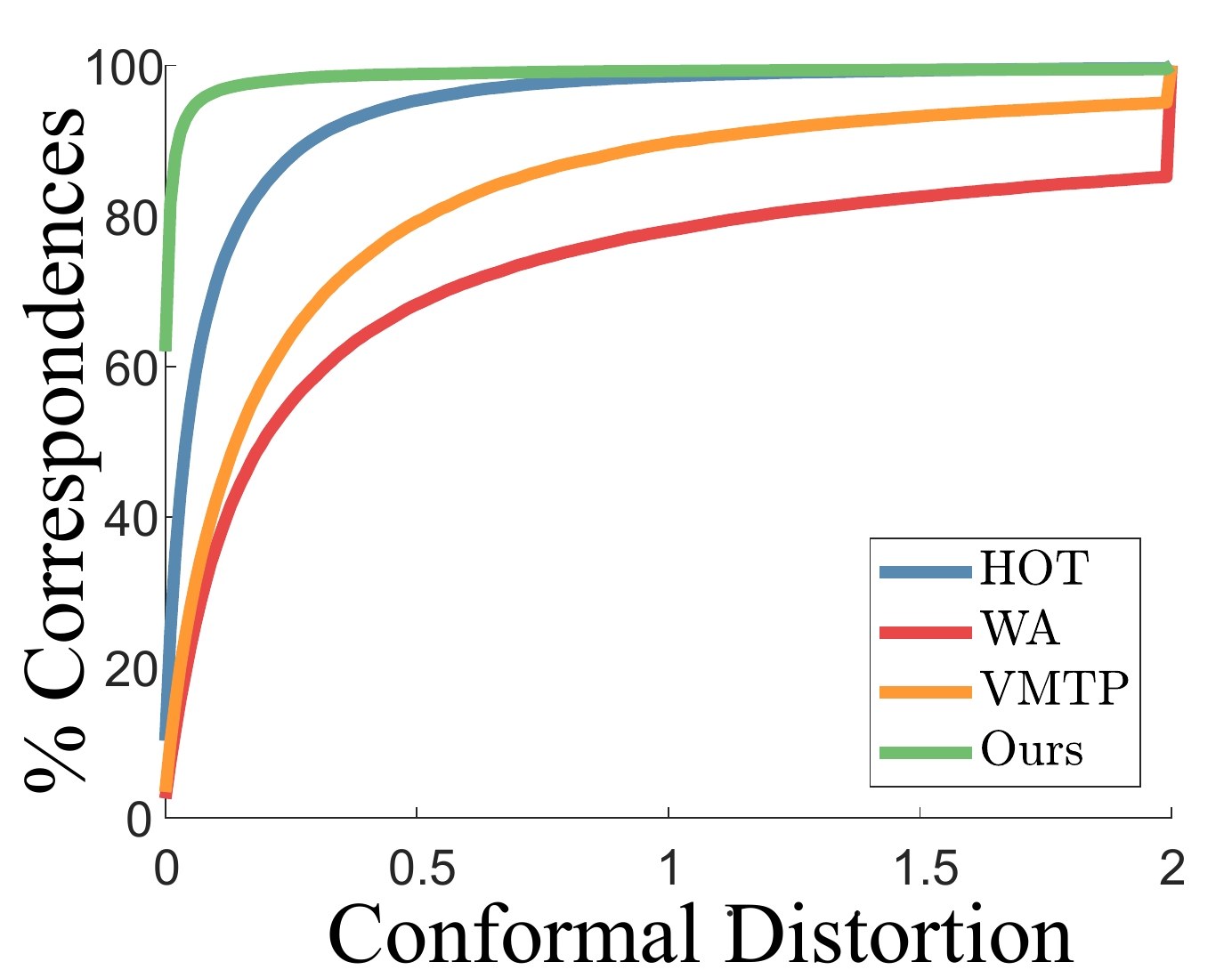}
		\caption{Quantitative measures for the meshes in Figure~\ref{fig:homers_texture}. Note that our method achieves a considerably better conformal distortion, while maintaining ground truth error comparable to existing methods.} 
		\label{fig:homers_graphs}
	\end{figure}
	
	\begin{figure*}[t]
		\centering
		\includegraphics[width=.16\linewidth]{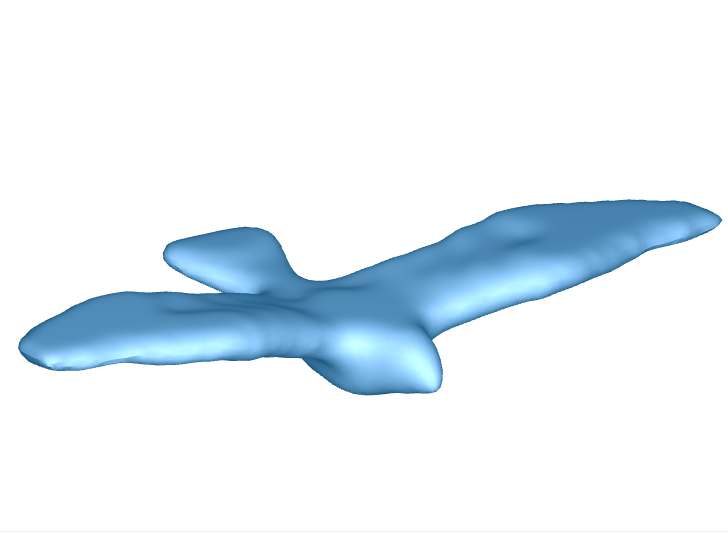}
		\includegraphics[width=.16\linewidth]{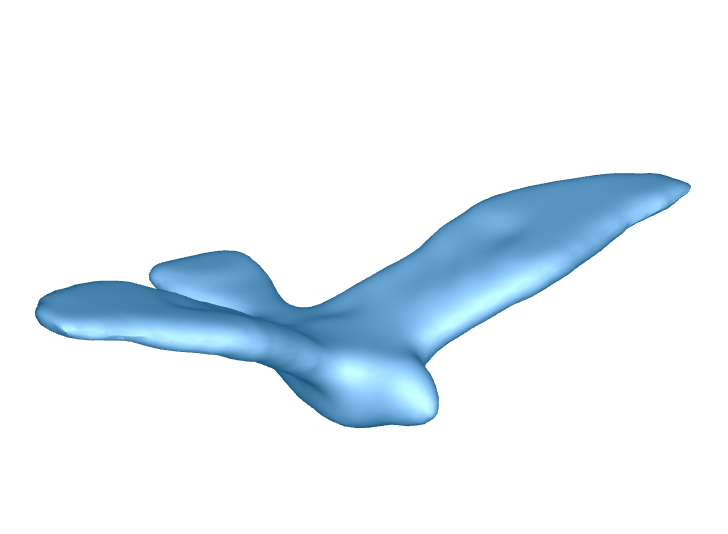}
		\includegraphics[width=.16\linewidth]{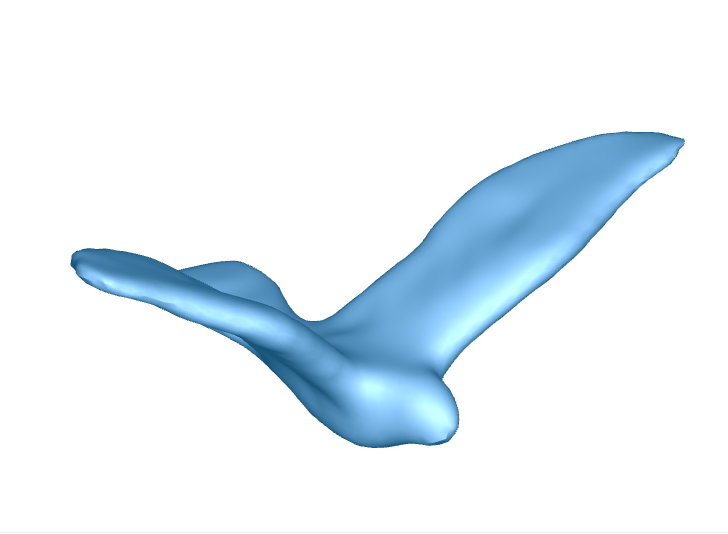}
		\includegraphics[width=.16\linewidth]{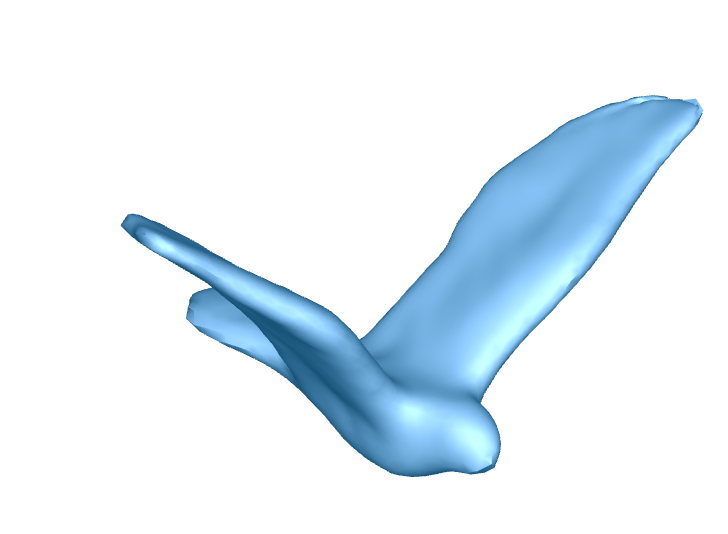}
		\includegraphics[width=.16\linewidth]{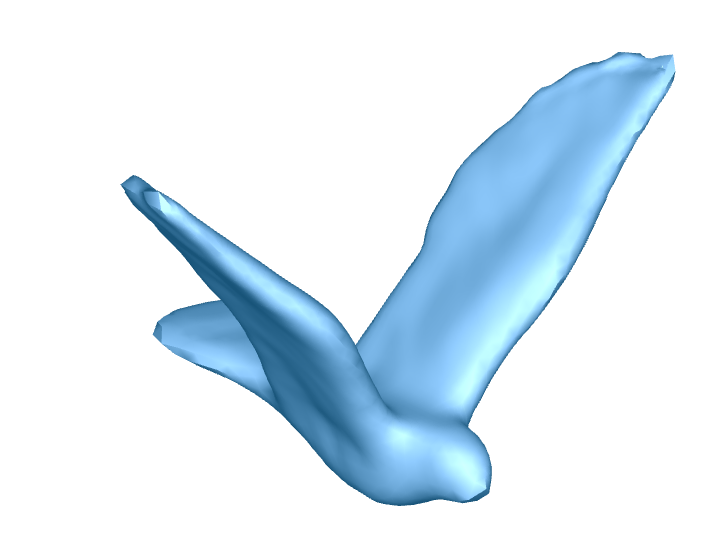}
		\includegraphics[width=.16\linewidth]{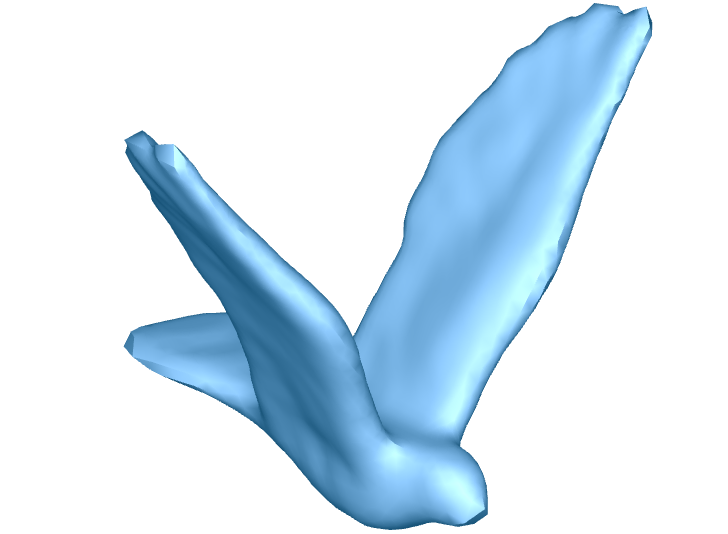}
		\caption{Shape interpolation using our computed correspondence as input for~\cite{HeRuSc14}.} 
		\label{fig:interpolation}
	\end{figure*}
	\subsection{Dataset: SHREC two pairs, input: functional map}
	The functional map~\cite{ovsjanikov2012functional} machinery is quite versatile, and allows to compute generalized maps in a variety of cases. Our method can also be used to extract a precise pointwise map from a given functional map. We use the SHREC dataset with its landmark data from the BIM benchmark~\cite{kim2011blended}, and use the landmarks to compute a functional map using the Wave Kernel Map and the Wave Kernel Signature~\cite{aubry2011wave}. Any other recent method for computing functional maps could be used as well. We provide the functional map as input to our approach and the recent map deblurring approach~\cite{ezuz2017deblurring} (DND), which is the only other method that recovers \emph{precise} maps from a functional map. Specifically, we used the consistency extension of DND with $\alpha=0.8$.
	Figure~\ref{fig:fmaps_texture} qualitatively visualizes the difference between the methods for two pairs of shapes. Note the map improvement on the handle of the cup and the legs of the cow. We also show graphs of the conformal distortion and ground-truth error of the landmarks.

	\subsection{Dataset: caricatures, input: landmarks}
	One of the advantages of our formulation is its simplicity, that leads to flexibility in adding additional components to the energy. For example, in some cases it can be beneficial to add \emph{weak} landmark constraints, to encourage feature points to remain in the neighborhood of the input landmarks.
	
	\begin{figure}[b!]
		\centering
		\includegraphics[height=.35\linewidth]{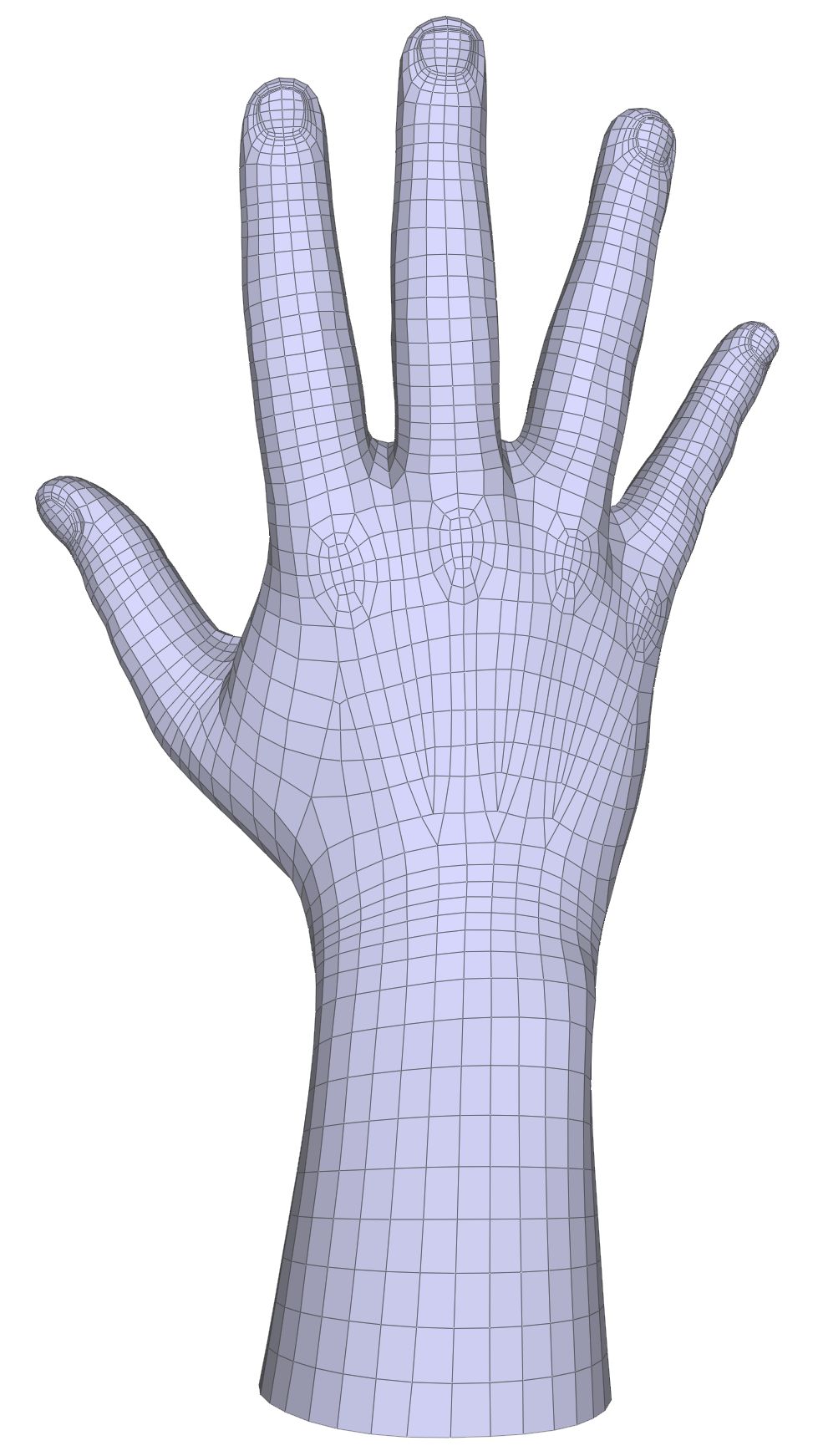}
		\includegraphics[height=.35\linewidth]{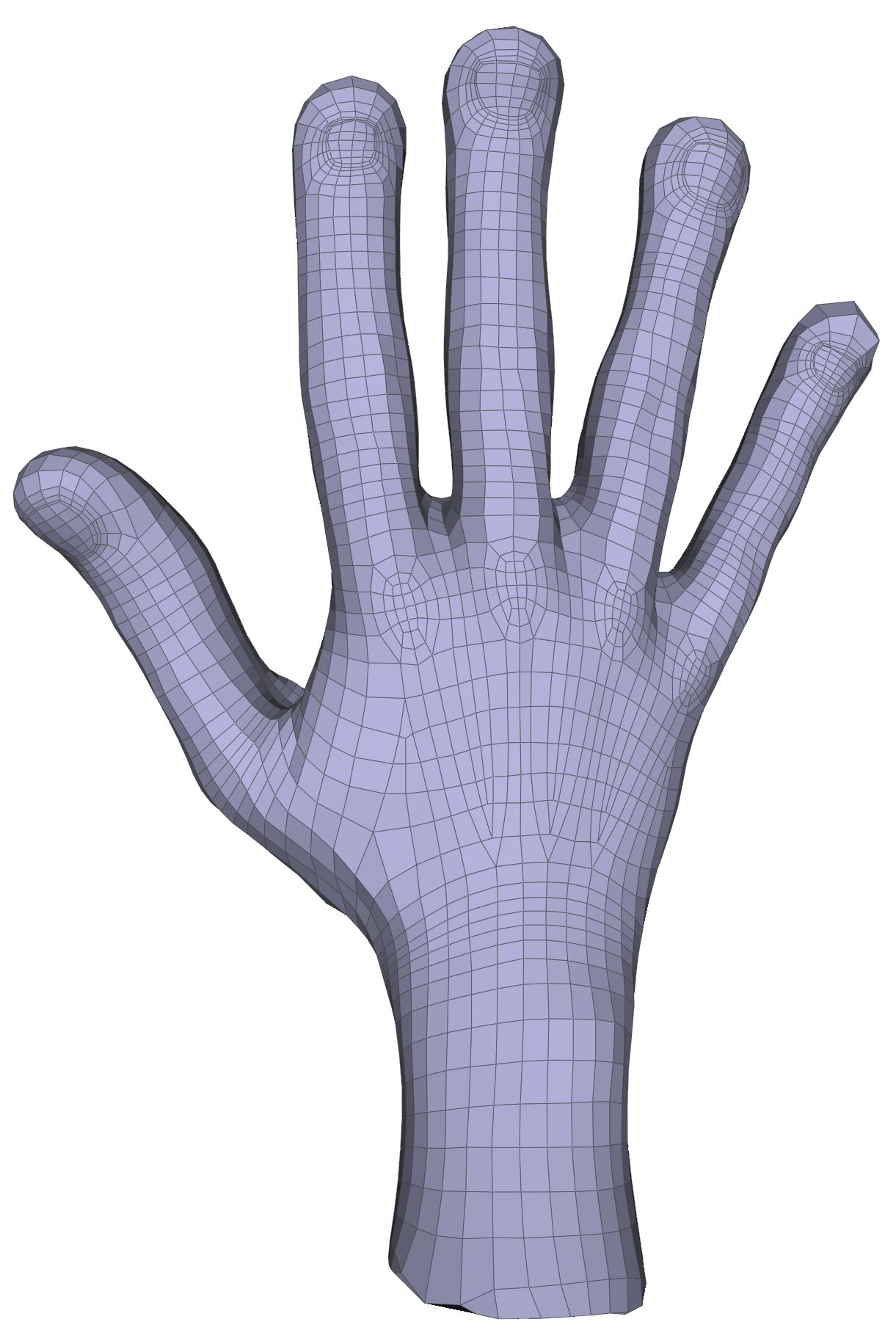}	
		\includegraphics[height=.35\linewidth]{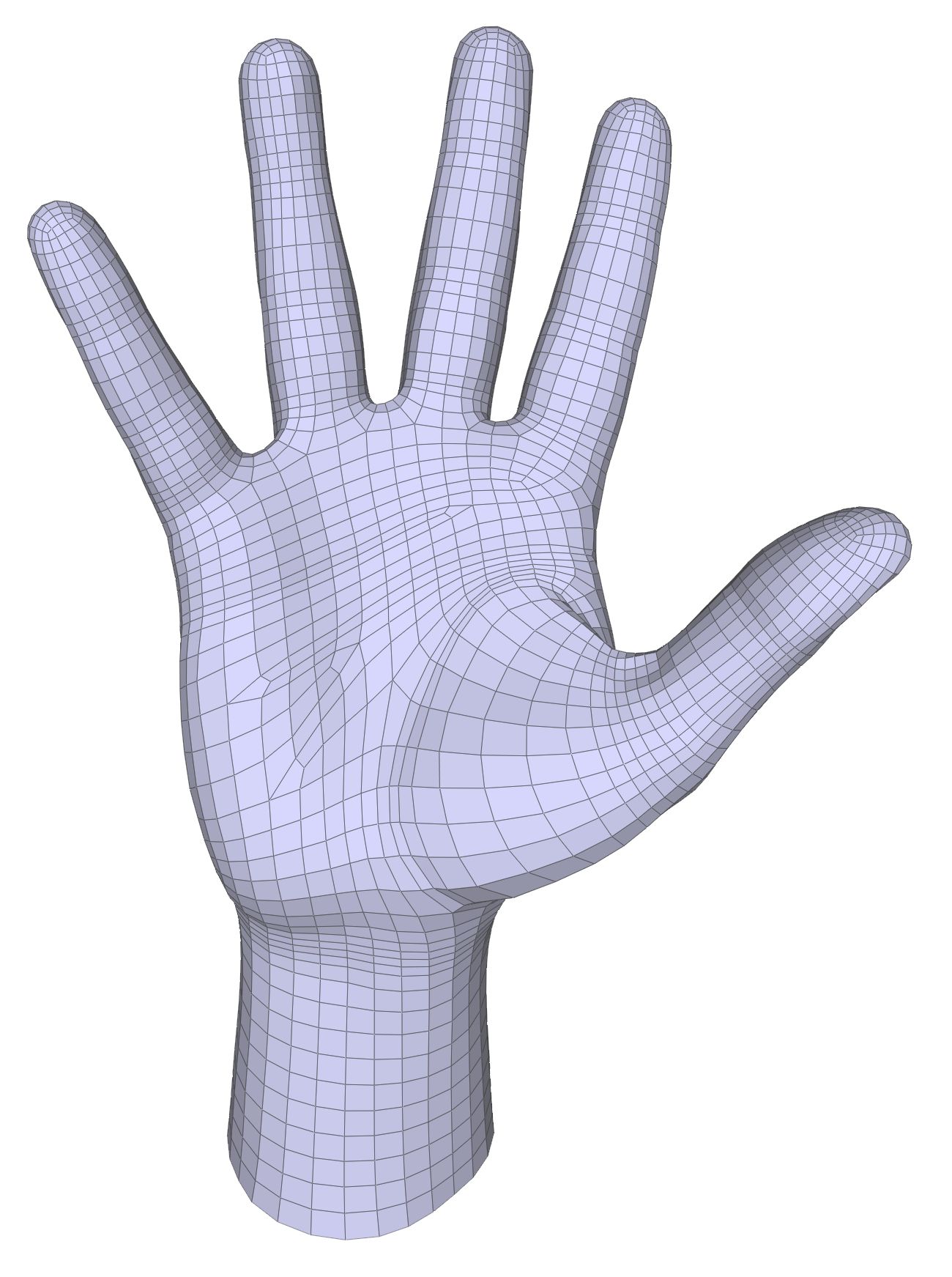}
		\includegraphics[height=.35\linewidth]{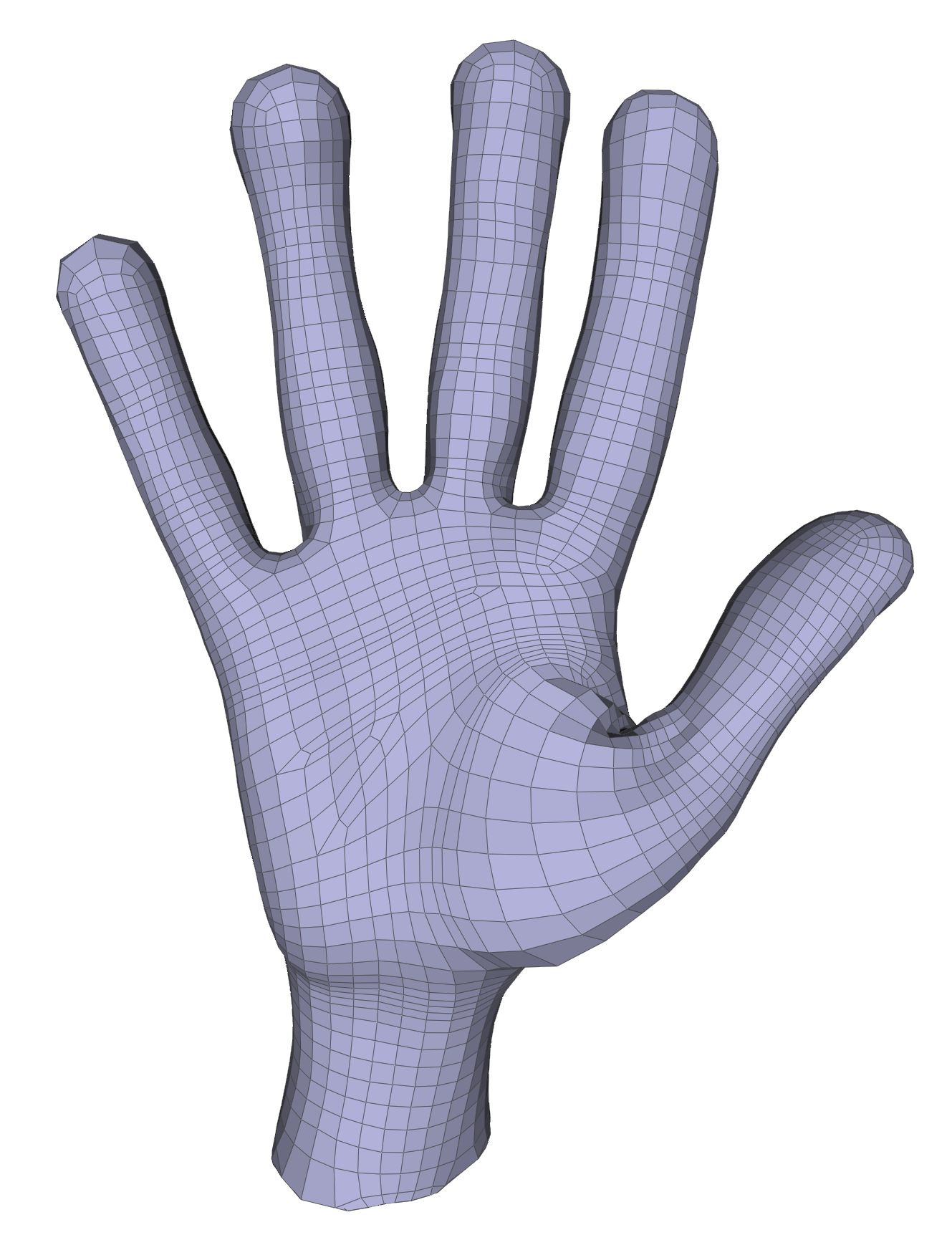}	
		\caption{Quad mesh transfer using our computed correspondence. Left: input quad mesh, right: output quad mesh. Note the preservation of the prominent edge flows in the quad mesh, such as the fingernails and knuckles.} 
		\label{fig:quad_hands_neut_lands_weak}
	\end{figure}
	
	\paragraph*{Weak landmark constraints.}
	Given pairs of matching landmarks $p_i\!\in\!\mV_1, q_i\!\in\!\mV_2, i = 1\dots r$, we add the following term to the energy:
	\begin{equation}
	\gamma \sum _{i=1} ^r A_1(p_i) \| X_{12} \left( p_i,: \right) - X_2 \left( q_i \right) \|^2_{M_2} + A_2(q_i) \| X_{21} \left( q_i,: \right) - X_1 \left( p_i \right) \|^2_{M_1}.
	\label{eq:weak_landmarks}
	\end{equation}
	The value of $\gamma$ depends on the reliability of the landmarks, if the landmarks are not accurate then $\gamma$ should be small.
	To demonstrate the effectiveness of this approach, we used as input the Homer and Max Planck models, and caricatures of these models generated using the method by~\citet{Sela20151}. The caricatures have the same triangulation as the original shapes. We remesh the caricatures to avoid bias, and project the original caricature to the remeshed model to generate the ground truth for validation. As input to all approaches, we picked 28 and 14 landmarks from the input map for the homer and Max models respectively. For our algorithm we added the energy in Equation~\eqref{eq:weak_landmarks} with $\gamma=1$. The qualitative and quantitative results are shown in figures~\ref{fig:homers_texture} and~\ref{fig:homers_graphs}, respectively. As in previous experiments, we achieve considerably better conformal distortion, with comparable ground truth error.

	\subsection{Application: Shape Interpolation}
	Existing shape interpolation methods require the input shapes to share the same connectivity, while real data rarely satisfies this requirement. Our mapping can be used to \emph{remesh} the target surface $M_2$ using the image of the vertices of $M_1$, given by $P_{12} V_2$, and the connectivity of $M_1$.
	We used our map between two birds from SHREC to demonstrate this application, starting from the BIM landmarks as described in section~\ref{sec:shrec_landmarks}. 
	After remeshing $M_2$, as a few faces had a zero area, we iteratively moved vertices of the degenerate faces to an average of their 1-ring neighbors until there were no degenerate faces. We then used the shape interpolation method by~\citet{HeRuSc14} to interpolate between the shapes, as shown in Figure~\ref{fig:interpolation}. Note that we correctly mapped the wings, head and tail of the birds, as is evident from the natural interpolation results.

	\subsection{Application: Quad Mesh Transfer}
	Finally, we demonstrate a potential application of our correspondence to quad mesh transfer of artist-generated quad meshes to scanned meshes. In this experiment, we start from a set of 41 landmark points, and use weak landmark constraints with $\gamma = 5\cdot 10^{-5}$.
	We choose to have a larger number of landmarks in this example to preserve the fine features such as fingernails and joints.
	The results are shown in Figure~\ref{fig:quad_hands_neut_lands_weak}, with two views of the input and output quad meshes on the left and right, respectively. Note, that the edge flow of the output quad mesh closely follows the features of the hand, and the special structures in the input, such as the fingernails and knuckles, are nicely preserved in the output mesh. Such a high quality transfer can only be achieved if the computed map has a low conformal distortion, which leads to well preserved quad shapes. 
	
	\section{Conclusion}
	
	We have suggested a novel correspondence method between non-isometric shapes, which considerably improves conformal distortion over existing techniques while preserving semantic features. Our approach is based on an alternating minimization of a quadratic energy, which is efficient, easy to implement, and flexible. In addition to demonstrating effectiveness on various benchmarks, we have shown applications to quad mesh transfer and shape interpolation. 
	
	Beyond its current utility, our formulation can serve as a framework for more involved energies on the differential of the map. For example, one can consider various metrics, such as non-isotropic, higher order, sparsity based and data-driven. We hope that our approach can be a stepping stone to generalizing the large body of literature on planar shape parameterization to the more general setting of shape correspondence.   
	
	\bibliographystyle{ACM-Reference-Format}
	\bibliography{draft}


\begin{thebibliography}{54}


\ifx \showCODEN    \undefined \def \showCODEN     #1{\unskip}     \fi
\ifx \showDOI      \undefined \def \showDOI       #1{#1}\fi
\ifx \showISBNx    \undefined \def \showISBNx     #1{\unskip}     \fi
\ifx \showISBNxiii \undefined \def \showISBNxiii  #1{\unskip}     \fi
\ifx \showISSN     \undefined \def \showISSN      #1{\unskip}     \fi
\ifx \showLCCN     \undefined \def \showLCCN      #1{\unskip}     \fi
\ifx \shownote     \undefined \def \shownote      #1{#1}          \fi
\ifx \showarticletitle \undefined \def \showarticletitle #1{#1}   \fi
\ifx \showURL      \undefined \def \showURL       {\relax}        \fi
\providecommand\bibfield[2]{#2}
\providecommand\bibinfo[2]{#2}
\providecommand\natexlab[1]{#1}
\providecommand\showeprint[2][]{arXiv:#2}

\bibitem[\protect\citeauthoryear{Aigerman, Kovalsky, and Lipman}{Aigerman
  et~al\mbox{.}}{2017}]%
        {aigerman2017spherical}
\bibfield{author}{\bibinfo{person}{Noam Aigerman}, \bibinfo{person}{Shahar~Z
  Kovalsky}, {and} \bibinfo{person}{Yaron Lipman}.}
  \bibinfo{year}{2017}\natexlab{}.
\newblock \showarticletitle{Spherical orbifold {T}utte embeddings}.
\newblock \bibinfo{journal}{{\em ACM Transactions on Graphics (TOG)\/}}
  \bibinfo{volume}{36}, \bibinfo{number}{4} (\bibinfo{year}{2017}),
  \bibinfo{pages}{90}.
\newblock


\bibitem[\protect\citeauthoryear{Aigerman and Lipman}{Aigerman and
  Lipman}{2015}]%
        {aigerman2015orbifold}
\bibfield{author}{\bibinfo{person}{Noam Aigerman} {and} \bibinfo{person}{Yaron
  Lipman}.} \bibinfo{year}{2015}\natexlab{}.
\newblock \showarticletitle{Orbifold {T}utte Embeddings}.
\newblock \bibinfo{journal}{{\em ACM Transactions on Graphics (TOG)\/}}
  \bibinfo{volume}{34} (\bibinfo{year}{2015}).
\newblock


\bibitem[\protect\citeauthoryear{Aigerman and Lipman}{Aigerman and
  Lipman}{2016}]%
        {aigerman2016hyperbolic}
\bibfield{author}{\bibinfo{person}{Noam Aigerman} {and} \bibinfo{person}{Yaron
  Lipman}.} \bibinfo{year}{2016}\natexlab{}.
\newblock \showarticletitle{Hyperbolic Orbifold {T}utte Embeddings}.
\newblock \bibinfo{journal}{{\em ACM Transactions on Graphics (TOG)\/}}
  \bibinfo{volume}{35} (\bibinfo{year}{2016}).
\newblock


\bibitem[\protect\citeauthoryear{Aigerman, Poranne, and Lipman}{Aigerman
  et~al\mbox{.}}{2015}]%
        {aigerman2015seamless}
\bibfield{author}{\bibinfo{person}{Noam Aigerman}, \bibinfo{person}{Roi
  Poranne}, {and} \bibinfo{person}{Yaron Lipman}.}
  \bibinfo{year}{2015}\natexlab{}.
\newblock \showarticletitle{Seamless surface mappings}.
\newblock \bibinfo{journal}{{\em ACM Transactions on Graphics (TOG)\/}}
  \bibinfo{volume}{34}, \bibinfo{number}{4} (\bibinfo{year}{2015}),
  \bibinfo{pages}{72}.
\newblock


\bibitem[\protect\citeauthoryear{Aubry, Schlickewei, and Cremers}{Aubry
  et~al\mbox{.}}{2011}]%
        {aubry2011wave}
\bibfield{author}{\bibinfo{person}{Mathieu Aubry}, \bibinfo{person}{Ulrich
  Schlickewei}, {and} \bibinfo{person}{Daniel Cremers}.}
  \bibinfo{year}{2011}\natexlab{}.
\newblock \showarticletitle{The Wave Kernel Signature: A Quantum Mechanical
  Approach to Shape Analysis}. In \bibinfo{booktitle}{{\em International
  Conference on Computer Vision Workshops (ICCV Workshops)}}. IEEE.
\newblock


\bibitem[\protect\citeauthoryear{Bronstein, Bronstein, and Kimmel}{Bronstein
  et~al\mbox{.}}{2006}]%
        {bronstein2006generalized}
\bibfield{author}{\bibinfo{person}{Alexander~M Bronstein},
  \bibinfo{person}{Michael~M Bronstein}, {and} \bibinfo{person}{Ron Kimmel}.}
  \bibinfo{year}{2006}\natexlab{}.
\newblock \showarticletitle{Generalized multidimensional scaling: a framework
  for isometry-invariant partial surface matching}.
\newblock \bibinfo{journal}{{\em Proceedings of the National Academy of
  Sciences\/}} \bibinfo{volume}{103}, \bibinfo{number}{5}
  (\bibinfo{year}{2006}), \bibinfo{pages}{1168--1172}.
\newblock


\bibitem[\protect\citeauthoryear{Burghard, Dieckmann, and Klein}{Burghard
  et~al\mbox{.}}{2017}]%
        {burghard2017embedding}
\bibfield{author}{\bibinfo{person}{Oliver Burghard}, \bibinfo{person}{Alexander
  Dieckmann}, {and} \bibinfo{person}{Reinhard Klein}.}
  \bibinfo{year}{2017}\natexlab{}.
\newblock \showarticletitle{Embedding shapes with Green's functions for global
  shape matching}.
\newblock \bibinfo{journal}{{\em Computers \& Graphics\/}}
  \bibinfo{volume}{68} (\bibinfo{year}{2017}), \bibinfo{pages}{1--10}.
\newblock


\bibitem[\protect\citeauthoryear{Chao, Pinkall, Sanan, and Schr{\"o}der}{Chao
  et~al\mbox{.}}{2010}]%
        {chao2010simple}
\bibfield{author}{\bibinfo{person}{Isaac Chao}, \bibinfo{person}{Ulrich
  Pinkall}, \bibinfo{person}{Patrick Sanan}, {and} \bibinfo{person}{Peter
  Schr{\"o}der}.} \bibinfo{year}{2010}\natexlab{}.
\newblock \showarticletitle{A simple geometric model for elastic deformations}.
\newblock \bibinfo{journal}{{\em ACM Transactions on Graphics (TOG)\/}}
  \bibinfo{volume}{29}, \bibinfo{number}{4} (\bibinfo{year}{2010}),
  \bibinfo{pages}{38}.
\newblock


\bibitem[\protect\citeauthoryear{Chen, Golovinskiy, and Funkhouser}{Chen
  et~al\mbox{.}}{2009}]%
        {Chen:2009:ABF}
\bibfield{author}{\bibinfo{person}{Xiaobai Chen}, \bibinfo{person}{Aleksey
  Golovinskiy}, {and} \bibinfo{person}{Thomas Funkhouser}.}
  \bibinfo{year}{2009}\natexlab{}.
\newblock \showarticletitle{A Benchmark for {3D} Mesh Segmentation}.
\newblock \bibinfo{journal}{{\em ACM Transactions on Graphics (Proc.
  SIGGRAPH)\/}} \bibinfo{volume}{28}, \bibinfo{number}{3} (\bibinfo{date}{Aug.}
  \bibinfo{year}{2009}).
\newblock


\bibitem[\protect\citeauthoryear{Chen, Saparov, Pang, and Funkhouser}{Chen
  et~al\mbox{.}}{2012}]%
        {chen2012schelling}
\bibfield{author}{\bibinfo{person}{Xiaobai Chen}, \bibinfo{person}{Abulhair
  Saparov}, \bibinfo{person}{Bill Pang}, {and} \bibinfo{person}{Thomas
  Funkhouser}.} \bibinfo{year}{2012}\natexlab{}.
\newblock \showarticletitle{Schelling points on 3D surface meshes}.
\newblock \bibinfo{journal}{{\em ACM Transactions on Graphics (TOG)\/}}
  \bibinfo{volume}{31}, \bibinfo{number}{4} (\bibinfo{year}{2012}),
  \bibinfo{pages}{29}.
\newblock


\bibitem[\protect\citeauthoryear{Cox and Cox}{Cox and Cox}{2000}]%
        {cox2000multidimensional}
\bibfield{author}{\bibinfo{person}{Trevor~F Cox} {and}
  \bibinfo{person}{Michael~AA Cox}.} \bibinfo{year}{2000}\natexlab{}.
\newblock \bibinfo{booktitle}{{\em Multidimensional scaling}}.
\newblock \bibinfo{publisher}{CRC Press}.
\newblock


\bibitem[\protect\citeauthoryear{Eells and Sampson}{Eells and Sampson}{1964}]%
        {eells1964harmonic}
\bibfield{author}{\bibinfo{person}{James Eells} {and} \bibinfo{person}{Joseph~H
  Sampson}.} \bibinfo{year}{1964}\natexlab{}.
\newblock \showarticletitle{Harmonic mappings of {R}iemannian manifolds}.
\newblock \bibinfo{journal}{{\em American Journal of Mathematics\/}}
  \bibinfo{volume}{86}, \bibinfo{number}{1} (\bibinfo{year}{1964}),
  \bibinfo{pages}{109--160}.
\newblock


\bibitem[\protect\citeauthoryear{Ezuz and Ben-Chen}{Ezuz and Ben-Chen}{2017}]%
        {ezuz2017deblurring}
\bibfield{author}{\bibinfo{person}{Danielle Ezuz} {and} \bibinfo{person}{Mirela
  Ben-Chen}.} \bibinfo{year}{2017}\natexlab{}.
\newblock \showarticletitle{Deblurring and Denoising of Maps between Shapes}.
  In \bibinfo{booktitle}{{\em Computer Graphics Forum}},
  Vol.~\bibinfo{volume}{36}. \bibinfo{pages}{165--174}.
\newblock


\bibitem[\protect\citeauthoryear{Geman and Yang}{Geman and Yang}{1995}]%
        {geman1995nonlinear}
\bibfield{author}{\bibinfo{person}{Donald Geman} {and} \bibinfo{person}{Chengda
  Yang}.} \bibinfo{year}{1995}\natexlab{}.
\newblock \showarticletitle{Nonlinear image recovery with half-quadratic
  regularization}.
\newblock \bibinfo{journal}{{\em IEEE Transactions on Image Processing\/}}
  \bibinfo{volume}{4}, \bibinfo{number}{7} (\bibinfo{year}{1995}),
  \bibinfo{pages}{932--946}.
\newblock


\bibitem[\protect\citeauthoryear{Giorgi, Biasotti, and Paraboschi}{Giorgi
  et~al\mbox{.}}{2007}]%
        {shrec07}
\bibfield{author}{\bibinfo{person}{Daniela Giorgi}, \bibinfo{person}{Silvia
  Biasotti}, {and} \bibinfo{person}{Laura Paraboschi}.}
  \bibinfo{year}{2007}\natexlab{}.
\newblock \bibinfo{title}{{SHREC}: Shape Retrieval Contest: Watertight Models
  Track}.
\newblock   (\bibinfo{year}{2007}).
\newblock


\bibitem[\protect\citeauthoryear{Gu, Wang, Chan, Thompson, and Yau}{Gu
  et~al\mbox{.}}{2004}]%
        {gu2004genus}
\bibfield{author}{\bibinfo{person}{Xianfeng Gu}, \bibinfo{person}{Yalin Wang},
  \bibinfo{person}{Tony~F Chan}, \bibinfo{person}{Paul~M Thompson}, {and}
  \bibinfo{person}{Shing-Tung Yau}.} \bibinfo{year}{2004}\natexlab{}.
\newblock \showarticletitle{Genus zero surface conformal mapping and its
  application to brain surface mapping}.
\newblock \bibinfo{journal}{{\em Trans. on Medical Imaging\/}}
  \bibinfo{volume}{23}, \bibinfo{number}{8} (\bibinfo{year}{2004}),
  \bibinfo{pages}{949--958}.
\newblock


\bibitem[\protect\citeauthoryear{Heeren, Rumpf, Schr{\"{o}}der, Wardetzky, and
  Wirth}{Heeren et~al\mbox{.}}{2014}]%
        {HeRuSc14}
\bibfield{author}{\bibinfo{person}{Behrend Heeren}, \bibinfo{person}{Martin
  Rumpf}, \bibinfo{person}{Peter Schr{\"{o}}der}, \bibinfo{person}{Max
  Wardetzky}, {and} \bibinfo{person}{Benedikt Wirth}.}
  \bibinfo{year}{2014}\natexlab{}.
\newblock \showarticletitle{Exploring the Geometry of the Space of Shells}.
\newblock \bibinfo{journal}{{\em Computer Graphics Forum\/}}
  \bibinfo{volume}{33}, \bibinfo{number}{5} (\bibinfo{year}{2014}),
  \bibinfo{pages}{247--256}.
\newblock


\bibitem[\protect\citeauthoryear{Heeren, Rumpf, Wardetzky, and Wirth}{Heeren
  et~al\mbox{.}}{2012}]%
        {heeren2012time}
\bibfield{author}{\bibinfo{person}{Behrend Heeren}, \bibinfo{person}{Martin
  Rumpf}, \bibinfo{person}{Max Wardetzky}, {and} \bibinfo{person}{Benedikt
  Wirth}.} \bibinfo{year}{2012}\natexlab{}.
\newblock \showarticletitle{Time-Discrete Geodesics in the Space of Shells}.
\newblock \bibinfo{journal}{{\em Computer Graphics Forum\/}}
  \bibinfo{volume}{31} (\bibinfo{year}{2012}).
\newblock


\bibitem[\protect\citeauthoryear{Hormann and Greiner}{Hormann and
  Greiner}{2000}]%
        {hormann2000mips}
\bibfield{author}{\bibinfo{person}{Kai Hormann} {and}
  \bibinfo{person}{G{\"u}nther Greiner}.} \bibinfo{year}{2000}\natexlab{}.
\newblock \bibinfo{booktitle}{{\em {MIPS}: An Efficient Global Parametrization
  Method}}.
\newblock \bibinfo{type}{{T}echnical {R}eport}. \bibinfo{institution}{DTIC
  Document}.
\newblock


\bibitem[\protect\citeauthoryear{Huang and Ovsjanikov}{Huang and
  Ovsjanikov}{2017}]%
        {huang2017adjoint}
\bibfield{author}{\bibinfo{person}{Ruqi Huang} {and} \bibinfo{person}{Maks
  Ovsjanikov}.} \bibinfo{year}{2017}\natexlab{}.
\newblock \showarticletitle{Adjoint Map Representation for Shape Analysis and
  Matching}. In \bibinfo{booktitle}{{\em Computer Graphics Forum}},
  Vol.~\bibinfo{volume}{36}. \bibinfo{pages}{151--163}.
\newblock


\bibitem[\protect\citeauthoryear{Izeki and Nayatani}{Izeki and
  Nayatani}{2005}]%
        {izeki2005combinatorial}
\bibfield{author}{\bibinfo{person}{Hiroyasu Izeki} {and} \bibinfo{person}{Shin
  Nayatani}.} \bibinfo{year}{2005}\natexlab{}.
\newblock \showarticletitle{Combinatorial harmonic maps and discrete-group
  actions on {H}adamard spaces}.
\newblock \bibinfo{journal}{{\em Geometriae Dedicata\/}} \bibinfo{volume}{114},
  \bibinfo{number}{1} (\bibinfo{year}{2005}), \bibinfo{pages}{147--188}.
\newblock


\bibitem[\protect\citeauthoryear{Kalogerakis, Hertzmann, and Singh}{Kalogerakis
  et~al\mbox{.}}{2010}]%
        {Kalogerakis:2010:labelMeshes}
\bibfield{author}{\bibinfo{person}{Evangelos Kalogerakis},
  \bibinfo{person}{Aaron Hertzmann}, {and} \bibinfo{person}{Karan Singh}.}
  \bibinfo{year}{2010}\natexlab{}.
\newblock \showarticletitle{{L}earning {3}{D} mesh segmentation and labeling}.
\newblock \bibinfo{journal}{{\em ACM Transactions on Graphics\/}}
  \bibinfo{volume}{29}, \bibinfo{number}{3} (\bibinfo{year}{2010}).
\newblock


\bibitem[\protect\citeauthoryear{Kim, Lipman, and Funkhouser}{Kim
  et~al\mbox{.}}{2011}]%
        {kim2011blended}
\bibfield{author}{\bibinfo{person}{Vladimir~G Kim}, \bibinfo{person}{Yaron
  Lipman}, {and} \bibinfo{person}{Thomas Funkhouser}.}
  \bibinfo{year}{2011}\natexlab{}.
\newblock \showarticletitle{{Blended Intrinsic Maps}}.
\newblock \bibinfo{journal}{{\em {ACM Transactions on Graphics (TOG)}\/}}
  \bibinfo{volume}{30} (\bibinfo{year}{2011}).
\newblock


\bibitem[\protect\citeauthoryear{Kovnatsky, Bronstein, Bresson, and
  Vandergheynst}{Kovnatsky et~al\mbox{.}}{2015}]%
        {kovnatsky2015functional}
\bibfield{author}{\bibinfo{person}{Artiom Kovnatsky},
  \bibinfo{person}{Michael~M Bronstein}, \bibinfo{person}{Xavier Bresson},
  {and} \bibinfo{person}{Pierre Vandergheynst}.}
  \bibinfo{year}{2015}\natexlab{}.
\newblock \showarticletitle{Functional Correspondence by Matrix Completion}. In
  \bibinfo{booktitle}{{\em Proceedings of Computer Vision and Pattern
  Recognition (CVPR)}}.
\newblock


\bibitem[\protect\citeauthoryear{Kovnatsky, Bronstein, Bronstein, Glashoff, and
  Kimmel}{Kovnatsky et~al\mbox{.}}{2013}]%
        {kovnatsky2013coupled}
\bibfield{author}{\bibinfo{person}{Artiom Kovnatsky},
  \bibinfo{person}{Michael~M Bronstein}, \bibinfo{person}{Alexander~M
  Bronstein}, \bibinfo{person}{Klaus Glashoff}, {and} \bibinfo{person}{Ron
  Kimmel}.} \bibinfo{year}{2013}\natexlab{}.
\newblock \showarticletitle{Coupled Quasi-harmonic Bases}.
\newblock \bibinfo{journal}{{\em Computer Graphics Forum\/}}
  \bibinfo{volume}{32} (\bibinfo{year}{2013}).
\newblock


\bibitem[\protect\citeauthoryear{L{\"{a}}hner, Vestner, Boyarski, Litany,
  Slossberg, Remez, Rodol{\`{a}}, Bronstein, Bronstein, Kimmel, and
  Cremers}{L{\"{a}}hner et~al\mbox{.}}{2017}]%
        {LahnerVBLSRRBBK17}
\bibfield{author}{\bibinfo{person}{Zorah L{\"{a}}hner},
  \bibinfo{person}{Matthias Vestner}, \bibinfo{person}{Amit Boyarski},
  \bibinfo{person}{Or Litany}, \bibinfo{person}{Ron Slossberg},
  \bibinfo{person}{Tal Remez}, \bibinfo{person}{Emanuele Rodol{\`{a}}},
  \bibinfo{person}{Alexander~M. Bronstein}, \bibinfo{person}{Michael~M.
  Bronstein}, \bibinfo{person}{Ron Kimmel}, {and} \bibinfo{person}{Daniel
  Cremers}.} \bibinfo{year}{2017}\natexlab{}.
\newblock \showarticletitle{Efficient Deformable Shape Correspondence via
  Kernel Matching}. In \bibinfo{booktitle}{{\em 3D Vision (3DV)}}.
\newblock


\bibitem[\protect\citeauthoryear{L{\'e}vy, Petitjean, Ray, and
  Maillot}{L{\'e}vy et~al\mbox{.}}{2002}]%
        {levy2002least}
\bibfield{author}{\bibinfo{person}{Bruno L{\'e}vy}, \bibinfo{person}{Sylvain
  Petitjean}, \bibinfo{person}{Nicolas Ray}, {and} \bibinfo{person}{J{\'e}rome
  Maillot}.} \bibinfo{year}{2002}\natexlab{}.
\newblock \showarticletitle{Least squares conformal maps for automatic texture
  atlas generation}. In \bibinfo{booktitle}{{\em ACM Transactions on Graphics
  (TOG)}}, Vol.~\bibinfo{volume}{21}. ACM, \bibinfo{pages}{362--371}.
\newblock


\bibitem[\protect\citeauthoryear{Mandad, Cohen-Steiner, Kobbelt, Alliez, and
  Desbrun}{Mandad et~al\mbox{.}}{2017}]%
        {mandad2017variance}
\bibfield{author}{\bibinfo{person}{Manish Mandad}, \bibinfo{person}{David
  Cohen-Steiner}, \bibinfo{person}{Leif Kobbelt}, \bibinfo{person}{Pierre
  Alliez}, {and} \bibinfo{person}{Mathieu Desbrun}.}
  \bibinfo{year}{2017}\natexlab{}.
\newblock \showarticletitle{Variance-Minimizing Transport Plans for
  Inter-surface Mapping}.
\newblock \bibinfo{journal}{{\em ACM Transactions on Graphics\/}}
  \bibinfo{volume}{36} (\bibinfo{year}{2017}), \bibinfo{pages}{14}.
\newblock


\bibitem[\protect\citeauthoryear{Maron, Dym, Kezurer, Kovalsky, and
  Lipman}{Maron et~al\mbox{.}}{2016}]%
        {maron2016point}
\bibfield{author}{\bibinfo{person}{Haggai Maron}, \bibinfo{person}{Nadav Dym},
  \bibinfo{person}{Itay Kezurer}, \bibinfo{person}{Shahar Kovalsky}, {and}
  \bibinfo{person}{Yaron Lipman}.} \bibinfo{year}{2016}\natexlab{}.
\newblock \showarticletitle{Point Registration Via Efficient Convex
  Relaxation}.
\newblock \bibinfo{journal}{{\em ACM Transactions on Graphics (TOG)\/}}
  \bibinfo{volume}{35} (\bibinfo{year}{2016}).
\newblock


\bibitem[\protect\citeauthoryear{Munsell, Dalal, and Wang}{Munsell
  et~al\mbox{.}}{2008}]%
        {munsell2008evaluating}
\bibfield{author}{\bibinfo{person}{Brent~C Munsell}, \bibinfo{person}{Pahal
  Dalal}, {and} \bibinfo{person}{Song Wang}.} \bibinfo{year}{2008}\natexlab{}.
\newblock \showarticletitle{Evaluating shape correspondence for statistical
  shape analysis: A benchmark study}.
\newblock \bibinfo{journal}{{\em IEEE Transactions on Pattern Analysis and
  Machine Intelligence\/}} \bibinfo{volume}{30}, \bibinfo{number}{11}
  (\bibinfo{year}{2008}), \bibinfo{pages}{2023--2039}.
\newblock


\bibitem[\protect\citeauthoryear{Nishikawa}{Nishikawa}{2000}]%
        {nishikawa-2000}
\bibfield{author}{\bibinfo{person}{S. Nishikawa}.}
  \bibinfo{year}{2000}\natexlab{}.
\newblock \bibinfo{booktitle}{{\em Variational Problems in Geometry}}.
\newblock \bibinfo{publisher}{AMS}.
\newblock


\bibitem[\protect\citeauthoryear{Nogneng and Ovsjanikov}{Nogneng and
  Ovsjanikov}{2017}]%
        {nogneng2017informative}
\bibfield{author}{\bibinfo{person}{Dorian Nogneng} {and} \bibinfo{person}{Maks
  Ovsjanikov}.} \bibinfo{year}{2017}\natexlab{}.
\newblock \showarticletitle{Informative descriptor preservation via
  commutativity for shape matching}. In \bibinfo{booktitle}{{\em Computer
  Graphics Forum}}, Vol.~\bibinfo{volume}{36}. \bibinfo{pages}{259--267}.
\newblock


\bibitem[\protect\citeauthoryear{Ovsjanikov, Ben-Chen, Solomon, Butscher, and
  Guibas}{Ovsjanikov et~al\mbox{.}}{2012}]%
        {ovsjanikov2012functional}
\bibfield{author}{\bibinfo{person}{Maks Ovsjanikov}, \bibinfo{person}{Mirela
  Ben-Chen}, \bibinfo{person}{Justin Solomon}, \bibinfo{person}{Adrian
  Butscher}, {and} \bibinfo{person}{Leonidas Guibas}.}
  \bibinfo{year}{2012}\natexlab{}.
\newblock \showarticletitle{Functional Maps: a Flexible Representation of Maps
  between Shapes}.
\newblock \bibinfo{journal}{{\em ACM Transactions on Graphics (TOG)\/}}
  \bibinfo{volume}{31} (\bibinfo{year}{2012}).
\newblock


\bibitem[\protect\citeauthoryear{Panozzo, Baran, Diamanti, and
  Sorkine-Hornung}{Panozzo et~al\mbox{.}}{2013}]%
        {panozzo2013weighted}
\bibfield{author}{\bibinfo{person}{Daniele Panozzo}, \bibinfo{person}{Ilya
  Baran}, \bibinfo{person}{Olga Diamanti}, {and} \bibinfo{person}{Olga
  Sorkine-Hornung}.} \bibinfo{year}{2013}\natexlab{}.
\newblock \showarticletitle{Weighted averages on surfaces}.
\newblock \bibinfo{journal}{{\em ACM Transactions on Graphics (TOG)\/}}
  \bibinfo{volume}{32}, \bibinfo{number}{4} (\bibinfo{year}{2013}),
  \bibinfo{pages}{60}.
\newblock


\bibitem[\protect\citeauthoryear{Pinkall and Polthier}{Pinkall and
  Polthier}{1993}]%
        {pinkall1993computing}
\bibfield{author}{\bibinfo{person}{Ulrich Pinkall} {and}
  \bibinfo{person}{Konrad Polthier}.} \bibinfo{year}{1993}\natexlab{}.
\newblock \showarticletitle{Computing discrete minimal surfaces and their
  conjugates}.
\newblock \bibinfo{journal}{{\em Experimental Mathematics\/}}
  \bibinfo{volume}{2}, \bibinfo{number}{1} (\bibinfo{year}{1993}),
  \bibinfo{pages}{15--36}.
\newblock


\bibitem[\protect\citeauthoryear{Rodol{\`a}, Moeller, and Cremers}{Rodol{\`a}
  et~al\mbox{.}}{2015}]%
        {rodola2015point}
\bibfield{author}{\bibinfo{person}{Emanuele Rodol{\`a}},
  \bibinfo{person}{Michael Moeller}, {and} \bibinfo{person}{Daniel Cremers}.}
  \bibinfo{year}{2015}\natexlab{}.
\newblock \showarticletitle{Point-wise Map Recovery and Refinement from
  Functional Correspondence}. In \bibinfo{booktitle}{{\em Proceedings of
  Vision, Modeling and Visualization (VMV)}}.
\newblock


\bibitem[\protect\citeauthoryear{Sahillio{\u g}lu and Yemez}{Sahillio{\u g}lu
  and Yemez}{2011}]%
        {sahillioǧlu2011coarse}
\bibfield{author}{\bibinfo{person}{Y Sahillio{\u g}lu} {and}
  \bibinfo{person}{Y{\"u}cel Yemez}.} \bibinfo{year}{2011}\natexlab{}.
\newblock \showarticletitle{Coarse-to-Fine Combinatorial Matching for Dense
  Isometric Shape Correspondence}. In \bibinfo{booktitle}{{\em Computer
  Graphics Forum}}, Vol.~\bibinfo{volume}{30}. \bibinfo{pages}{1461--1470}.
\newblock


\bibitem[\protect\citeauthoryear{Sela, Aflalo, and Kimmel}{Sela
  et~al\mbox{.}}{2015}]%
        {Sela20151}
\bibfield{author}{\bibinfo{person}{Matan Sela}, \bibinfo{person}{Yonathan
  Aflalo}, {and} \bibinfo{person}{Ron Kimmel}.}
  \bibinfo{year}{2015}\natexlab{}.
\newblock \showarticletitle{Computational caricaturization of surfaces}.
\newblock \bibinfo{journal}{{\em Computer Vision and Image Understanding\/}}
  \bibinfo{volume}{141} (\bibinfo{year}{2015}), \bibinfo{pages}{1 -- 17}.
\newblock
\showISSN{1077-3142}


\bibitem[\protect\citeauthoryear{Shi, Zeng, Su, Jiang, Damasio, Lu, Wang, Yau,
  and Gu}{Shi et~al\mbox{.}}{2017}]%
        {shi2017hyperbolic}
\bibfield{author}{\bibinfo{person}{Rui Shi}, \bibinfo{person}{Wei Zeng},
  \bibinfo{person}{Zhengyu Su}, \bibinfo{person}{Jian Jiang},
  \bibinfo{person}{Hanna Damasio}, \bibinfo{person}{Zhonglin Lu},
  \bibinfo{person}{Yalin Wang}, \bibinfo{person}{Shing-Tung Yau}, {and}
  \bibinfo{person}{Xianfeng Gu}.} \bibinfo{year}{2017}\natexlab{}.
\newblock \showarticletitle{Hyperbolic harmonic mapping for surface
  registration}.
\newblock \bibinfo{journal}{{\em IEEE Transactions on Pattern Analysis and
  Machine Intelligence\/}} \bibinfo{volume}{39}, \bibinfo{number}{5}
  (\bibinfo{year}{2017}), \bibinfo{pages}{965--980}.
\newblock


\bibitem[\protect\citeauthoryear{Shtern and Kimmel}{Shtern and Kimmel}{2014}]%
        {shtern2014iterative}
\bibfield{author}{\bibinfo{person}{Alon Shtern} {and} \bibinfo{person}{Ron
  Kimmel}.} \bibinfo{year}{2014}\natexlab{}.
\newblock \showarticletitle{Iterative Closest Spectral Kernel Maps}. In
  \bibinfo{booktitle}{{\em 3D Vision (3DV)}}. IEEE.
\newblock


\bibitem[\protect\citeauthoryear{Solomon, Peyr{\'e}, Kim, and Sra}{Solomon
  et~al\mbox{.}}{2016}]%
        {solomon2016entropic}
\bibfield{author}{\bibinfo{person}{Justin Solomon}, \bibinfo{person}{Gabriel
  Peyr{\'e}}, \bibinfo{person}{Vladimir~G. Kim}, {and} \bibinfo{person}{Suvrit
  Sra}.} \bibinfo{year}{2016}\natexlab{}.
\newblock \showarticletitle{Entropic Metric Alignment for Correspondence
  Problems}.
\newblock \bibinfo{journal}{{\em ACM Transactions on Graphics (TOG)\/}}
  \bibinfo{volume}{35} (\bibinfo{year}{2016}).
\newblock


\bibitem[\protect\citeauthoryear{Sorkine and Alexa}{Sorkine and Alexa}{2007}]%
        {sorkine2007rigid}
\bibfield{author}{\bibinfo{person}{Olga Sorkine} {and} \bibinfo{person}{Marc
  Alexa}.} \bibinfo{year}{2007}\natexlab{}.
\newblock \showarticletitle{As-rigid-as-possible surface modeling}. In
  \bibinfo{booktitle}{{\em Computer Graphics Forum}}, Vol.~\bibinfo{volume}{4}.
\newblock


\bibitem[\protect\citeauthoryear{Sumner and Popovi{\'c}}{Sumner and
  Popovi{\'c}}{2004}]%
        {sumner2004deformation}
\bibfield{author}{\bibinfo{person}{Robert~W Sumner} {and}
  \bibinfo{person}{Jovan Popovi{\'c}}.} \bibinfo{year}{2004}\natexlab{}.
\newblock \showarticletitle{Deformation Transfer for Triangle Meshes}.
\newblock \bibinfo{journal}{{\em ACM Transactions on Graphics (TOG)\/}}
  \bibinfo{volume}{23} (\bibinfo{year}{2004}).
\newblock


\bibitem[\protect\citeauthoryear{Tam, Cheng, Lai, Langbein, Liu, Marshall,
  Martin, Sun, and Rosin}{Tam et~al\mbox{.}}{2013}]%
        {tam2013registration}
\bibfield{author}{\bibinfo{person}{Gary~KL Tam}, \bibinfo{person}{Zhi-Quan
  Cheng}, \bibinfo{person}{Yu-Kun Lai}, \bibinfo{person}{Frank~C Langbein},
  \bibinfo{person}{Yonghuai Liu}, \bibinfo{person}{David Marshall},
  \bibinfo{person}{Ralph~R Martin}, \bibinfo{person}{Xian-Fang Sun}, {and}
  \bibinfo{person}{Paul~L Rosin}.} \bibinfo{year}{2013}\natexlab{}.
\newblock \showarticletitle{Registration of 3D point clouds and meshes: a
  survey from rigid to nonrigid}.
\newblock \bibinfo{journal}{{\em IEEE Transactions on Visualization and
  Computer Graphics\/}} \bibinfo{volume}{19}, \bibinfo{number}{7}
  (\bibinfo{year}{2013}), \bibinfo{pages}{1199--1217}.
\newblock


\bibitem[\protect\citeauthoryear{Tsui, Fenton, Vuong, Hass, Koehl, Amenta,
  Coeurjolly, DeCarli, and Carmichael}{Tsui et~al\mbox{.}}{2013}]%
        {tsui2013globally}
\bibfield{author}{\bibinfo{person}{Alex Tsui}, \bibinfo{person}{Devin Fenton},
  \bibinfo{person}{Phong Vuong}, \bibinfo{person}{Joel Hass},
  \bibinfo{person}{Patrice Koehl}, \bibinfo{person}{Nina Amenta},
  \bibinfo{person}{David Coeurjolly}, \bibinfo{person}{Charles DeCarli}, {and}
  \bibinfo{person}{Owen Carmichael}.} \bibinfo{year}{2013}\natexlab{}.
\newblock \showarticletitle{Globally optimal cortical surface matching with
  exact landmark correspondence}. In \bibinfo{booktitle}{{\em Information
  Processing in Medical Imaging}}, Vol.~\bibinfo{volume}{23}. NIH Public
  Access, \bibinfo{pages}{487}.
\newblock


\bibitem[\protect\citeauthoryear{Urakawa}{Urakawa}{1993}]%
        {urakawa-1993}
\bibfield{author}{\bibinfo{person}{H. Urakawa}.}
  \bibinfo{year}{1993}\natexlab{}.
\newblock \bibinfo{booktitle}{{\em Calculus of Variations and Harmonic Maps}}.
\newblock \bibinfo{publisher}{AMS}.
\newblock


\bibitem[\protect\citeauthoryear{Van~Kaick, Zhang, Hamarneh, and
  Cohen-Or}{Van~Kaick et~al\mbox{.}}{2011}]%
        {van2011survey}
\bibfield{author}{\bibinfo{person}{Oliver Van~Kaick}, \bibinfo{person}{Hao
  Zhang}, \bibinfo{person}{Ghassan Hamarneh}, {and} \bibinfo{person}{Daniel
  Cohen-Or}.} \bibinfo{year}{2011}\natexlab{}.
\newblock \showarticletitle{A survey on shape correspondence}. In
  \bibinfo{booktitle}{{\em Computer Graphics Forum}},
  Vol.~\bibinfo{volume}{30}. \bibinfo{pages}{1681--1707}.
\newblock


\bibitem[\protect\citeauthoryear{Vestner, Litman, Rodol{\`a}, Bronstein, and
  Cremers}{Vestner et~al\mbox{.}}{2017}]%
        {vestner2017pmf}
\bibfield{author}{\bibinfo{person}{M. Vestner}, \bibinfo{person}{R. Litman},
  \bibinfo{person}{E. Rodol{\`a}}, \bibinfo{person}{A. Bronstein}, {and}
  \bibinfo{person}{D. Cremers}.} \bibinfo{year}{2017}\natexlab{}.
\newblock \showarticletitle{Product Manifold Filter: Non-Rigid Shape
  Correspondence via Kernel Density Estimation in the Product Space}. In
  \bibinfo{booktitle}{{\em Proceedings of Computer Vision and Pattern
  Recognition (CVPR)}}.
\newblock


\bibitem[\protect\citeauthoryear{Von-Tycowicz, Schulz, Seidel, and
  Hildebrandt}{Von-Tycowicz et~al\mbox{.}}{2015}]%
        {von2015real}
\bibfield{author}{\bibinfo{person}{Christoph Von-Tycowicz},
  \bibinfo{person}{Christian Schulz}, \bibinfo{person}{Hans-Peter Seidel},
  {and} \bibinfo{person}{Klaus Hildebrandt}.} \bibinfo{year}{2015}\natexlab{}.
\newblock \showarticletitle{Real-time Nonlinear Shape Interpolation}.
\newblock \bibinfo{journal}{{\em ACM Transactions on Graphics (TOG)\/}}
  \bibinfo{volume}{34} (\bibinfo{year}{2015}).
\newblock


\bibitem[\protect\citeauthoryear{Wang, Yang, Yin, and Zhang}{Wang
  et~al\mbox{.}}{2008}]%
        {wang2008new}
\bibfield{author}{\bibinfo{person}{Yilun Wang}, \bibinfo{person}{Junfeng Yang},
  \bibinfo{person}{Wotao Yin}, {and} \bibinfo{person}{Yin Zhang}.}
  \bibinfo{year}{2008}\natexlab{}.
\newblock \showarticletitle{A new alternating minimization algorithm for total
  variation image reconstruction}.
\newblock \bibinfo{journal}{{\em SIAM Journal on Imaging Sciences\/}}
  \bibinfo{volume}{1}, \bibinfo{number}{3} (\bibinfo{year}{2008}),
  \bibinfo{pages}{248--272}.
\newblock


\bibitem[\protect\citeauthoryear{Xu, Kim, Huang, Mitra, and Kalogerakis}{Xu
  et~al\mbox{.}}{2016}]%
        {xu2016data}
\bibfield{author}{\bibinfo{person}{Kai Xu}, \bibinfo{person}{Vladimir~G Kim},
  \bibinfo{person}{Qixing Huang}, \bibinfo{person}{Niloy Mitra}, {and}
  \bibinfo{person}{Evangelos Kalogerakis}.} \bibinfo{year}{2016}\natexlab{}.
\newblock \showarticletitle{Data-driven shape analysis and processing}. In
  \bibinfo{booktitle}{{\em SIGGRAPH ASIA 2016 Courses}}. ACM,
  \bibinfo{pages}{4}.
\newblock


\bibitem[\protect\citeauthoryear{Xu and Yin}{Xu and Yin}{2013}]%
        {xu2013block}
\bibfield{author}{\bibinfo{person}{Yangyang Xu} {and} \bibinfo{person}{Wotao
  Yin}.} \bibinfo{year}{2013}\natexlab{}.
\newblock \showarticletitle{A block coordinate descent method for regularized
  multiconvex optimization with applications to nonnegative tensor
  factorization and completion}.
\newblock \bibinfo{journal}{{\em SIAM Journal on Imaging Sciences\/}}
  \bibinfo{volume}{6}, \bibinfo{number}{3} (\bibinfo{year}{2013}),
  \bibinfo{pages}{1758--1789}.
\newblock


\bibitem[\protect\citeauthoryear{Zheng, Wen, Lei, Ma, and Gu}{Zheng
  et~al\mbox{.}}{2017}]%
        {zheng2017surface}
\bibfield{author}{\bibinfo{person}{Xiaopeng Zheng}, \bibinfo{person}{Chengfeng
  Wen}, \bibinfo{person}{Na Lei}, \bibinfo{person}{Ming Ma}, {and}
  \bibinfo{person}{Xianfeng Gu}.} \bibinfo{year}{2017}\natexlab{}.
\newblock \showarticletitle{Surface Registration via Foliation}. In
  \bibinfo{booktitle}{{\em Proceedings of the IEEE Conference on Computer
  Vision and Pattern Recognition}}. \bibinfo{pages}{938--947}.
\newblock


\bibitem[\protect\citeauthoryear{Zoran and Weiss}{Zoran and Weiss}{2011}]%
        {zoran2011learning}
\bibfield{author}{\bibinfo{person}{Daniel Zoran} {and} \bibinfo{person}{Yair
  Weiss}.} \bibinfo{year}{2011}\natexlab{}.
\newblock \showarticletitle{From learning models of natural image patches to
  whole image restoration}. In \bibinfo{booktitle}{{\em IEEE International
  Conference on Computer Vision (ICCV)}}. IEEE, \bibinfo{pages}{479--486}.
\newblock


\end{thebibliography}
	
	\pagebreak
	
	\appendix{Appendix A. Reversibility}
	\label{sec:appendix}
	\begin{prop}
		
		Let $(M_1,g_1),(M_2,g_2)$ be two smooth compact Riemannian surfaces, with geodesic distance functions $d_{M_i}(\cdot,\cdot)$ given by the metrics $g_i$, respectively, and let $\phi_{12}\!:\!M_1\!\to\!M_2$, $\phi_{21}\!:\!M_2\!\to\!M_1$ be smooth maps. 	
		If there exists $\epsilon \geq 0$ such that:
		\begin{equation*}
		d_{M_1} \left( p_1, \phi_{21} \left( \phi_{12} \left( p_1 \right) \right) \right) \leq \epsilon, \quad \forall p_1\in M_1, 
		\end{equation*}
		then:
		\begin{enumerate}
			\item 
			$\phi_{12} \left( p_1 \right) = \phi_{12} \left( q_1 \right) \Rightarrow d_{M_1} \left( p_1, q_1 \right) \leq 2\epsilon, \quad \forall p_1,q_1\!\in\!M_1$.
			
			\item 
			$\forall p_1\!\in\!M_1 \,\, \exists p_2\!\in\!M_2 \quad \text{s.t.} \quad  d_{M_1} \left( p_1, \phi_{21} \left( p_2 \right) \right) \leq \epsilon.$
			
		\end{enumerate} 	
	\end{prop}
	As a corollary of Proposition 1 we have that if the reversibility energy defined in Equation~\eqref{eq:consistency_energy} is zero, then the maps $\phi_{12},\phi_{21}$ are both injective and surjective. 
	
	\paragraph*{Proof.}
	Let $p_1,q_1\!\in\!M_1$, and set $p_2\!=\!\phi_{12}\left( p_1 \right), q_2\!=\!\phi_{12} \left( q_1 \right)$. Further, set $\hat{p}_1\!=\!\phi_{21} \left(p_2 \right)$, and $\hat{q}_1\! = \!\phi_{21} \left(q_2 \right)$. 
	\begin{enumerate}
		\item From the triangle inequality we have that $d_{M_1}(p_1,q_1) \leq d_{M_1}(p_1,\hat{q}_1) + d_{M_1}(\hat{q}_1,q_1)$. From the assumption (1) we have that $p_2 = q_2$ and therefore $\hat{p}_1\! =\! \hat{q}_1$. Thus, we have $d_{M_1}(p_1,\hat{q}_1)\! =\! d_{M_1}(p_1,\hat{p}_1)\! \leq\! \epsilon$ and $d_{M_1}(\hat{q}_1,q_1)\! \leq\! \epsilon$, which gives the required result.
		
		\item This follows trivially from the assumption of the Proposition if we set $p_2 = \phi_{12}(p_1)$.
		
	\end{enumerate}

\end{document}